\documentclass[article]{aa}

\usepackage{graphicx}
\usepackage{mwe}
\usepackage{longtable}
\usepackage{siunitx}
\usepackage[table]{xcolor}
\usepackage{textcomp}
\usepackage{chemist}
\usepackage{multirow}
\usepackage{caption}
\usepackage{subcaption}
\usepackage[flushleft]{threeparttable}
\usepackage{float}
\usepackage{txfonts}
\usepackage{array}
\usepackage{siunitx}

\usepackage{natbib,twoopt}

\usepackage[colorlinks=true, pdfstartview=FitV, linkcolor=blue, citecolor=blue, urlcolor=blue, breaklinks=true]{hyperref} 
\bibpunct{(}{)}{;}{a}{}{,}             
\makeatletter
  \newcommandtwoopt{\citeads}[3][][]{\href{http://adsabs.harvard.edu/abs/#3}%
    {\def\hyper@linkstart##1##2{}%
     \let\hyper@linkend\@empty\citealp[#1][#2]{#3}}}
  \newcommandtwoopt{\citepads}[3][][]{\href{http://adsabs.harvard.edu/abs/#3}%
    {\def\hyper@linkstart##1##2{}%
     \let\hyper@linkend\@empty\citep[#1][#2]{#3}}}
  \newcommandtwoopt{\citetads}[3][][]{\href{http://adsabs.harvard.edu/abs/#3}%
    {\def\hyper@linkstart##1##2{}%
     \let\hyper@linkend\@empty\citet[#1][#2]{#3}}}
  \newcommandtwoopt{\citeyearads}[3][][]%
    {\href{http://adsabs.harvard.edu/abs/#3}
    {\def\hyper@linkstart##1##2{}%
     \let\hyper@linkend\@empty\citeyear[#1][#2]{#3}}}
\makeatother

\usepackage{color}

\usepackage{color}

\usepackage{cleveref}

\usepackage{chemist}
\usepackage{orcidlink}
\begin{document}
\title{Circumstellar emission of Cepheids across the instability strip:\\
Mid-infrared observations with VLTI/MATISSE\thanks{Based on observations made with ESO telescopes at Paranal observatory  under  program  104.D-0554(B), 106.21RL.001, 106.21RL.002, 106.21RL.003.} }

\titlerunning{Cepheids across the instability strip with VLTI/MATISSE}
\authorrunning{Hocd\'e et al. }
\author{V.~Hocd\'e \inst{1}
\orcidlink{0000-0002-3643-0366}
\and A.~Matter \inst{2}
\and N.~Nardetto \inst{2}
\and A.~Gallenne \inst{3,4}
\and P.~Kervella \inst{5}
\and A.~M\'erand \inst{6}
\and G.~Pietrzyński \inst{1}
\and W.~Gieren \inst{7}
\and J.~Leftley \inst{2}
\and S.~Robbe-Dubois \inst{2}
\and B.~Lopez \inst{2}
\and M.~C.~Bailleul \inst{2}
\and G.~Bras \inst{5}
\and R.~Smolec \inst{1}
\and P.~Wielgórski \inst{1}
\and G.~Hajdu \inst{1}
\and A.~Afanasiev \inst{5}
}

\institute{Nicolaus Copernicus Astronomical Centre, Polish Academy of Sciences, Bartycka 18, 00-716 Warszawa, Poland\\
email : \texttt{vhocde@camk.edu.pl}
\and Universit\'e Côte d'Azur, Observatoire de la C\^ote d'Azur, CNRS, Laboratoire Lagrange, France,
\and Instituto de Astrofísica, Departamento de Ciencias Físicas, Facultad de Ciencias Exactas, Universidad Andrés Bello, Fernández
Concha 700, Las Condes, Santiago, Chile
\and French-Chilean Laboratory for Astronomy, IRL 3386, CNRS, Casilla 36-D, Santiago, Chile
\and LESIA, Observatoire de Paris, Université PSL, CNRS,  Sorbonne Université, Université Paris-Cité, 5 Place Jules Janssen,92195 Meudon, France,
\and European Southern Observatory, Karl-Schwarzschild-Str. 2, 85748 Garching, Germany
\and Universidad de Concepcion, Departamento de Astronomia, Casilla 160C, Chile
}

\date{Received 31 October 2024 ; Accepted 10 December 2024}
\abstract{The circumstellar envelopes (CSE) of Cepheids are still not well characterized despite their potential impact on distance determination via both the period-luminosity relation and the parallax-of-pulsation method.}{This paper aims to investigate Galactic Cepheids across the instability strip in the mid-infrared with MATISSE/VLTI in order to constrain the geometry and physical nature (gas and/or dust) of their CSEs.}{
We secured observations of eight Galactic Cepheids from short up to long period of pulsation, with MATISSE/VLTI in $L$, $M$ and $N$-bands. For each star we calibrate the flux measurements to potentially detect dust spectral signature in the spectral energy distribution (SED). We then analyze the closure phase and the visibilities in $L$, $M$ and $N$-bands. The parallax-of-pulsation code \texttt{SPIPS} is used in order to derive the infrared (IR) excess and the expected star angular diameter at the date of MATISSE observations. We also computed test cases of radiative transfer model of dusty envelopes with \texttt{DUSTY} to compare with the visibilities in the $N$-band.} {The SED analysis in the mid-IR confirms the absence of dust spectral signature for all the star sample. For each star in $L$, $M$ and $N$-band we observe closure phases which are consistent with centro-symmetric geometry for the different targets. Finally, the visibilities in $L$, $M$ and $N$ bands are in agreement with the expected star angular diameter. Although we do not resolve any circumstellar emission, the observations are compatible with the presence of compact CSEs within the uncertainties. We provide 2$\,\sigma$ upper limits on the CSE flux contribution based on model residuals for several CSE radius, which yield to exclude models simultaneously large and bright ($R_\mathrm{CSE}\approx10\,R_\star$ and $f_\mathrm{CSE}\approx10\%$) for all the stars of the sample. Last, the visibilities in the $N$-band rule out CSE models with significant amount of different type of dust with an optical depth $\tau_V \gtrsim 0.001$.}
{The MATISSE observations of eight Cepheids with different pulsation period (from 7 up to 38$\,$day) and evolution stage, provide for the first time a comprehensive picture of Cepheids from mid-IR interferometry. We present additional evidences that circumstellar dust emission is negligible or absent around Cepheids for a wide range of stellar parameters in the instability strip. Further interferometric observations in the visible and the near-infrared will be necessary to disentangle the star and the CSE, which is crucial to constrain the CSE contribution and its possible gaseous nature.}

\maketitle

\section{Introduction}\label{s_Introduction}
Cepheids demonstrate a well-known relationship between their luminosity and pulsation period, which is commonly referred to as the Period-Luminosity (PL) relation, and famously recognized as the Leavitt law \citep{leavitt08,leavitt12}. The PL relation was first applied to prove the extragalactic nature of spiral nebulae \citep{Hubble1925NGC6822,Hubble1926M33,HUbble1929M31}, and then led to the discovery of the expansion of the Universe characterized by the Hubble-Lemaître constant $H_0$ \citep{lemaitre27,hubble29}. The extragalactic distance scale, which largely relies on the Cepheid PL relation, coupled with the observation of extragalactic Cepheids with the Hubble Space Telescope yielded a precise determination of the accelerated expansion of the Universe \citep{Riess2022}.

However, the calibration of the PL relation is still affected by uncertainties at the percent level, and is one of the largest contributors to the error budget on the Hubble-Lemaître constant $H_0$ \citep{Riess2022,Riess2024_new}. Despite evidences for infrared (IR) emission from CircumStellar Envelope (CSE) of Cepheids, its impact as a possible bias or dispersion on the PL relation is not yet taken into account, or is assumed to be negligible in the near-IR bands \citep{di2024hubble,Anderson2024book}. Indeed, the physical nature of CSE emission is still poorly constrained, therefore no satisfying model of spectral energy distribution (SED) is currently available. Moreover, the presence of CSEs around Cepheids, which could depend on fundamental stellar parameters, remains to be understood. This is of paramount importance to ensure that  no systematic errors are introduced into the PL relation, and, consequently, on the measurement of $H_0$. The need for a distance scale free from systematics is essential in the era of the James Webb Space Telescope (JWST), which is currently observing Cepheids in distant galaxies to refine the extragalactic distance scale \citep{Freedman2024IAU,Riess2024}. Up to now, several studies have shown the potential impact of CSE emission on the PL relation \citep{Neilson2009LMC,Neilson2010} and the Baade-Wesselink method to derive individual distance of Cepheids \citep{merand07,merand15,Gallenne2021,Nardetto2023,Hocde2024}. Therefore, it is essential to understand the CSEs characteristics as a function of the Cepheid properties to correct possible biases on photometric measurements.

Moreover, the presence of circumstellar material can provide valuable insights into the mass loss history of Cepheids. This holds particular significance within the context of the mass discrepancy observed in Cepheids, where derived stellar masses can exhibit discrepancies of up to 20\% when compared to both pulsation models, such as those discussed in \cite{Caputo2005, bono06, keller08}, and dynamical mass measurements, as outlined in \cite{Pietrzynski2010}.
The first attempts to detect traces of mass-loss were provided by mid- and far-infrared photometric and imaging observations with IRAS \citep{mcalary1986,Deasy1986,Deasy1988} and later with \textit{Spitzer} and \textit{Herschel} \citep{Neilson2009LMC,Marengo2010Spitzer,MARENGO2010,Barmby2011,Gallenne2012,Hocde2020a}. Radio observations using HI 21 cm line emission with the VLA also revealed extended gas environment around $\delta$~Cep and T~Mon \citep{matthews12,Matthews2016}. However, at the exception of few cases where Cepheids are embedded in large and cold interstellar dust materials (for example RS Puppis and SU Cas), these observations have shown that extended dusty environment can be excluded for the vast majority of Cepheids.

Another method for identifying IR excess from CSEs involves the comparison of observations with static atmospheric models. Several studies consistently reported a lack of substantial observational support for IR excess in a significant population of Cepheids \citep{Schmidt2015, Gro2020, Groenewegen2023}. However, these studies used mean photometry along the pulsation cycle, which might be problematic to detect an IR excess of few percents compared to the photosphere, or a variable CSE emission. In contrast, \cite{Gallenne2012} interpolated photometric data to the specific phase of mid-IR observations to fit a consistent atmosphere model and found an IR excess for most of the Cepheid in their sample. In order to achieve sufficient precision, the parallax-of-pulsation code \texttt{SPIPS} \citep{Merand2015} is well-suited for detecting subtle IR excesses due to its capability to adjust atmospheric models along the pulsation cycle, aligning them with an extensive array of observations. In particular, \cite{Gallenne2021} demonstrated that fitting a CSE model in the parallax-of-pulsation model \texttt{SPIPS} significantly improves the goodness-of-fit for 13 out of 45 Cepheids. Notably, they found that these CSE models present a mean excess of $0.08\pm$0.04$\,$mag in the $K$-band ($\lambda=2.2\,\mu$m).

High-angular resolution provided by long-baseline interferometry is ideal to directly resolve the emission extended at only several radii of the Cepheids.
The first CSE was resolved for the first time in the $K$-band owing to the Very Large Telescope Interferometer (VLTI/VINCI) around $\ell$ Car \citep{kervella06a}, then around Polaris, $\delta$ Cep and Y~Oph at the Center for High Angular Resolution Astronomy (CHARA/FLUOR) \citep{kervella06a,merand06,merand07}. In the $K$-band, the diameter of the envelope appears to be at least 2 stellar radii and the flux contribution is up to 5\% of the continuum. A similar result was achieved in the $L$-band (3.5$\,\mu$m) around $\ell$~Car \citep{Hocde2021} with VLTI/MATISSE. CSEs were also resolved in the $N$-band (10$\,\mu$m) around four Cepheids with VLTI/VISIR and VLTI/MIDI \citep{kervella09,gallenne13b}. Interestingly, a CSE was also resolved in the visible band with the CHARA/VEGA instrument around $\delta$~Cep \citep{nardetto16}.

The size of CSEs directly relates to their nature, which was first assumed to be dusty. Several radiative transfer studies were based on that assumption \citep{gallenne13b,Gro2020,Hocde2020a,Groenewegen2023}. 
Cepheids of long pulsation period might have a dusty envelope as expected from unusual chemical composition \citep{Kovtyukh2024}. However, near-IR emission and a CSE close to the star suggest an environment that is too hot for dust condensation, with temperatures exceeding 2000 K. Such CSEs are likely to be composed of ionized gas, emitting continuous free-free emission as outlined by radiative transfer models \citep{Hocde2020a}. This region could be linked to the chromospheric activity of the star for which the radius was derived to be about $R_\mathrm{chromo}$=1.50$\,R_\star$ \citep{Hocde2020b}. The presence of hot gas material at few stellar radii above the photosphere might be caused by pulsation driven shocks \citep{fokin96,neilson11,Moschou2020,Fraschetti2023}.

Further observations are needed to constrain the size of the CSE which is essential to model the physical process at play. While the first interferometric detections suggested a potential correlation between envelope emission and the pulsation period \citep{merand07}, recent photometric analysis showed that the IR emission is not correlated to the pulsation period nor the evolution's stage of the Cepheid \citep{Gallenne2021}. We however expect that the efficiency of the mass-loss mechanism does depend on the Cepheid characteristics as shown by \cite{neilson2008}, especially in the case of pulsation driven mass-loss, since the shock intensity and the surface gravity must play a significant role.

In this paper, we report on the study of CSE characteristics of eight Cepheids in the $L$ (2.8-4.0$\,\mu$m), $M$ (4.5 - 5$\,\mu$m) and $N$~ bands (8 - 13$\,\mu$m) thanks to the Multi AperTure mid-Infrared SpectroScopic Experiment \citep[VLTI/MATISSE,][]{Lopez2014,Allouche2016,Robbe2018,Lopez2022}.
To this end, we observed Cepheids with different pulsation period across the instability strip:  X~Sgr~($7.01\,$d), U~Aql~(7.02$\,$d), $\eta$~Aql~(7.17$\,$d), $\beta$ Dor~(9.84$\,$d), $\zeta$~Gem~(10.15$\,$d),  TT~Aql~(13.75$\,$d), T~Mon~(27.03$\,$d) and U~Car~(38.87$\,$d). 

In the following, we first present the observations, the data reduction and the calibration process of VLTI/MATISSE data in Sect.~\ref{sect:matisse_obs}. From the MATISSE data alone, it is not possible to disentangle the Cepheid and the CSE, as we do not resolve the stars up to the first null of the visibility function, which is essential for accurately determining the stellar angular diameter. Instead, we model the Cepheid photosphere along the pulsation cycle in Sect.~\ref{sect:spips} in order to derive the expected visibilities and flux of each star at the specific date of MATISSE observations. Then, we study successively the calibrated flux in Sect.~\ref{sect:flux_cal}, the closure phase in Sect.~\ref{sect:T3} and the squared visibilities in Sect.~\ref{sect:vis}. We discuss and summarize our results in Sect.~\ref{sect:conclusion}.

\section{VLTI/MATISSE interferometric observations}\label{sect:matisse_obs}
MATISSE operates as the beam combiner for the $L$, $M$, and $N$ bands within the Very Large Telescope Interferometer (VLTI) facility at the Paranal Observatory. The VLTI comprises a combination of four 1.8-meter auxiliary telescopes (ATs) or four 8-meter unit telescopes (UTs) which offer an array of baseline lengths spanning from 11 meters to 150 meters. As a spectro-interferometer, MATISSE captures both dispersed fringes and spectra. MATISSE employs a standard observing mode known as the "hybrid" mode, which incorporates two distinct photometric measurement methods. For the $L$ and $M$ bands, it utilizes SIPHOT to simultaneously measure photometry alongside the dispersed interference fringes. Conversely, for the $N$ band, it employs HIGH SENS to measure photometric flux independently after the interferometric observations.

The observational campaign was conducted between 2020 and 2022 during open and guaranteed time observation (GTO, T~Mon is part of this program). These observations were executed using the UT quadruplet, resulting in ground baseline lengths ranging from 58 to 132 meters. Observations with UTs are required to observe in the $N$-band since the flux of the brightest Cepheids is only a few Jansky. The log of the MATISSE observations  is  given  in  Table~\ref{tab:log}.

We processed the raw data using the most recent version of the MATISSE data reduction software (DRS) version 2.0.2, which is publicly available \footnote{The MATISSE reduction pipeline is publicly available at \url{http://www.eso.org/sci/software/pipelines/matisse/.}}. The data reduction steps are explained in details in \citet{Millour2016}. The MATISSE absolute visibilities are estimated by dividing the measured correlated flux by the photometric flux. In our subsequent analysis, we excluded the 4.0 to 4.5$\,\mu$m spectral region, where the atmosphere is not transmissive, and also the noisy edges of the atmospheric spectral bands (which might slightly differ from a star to another). Thus, we analyzed the following spectral region in $L$ ($ 3.1-3.9\,\mu$m), $M$~($4.5-4.9\,\mu$m) and $N$ ($8.1-12\,\mu$m) for both the absolute visibility and flux. The only exception is U Aql for which we acquired the correlated flux in the $N$-band.

\subsection{Atmospheric conditions}
In the $L$ and $M$ bands, the brightness of the Cepheids sample (around 10$\,$Jy) significantly exceeds the UT sensitivity in low-resolution mode (approximately $0.1\,$Jy), even in sub-optimal atmospheric conditions. In contrast, under the most adverse conditions, atmospheric factors could impact the visibility of the faintest targets in the $N$-band. As stated by \cite{Petrov2020} the coherence time has the most significant impact on the instrumental visibility.  Consequently, caution is required for coherence time below 4$\,$ms in particular for X Sgr in our sample ($\tau_0\approx2\,$ms) and therefore we do not display its visibilities. Moreover, according to \cite{Lopez2022}, in the $L$ band, the seeing-dependent Strehl ratio dominates the visibility calibration errors, while in the $N$ band, as in MIDI, MATISSE is sensitive to the temporal fluctuations of the thermal background.  Overall, our star sample was observed in excellent conditions with a seeing below 0.8 arsecond and a coherence time $\tau_0>4\,$ms (see Table \ref{tab:log}).

\subsection{Choice of calibrators}\label{sect:cal_sci}
For most of the Cepheids, two calibrators were observed in order to ensure a reliable calibration in the $LM$ and $N$ band (see Table~\ref{tab:log}). We chose calibrators whose flux is comparable to scientific targets in both the $L$ and $N$-bands, i.e. $\approx$10 to 1$\,$Jy respectively, as shown in Table \ref{Tab.cal}. These flux requirements induce the use of partially resolved calibrators at the 132-m longest baseline. Calibrators were selected using the SearchCal tool of the JMMC\footnote{SearchCal is publicly available at \url{https://www.jmmc.fr/english/tools/proposal-preparation/search-cal/}}, with a high confidence level on the derived angular diameter
\citep[$\chi^2\leq5$, see Appendix A.2 in][]{Chelli2016}. We also ensured that these calibrators are free from any IR emission flags as defined from the Mid-infrared stellar Diameters and Fluxes compilation Catalogue \citep[MDFC,][]{cruz2019}. The two calibrators of T~Mon, namely 30~Gem and 18~Mon, were chosen in order to compare consistently the results with previous observations with MIDI/VLTI \citep[hereafter G13]{gallenne13b}. The $L$-band uniform-disk (UD) angular diameters for the standard stars, as well as the corresponding $L$ and $N$-band fluxes and the spectral types are given in Table \ref{Tab.cal}. Last, we selected calibrators with a relatively small angular separation from the Cepheid (<15$\,$degrees) in order to ensure a similar sky and instrumental transmission function, as well as a similar airmass for a reliable flux calibration.

Subsequently to the calibrator observations, we verified the consistency of their angular diameter with values from the JSDC catalogue \citep{JSDC2014}. This is important to investigate possible systematic effects on the angular diameter measurements of the Cepheid. To this end, we followed the method employed by \cite{Robbe2022}. The observed difference between the fitted angular diameter $\theta_\mathrm{fit}$ and the angular diameter from the JSDC $\theta_\mathrm{JSDC}$ in the $L$ band is indicated in Table~\ref{tab:log}. In all cases, we find a difference lower than $5\%$, which allows to calibrate accurately the Cepheid visibilities.

\begin{table*}
    \centering
    \begin{threeparttable}
        \caption{\label{Tab.cal} \small Standard stars properties for visibility and flux}
        \begin{tabular}{l|c|c|c|l|c}
            \hline
            \hline
            Calibrator & $\theta_\mathrm{UD}$(mas)[$L$] & $F_L$ (Jy) & $F_N$(Jy) & Sp. Type &Flux calibration\\
            \hline

            HD 47156 & 1.096$\pm$0.108 & 13.7$\pm$2.2 & 2.6$\pm$0.8 & K0 &\\

            HD 184574 & 0.851$\pm$0.076 & 11.0$\pm$1.5 & 1.4$\pm$0.4 & K0 III& \\
            HD 189533 & 0.913$\pm$0.087 & 11.2$\pm$2.6 & 1.5$\pm$0.4 & G8 II &\\
            HD 188390 & 1.064$\pm$0.102 & 11.7$\pm$4.2 & 1.5$\pm$0.4 & K3/4III& \\

            $\alpha$ Char & 0.938$\pm$0.095 & 19.5$\pm$1.0 & 3.1$\pm$0.5 & F5 V &\\
            HD 189695& 2.014$\pm$0.159 & 38.5$\pm$5.4 & 5.8$\pm$1.6 & K5 III &\cite{Cohen1999}\\
            30 Gem & 2.082$\pm$0.194 & 56.5$\pm$3.4 & 8.5$\pm$0.7 & K0 III& \cite{Cohen1999}\\
            18 Mon & 2.075$\pm$0.165 & 46.8$\pm$9.4 & 7.0$\pm$2.1 & K0 III& \cite{Cohen1999}\\
            87 Gem & 1.977$\pm$0.175 & 36.1$\pm$1.4 & 5.6$\pm$1.9 & K5 III &\cite{Cohen1999}\\

            HD 183925 & 1.449$\pm$0.128 & 30.4$\pm$8.7 & 3.3$\pm$1.1 & K5 III& ATLAS9, $T_\mathrm{eff}=4115^{+113}_{-125}$\\
            HD 101162 & 0.913$\pm$0.083 & 11.2$\pm$0.5 & 1.5$\pm$0.4 & K0 III &ATLAS9, $T_\mathrm{eff}=4811^{+4.5}_{-3}$\\
            HD 59219 & 1.410$\pm$0.119 & 23.6$\pm$0.9 & 3.5$\pm$0.9 & K0 III&ATLAS9, $T_\mathrm{eff}=4834^{+78}_{-113}$ \\

            \hline
            \hline
        \end{tabular}
        \begin{tablenotes}
            \item[] \textbf{Notes:} $\theta_\mathrm{UD}$ is the uniform disk (UD) angular diameter in $L$ band from the JSDC V2 catalogue \citep{bourges14}. $F_L$ and $F_N$ are the flux in the $L$ and $N$ bands from the Mid-infrared stellar Diameters and Fluxes compilation Catalogue \citep[MDFC,][]{cruz2019}. Templates for the flux calibration are either from \cite{Cohen1999} or constructed from ATLAS9 atmosphere models with effective temperature as measured by \textit{Gaia} (see Sect.~\ref{sect:flux_cal}).
        \end{tablenotes}
    \end{threeparttable}
\end{table*}

\section{Accurate star photosphere modeling with SPIPS}\label{sect:spips}

\subsection{SPIPS modeling of the star sample}
As outlined in the introduction, we need an accurate model of the photosphere and distance of the Cepheid in order to compare with the MATISSE flux and angular diameter measurements. For this purpose, we used \texttt{SPIPS} (SpectroPhoto-Interferometric modeling of Pulsating Stars) introduced by \cite{Merand2015}, which enables to interpolate atmosphere model of the Cepheids over the observations along the pulsation cycle. This is a parallax-of-pulsation code whose principle is based on the Baade-Wesselink (BW) method \citep{baade26,wesselink46}. In this method, the distance is derived by comparing the variation of both the radius and the angular diameter. The radius variation is derived from the pulsation velocity of the star by the mean of radial velocity measurements and the assumption of the projection factor  \citep[$p$-factor, see][ and references therein]{Nardetto2023}.
On the other hand, the angular diameter variation is measured from interferometry or based on surface brightness through the empirical Surface Brightness Color Relarions(SBCR).

In contrast to classical BW method which derives the angular diameter using SBCR in only two photometric bands, \texttt{SPIPS} is a sophisticated version which includes photometric, interferometric, effective temperature and radial velocity measurements in a robust all-at-once model fit. Assuming quasi-static photosphere of the Cepheids, it makes use of atmospheric ATLAS9 models from \cite{castelli2003}, with solar metallicity and a standard turbulent velocity of 2~km/s, to fit observational data along the pulsation cycle. \texttt{SPIPS} was used in different studies including Cepheids and RR~Lyrae stars \citep{breitfelder16,Gallenne2017,Hocde2020a,Trahin2021,Gallenne2021,Bras2024}.  

Since \texttt{SPIPS} is a parallax-of-pulsation code one must fix either the distance or the $p$~factor which are fully correlated.  We adopted \textit{Gaia} DR3 parallaxes \citep{gaia2016,Gaia2022} only in the case of TT~Aql and U~Car because they meet the parallax quality criterion defined as RUWE<1.4 and do not saturate the detector with $G>6\,$mag. For these two stars we also corrected the parallax by the \textit{Gaia} zero-point\footnote{Code available at \url{https://pypi.org/project/gaiadr3-zeropoint/}} \citep[][]{Lindegren2021}. For all other stars we fix the $p$-factor to 1.27 following the mean values obtained for Cepheids \citep{Trahin2021} and we adjust the distance. We note that fixing either the distance or the $p$-factor has no influence on the accuracy of the angular diameter and flux level, which are well constrained by the interferometric and photometric observations. For each star we retrieved radial velocity, effective temperature, angular diameter and photometric measurements in various bands. The PIONIER interferometric observations in the $H$-band (1.65$\mu$m) available for most of the stars in our sample allow to constrain tightly the star angular diameter. We did not collect extensive photometric datasets but instead used data from common sources whenever possible. Ideally, datasets must be obtained from the same instrument and observed during a short time baseline to avoid any bias when applying the BW method \citep{Wielgorski2024,Zgirski2024}. This approach also ensures that the overall \texttt{SPIPS} fitting and IR excess measurements across different stars are more easily comparable. 

In the case of U Aql, a special correction is needed since it is a known spectroscopic binary with an orbital period of 1862.8$\pm1.1\,$day \citep{Hocde2024b}. We corrected radial velocity measurements for orbital motion following \cite{Gallenne2018,Gallenne2019a}. Among our star sample, X~Sgr,  $\eta$~Aql, $\beta$~Dor and $\zeta$~Gem represent the most complete datasets. We present \texttt{SPIPS} results of $\eta$~Aql and X Sgr in Fig.~\ref{fig:spipsou} and for others stars in Appendix \ref{app:spips} together with the source of the observations used in Table \ref{Tab.SPIPS}. Although we assumed different $p$-factors and used different datasets, our \texttt{SPIPS} fitting are in good agreement with previous work from \cite{Trahin2021} and \cite{Gallenne2021} (see Table \ref{tab:comparison}).

\subsection{Modeling the infrared excess}\label{sect:ir}
\begin{figure*}[h!]
\centering
\begin{subfigure}{0.9\textwidth}
\includegraphics[width=\linewidth]{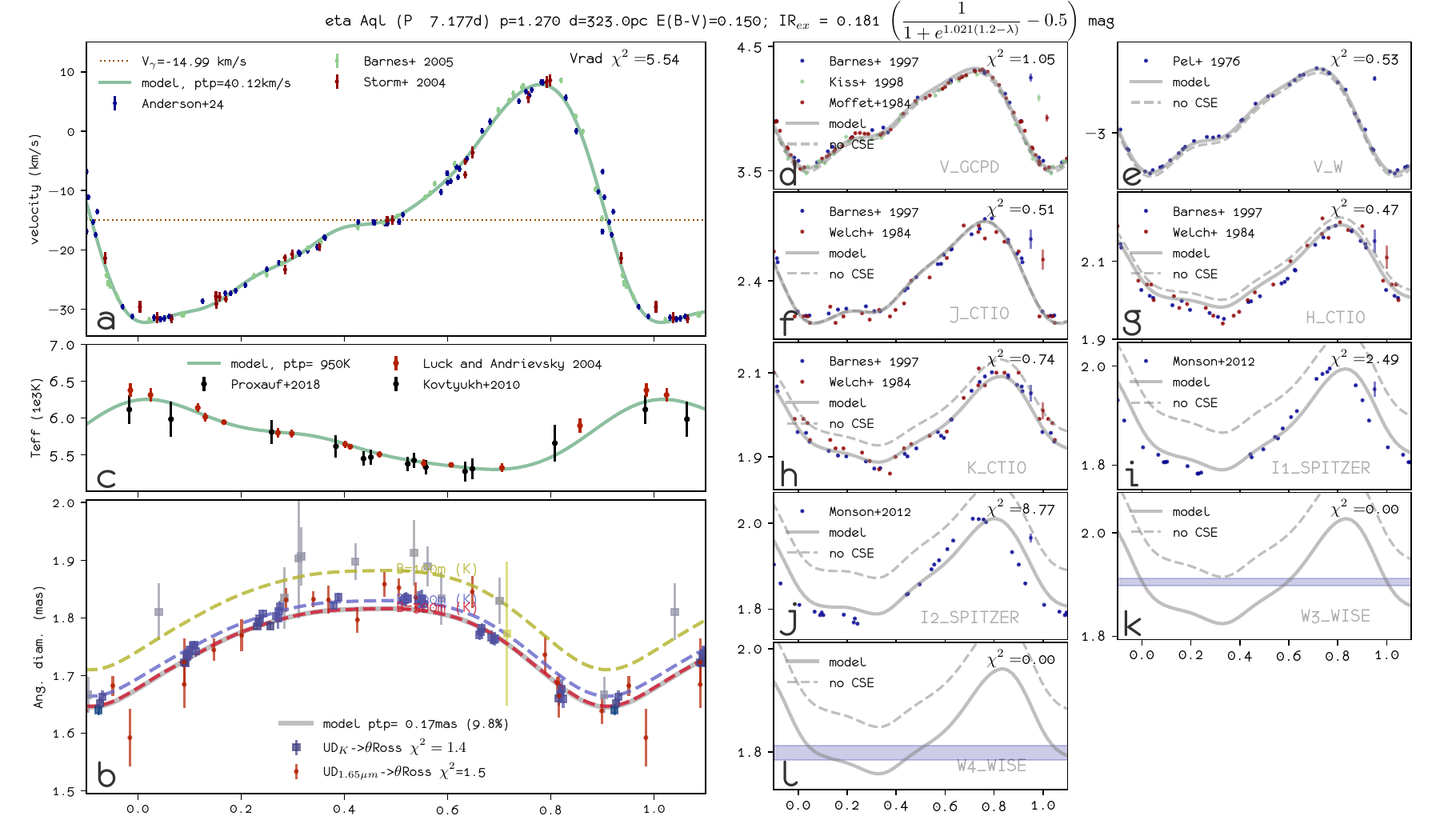}
\caption{}  
\end{subfigure}\vskip\baselineskip
\begin{subfigure}{0.9\textwidth}
\includegraphics[width=\linewidth]{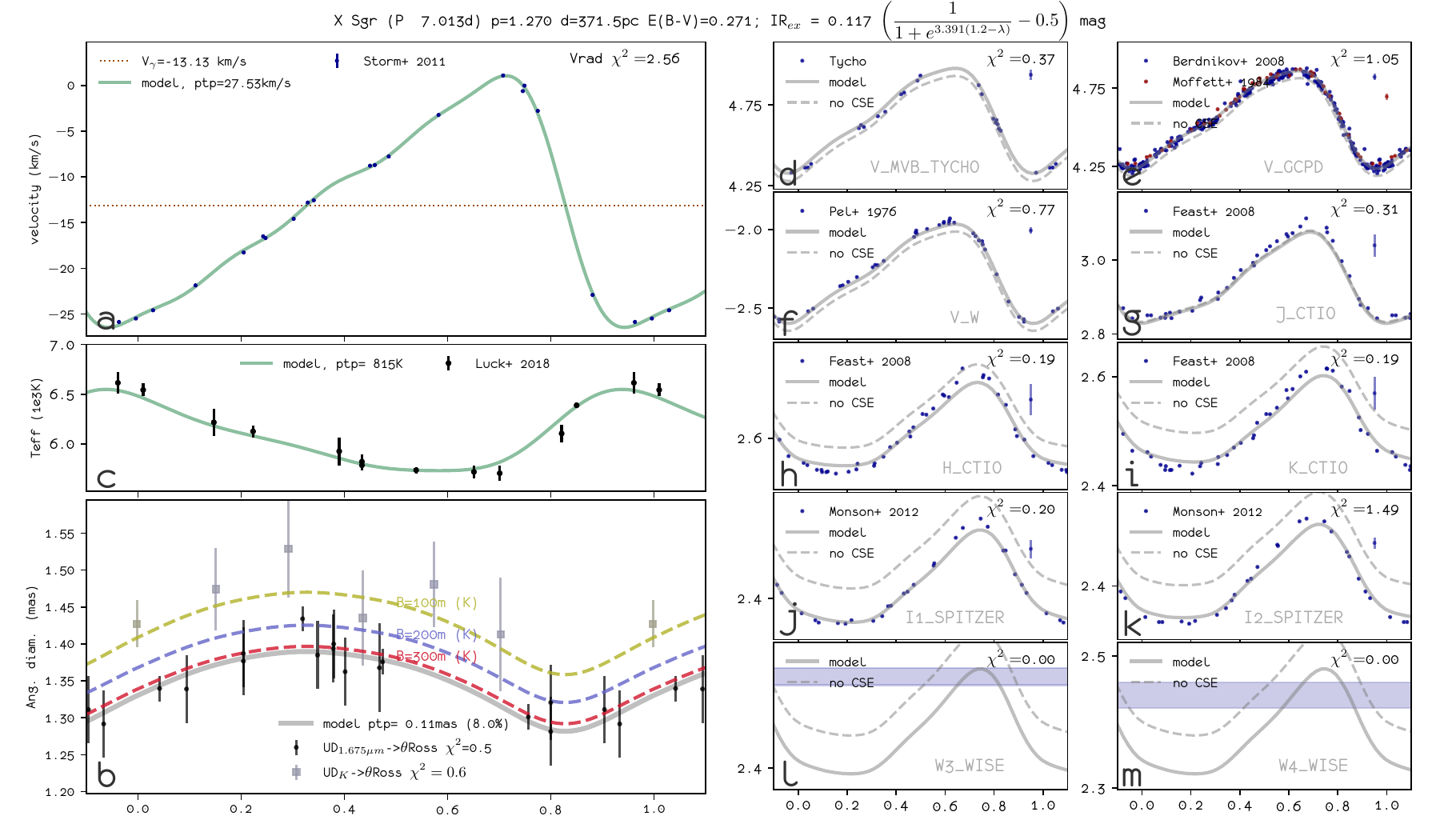}
\caption{}  
\end{subfigure}
\caption{\small The \texttt{SPIPS} results of $\eta$ Aql (a) and X Sgr (b) as a function of the pulsation phase. Above the figures, the $p$-factor is indicated, along with the fitted distance $d$, the fitted color excess $E(B-V)$, and the parametric CSE model. The gray thick line corresponds to the best \texttt{SPIPS} model, which is composed of the latter model without CSE plus an IR excess model. In the angular diameter panels, the gray curve corresponds to limb-darkened (LD) angular diameters. We provide references for the observations in Table \ref{Tab.SPIPS} of the Appendix.} \label{fig:spipsou}
\end{figure*}

\texttt{SPIPS} also takes into account the presence of CSE for modeling the synthetic photometry and interferometric observations from different instruments. As shown by \cite{Gallenne2021}, the introduction of a parametric CSE model significantly improves the fitting of the \texttt{SPIPS} model for a large number of Cepheids. In our sample, X~Sgr and $\eta$~Aql are two stars for which the impact of the CSE can be seen from interferometric observations as we can see from the angular diameter panels for each star in Fig.~\ref{fig:spipsou}. This model consists of an IR excess modeled by a parametric power-law. In this study, we rather employed a logistic function with two degree of freedom to model the IR excess analytically as proposed by \cite{Hocde2024}. This is based on physical justifications of ionized gas CSE around the star leading to a spectral index of $S_\nu \propto \nu^2$ \citep{Hocde2020a}. Indeed, this parametric model allows absorption at shorter wavelength as well as saturation of IR excess at longer wavelength,  in agreement with ionized hydrogen opacity models: 
\begin{equation}\label{eq:ir_excess}
    \Delta \mathrm{mag}_\lambda(\alpha,\beta,\lambda_0)=\alpha \left(\frac{1}{1+e^{\beta(\lambda_0-\lambda)}}-\frac{1}{2}\right)\,\mathrm{mag}
,\end{equation}
where $\alpha>0$ and $\beta>0$ represent the intensity and the slope of the logistic function, respectively, and $\lambda_0$ is the pivot wavelength for which $\Delta \mathrm{mag}_{\lambda_0}=0\,$mag. Although it is possible to adjust $\lambda_0$, in the following we rather fix $\lambda_0=1.2\,\mu$m for every star, similarly to the parametric power-law originally included in \cite{Merand2015} and we keep only $\alpha$ and $\beta$ as free parameters.

\begin{figure*}[]
    \centering
    \begin{subfigure}[b]{0.24\textwidth}
        \includegraphics[width=\linewidth]{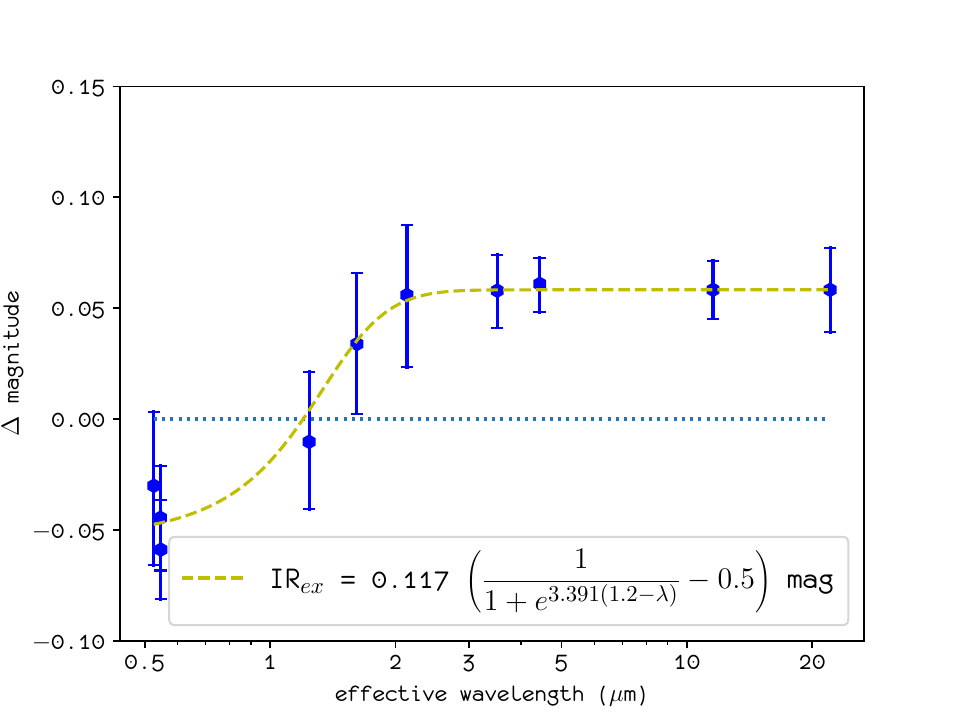}
        \caption{X Sgr, 7.01$\,$d}
        \label{fig:x_sgr_ir}
    \end{subfigure}
    \begin{subfigure}[b]{0.24\textwidth}
        \includegraphics[width=\linewidth]{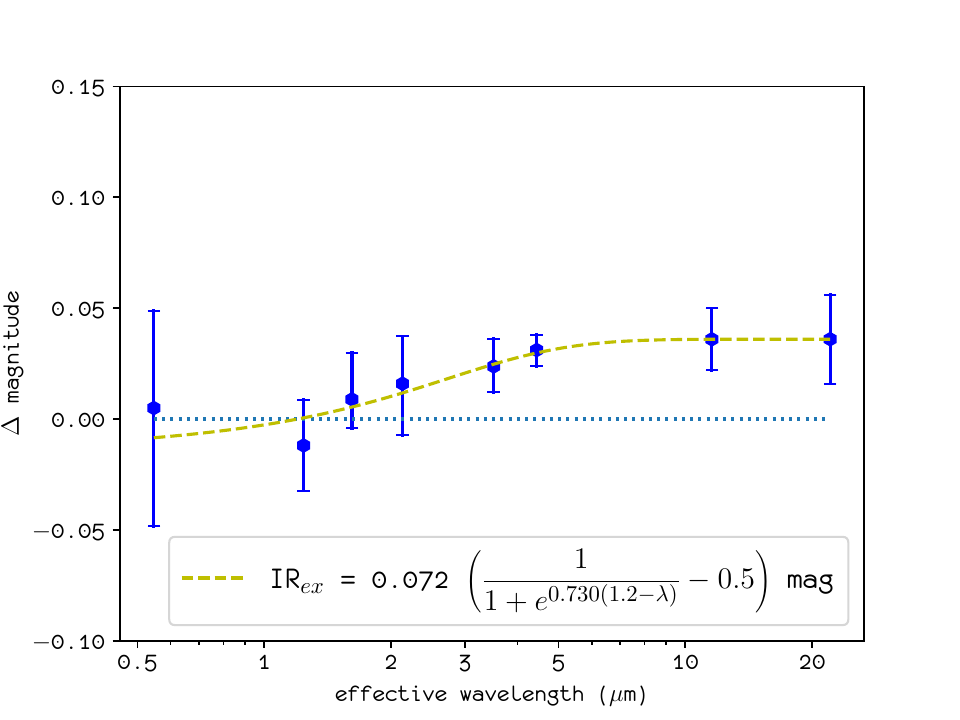}
        \caption{U Aql, 7.02$\,$d}
        \label{fig:u_aql_ir}
    \end{subfigure}
    \begin{subfigure}[b]{0.24\textwidth}
        \includegraphics[width=\linewidth]{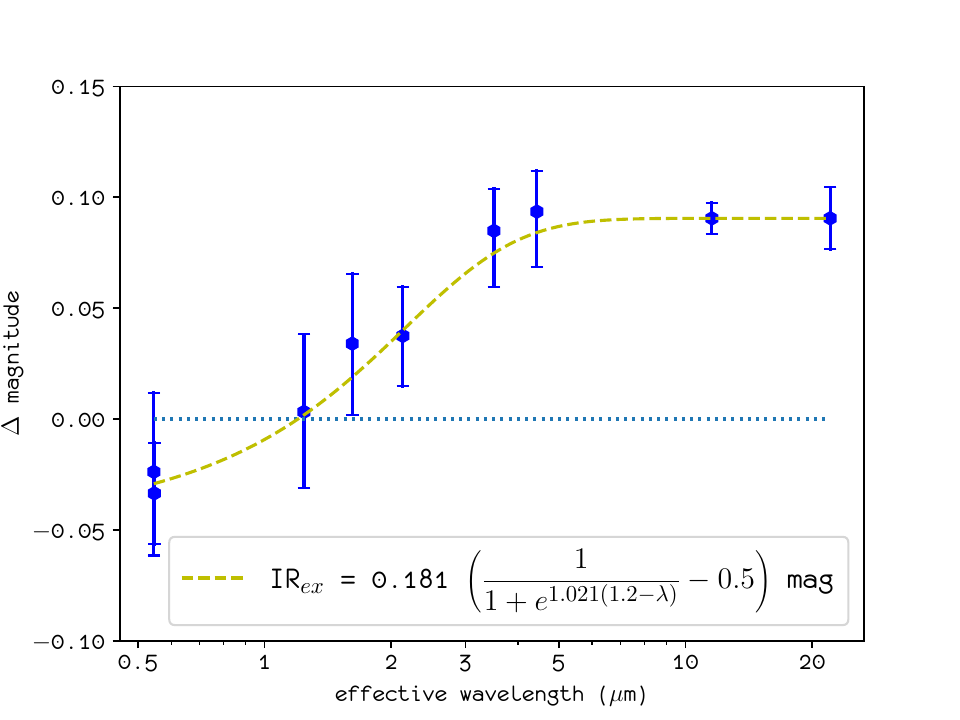}
        \caption{$\eta$ Aql, 7.17$\,$d}
        \label{fig:eta_aql_ir}
    \end{subfigure}
    \begin{subfigure}[b]{0.24\textwidth}
        \includegraphics[width=\linewidth]{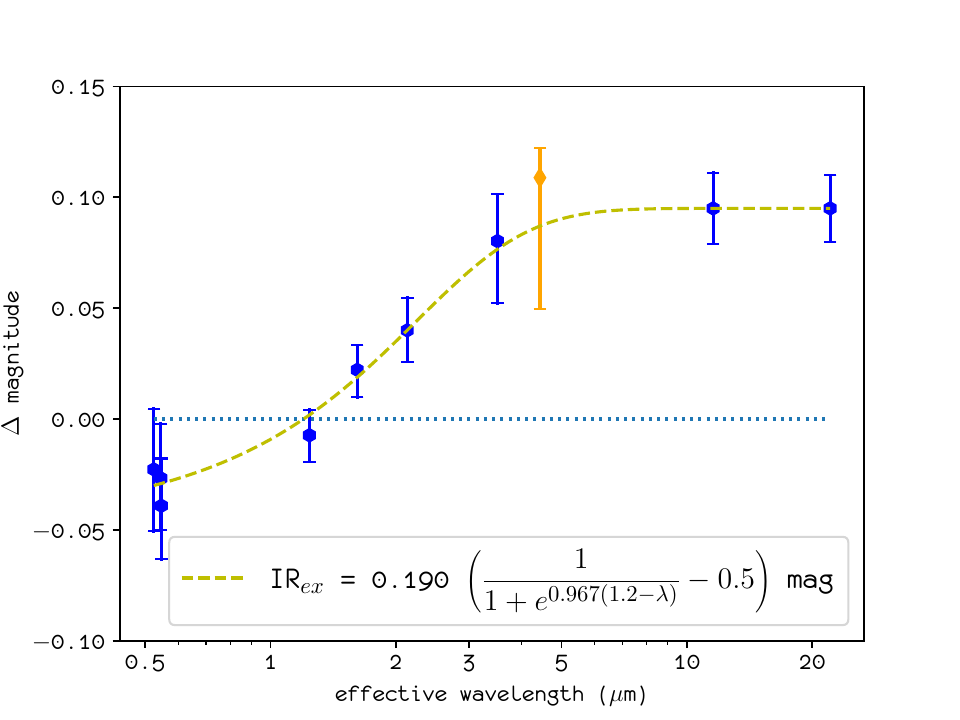}
        \caption{$\beta$ Dor, 9.84$\,$d}
        \label{fig:beta_dor_ir}
    \end{subfigure}

    \begin{subfigure}[b]{0.24\textwidth}
        \includegraphics[width=\linewidth]{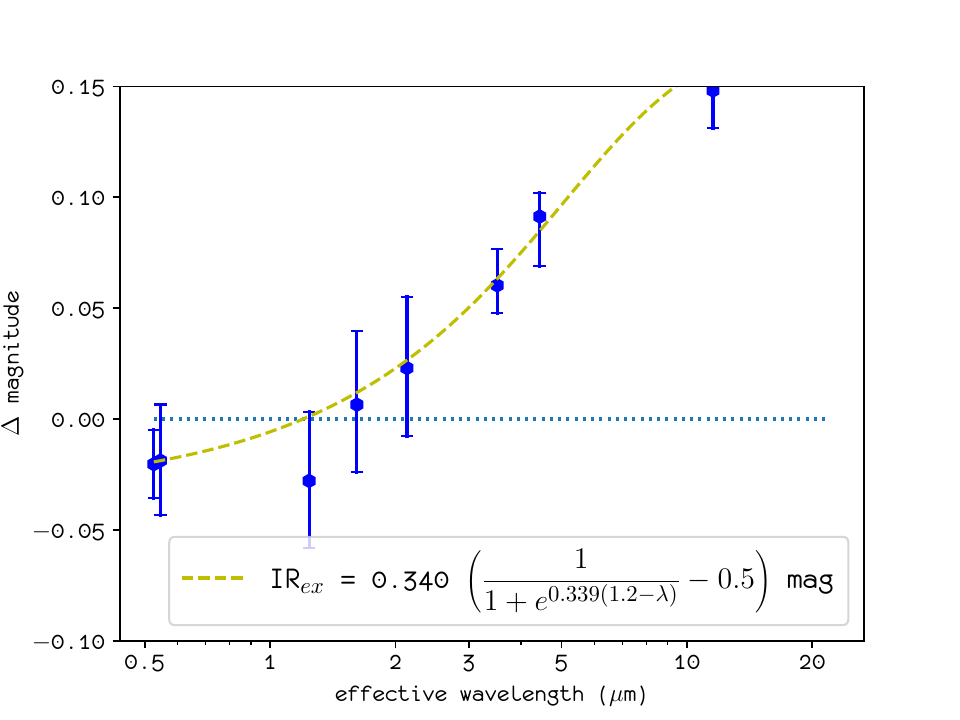}
        \caption{$\zeta$ Gem, 10.15$\,$d}
        \label{fig:zeta_gem_ir}
    \end{subfigure}
    \begin{subfigure}[b]{0.24\textwidth}
        \includegraphics[width=\linewidth]{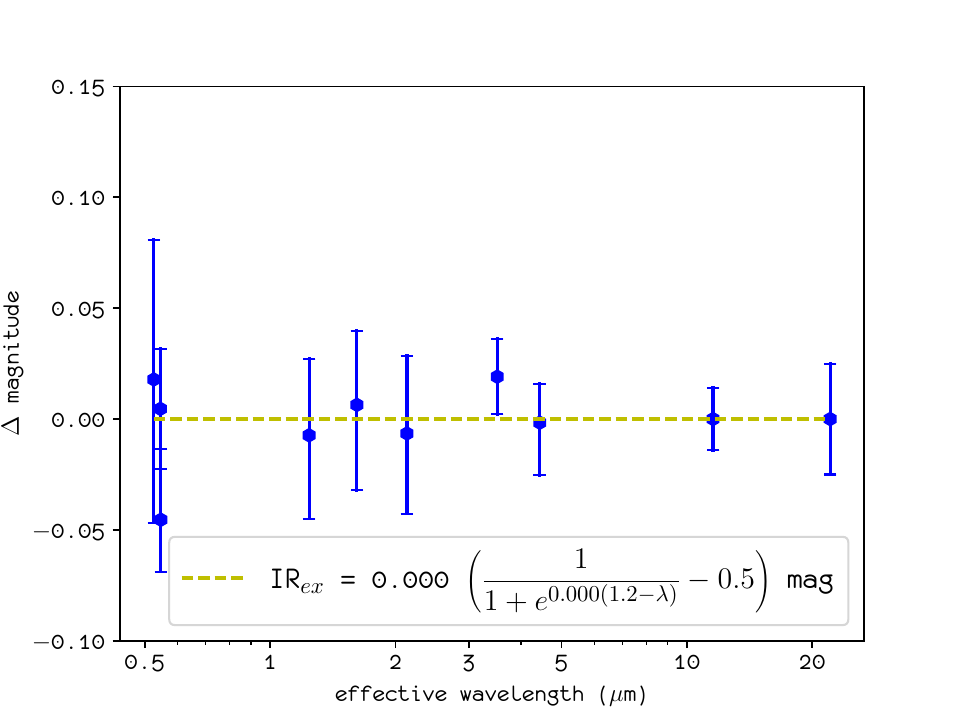}
        \caption{TT Aql, 13.75$\,$d}
        \label{fig:tt_aql_ir}
    \end{subfigure}
    \begin{subfigure}[b]{0.24\textwidth}
        \includegraphics[width=\linewidth]{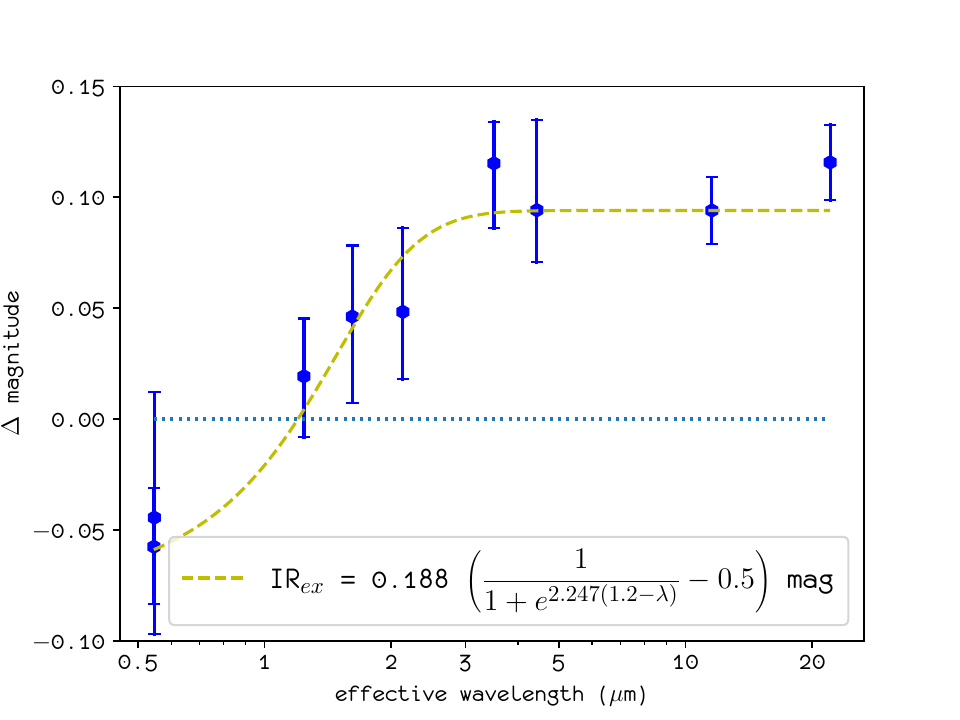}
        \caption{T Mon, 27.03$\,$d}
        \label{fig:t_mon_ir}
    \end{subfigure}
    \begin{subfigure}[b]{0.24\textwidth}
        \includegraphics[width=\linewidth]{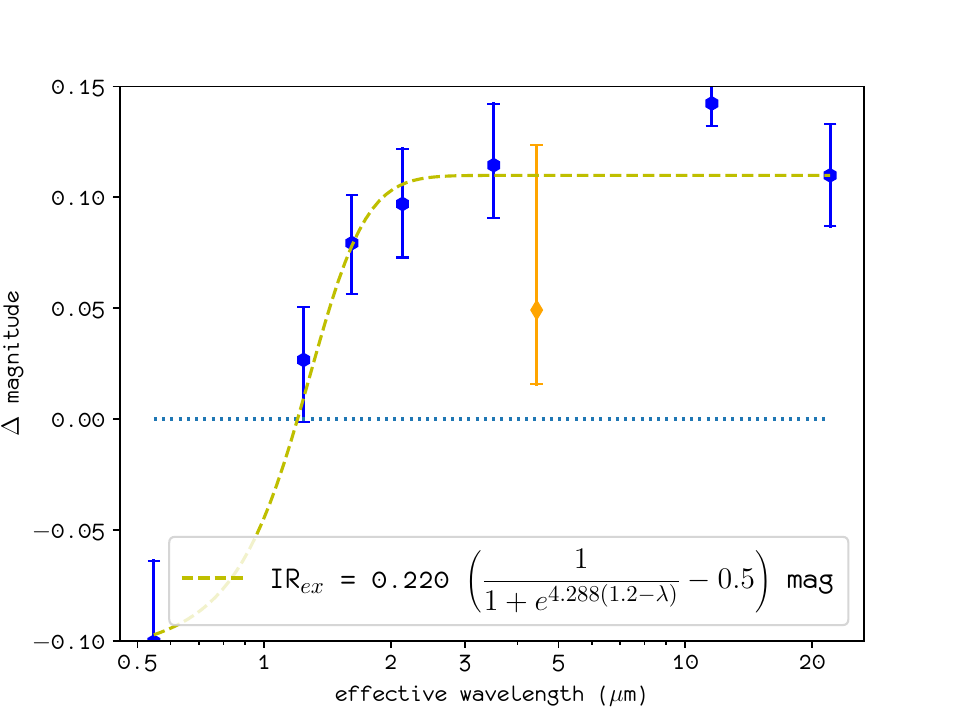}
        \caption{U Car, 38.87$\,$d}
        \label{fig:u_car_ir}
    \end{subfigure}

    \caption{IR excess derived by \texttt{SPIPS} for each star of the sample (see also Table~\ref{tab:spips_mag}). The excess is presented as the magnitude difference between the atmospheric model and the median photometric data, where positive and negative values indicate emission and absorption, respectively. The fitted IR excess (dashed green line) is modeled with a logistic function presented at the bottom of each figure. The orange bar in Figure (h) for U Car corresponds to the I2 \textit{Spitzer} band, ignored in the fit due to CO band-head absorption at 4.6 micron \citep{Marengo2010Spitzer,scowcroft2016, Gallenne2021}.}
    \label{fig:ir_excess}
\end{figure*}

In the \texttt{SPIPS} modeling, the impact of the CSE emission on the measured angular diameter is derived from a spherical layer with no geometrical thickness following the CSE model from \cite{Perrin2005}. The radius of the spherical layer is fixed arbitrarily to 2.5$\,$R$_\star$ in the \texttt{SPIPS} model, consistently to CSE size resolved in the near-IR \citep{merand06,merand07}. We present the IR excess derived for each star in Fig.~\ref{fig:ir_excess}. For most of the stars, we derived a slight deficit in the visible domain of the order of 0.05$\,$mag while the excess in the mid-IR in the MATISSE spectral bands is between 0.00 to 0.15$\,$mag (see Table \ref{tab:spips_mag}). TT Aql is the only star for which we derived no IR excess, and, thus, the observation of the star are well represented by an atmosphere model without any envelope contribution. According to the IR excess study with \texttt{SPIPS} from \cite{Gallenne2021}, we note that for only three stars ($\eta$~Aql, $\zeta$~Gem and U~Car) the CSE model significantly improves the \texttt{SPIPS} goodness-of-fit. In the remaining cases, the CSE model does not notably improve the fit, suggesting a minimal or negligible contribution from the CSE (such as TT~Aql). We note that allowing an absorption in the visible slightly reduces the IR excess as compared to other works. Overall the IR excess we derived in Table \ref{tab:spips_mag} are also consistent with those obtained by \cite{Gallenne2021} in the different bands.

\begin{table}[h]
    \centering
    \begin{threeparttable}
      \caption{IR excess derived by \texttt{SPIPS}.}\label{tab:spips_mag}
        \begin{tabular}{l|c|c|c}
            \hline
            \hline
            Star & $\Delta$mag($K$) & $\Delta$mag($L$) & $\Delta$mag($N$) \\ \hline
            X Sgr       &   0.056$\pm$0.032 & 0.058$\pm$0.017 & 0.058$\pm$0.013 \\ 
            U Aql       &   0.016$\pm$0.022 & 0.024$\pm$0.012 & 0.036$\pm$0.014 \\ 
            $\eta$ Aql  &   0.037$\pm$0.022 & 0.085$\pm$0.023 & 0.091$\pm$0.007 \\ 
            $\beta$ Dor &   0.040$\pm$0.014 & 0.080$\pm$0.025 & 0.095$\pm$0.016 \\ 
            $\zeta$ Gem &   0.023$\pm$0.032 & 0.060$\pm$0.014 & 0.148$\pm$0.017 \\ 
            TT Aql      & $-0.006\pm$0.035 & 0.019$\pm$0.017 & 0.000$\pm$0.014 \\ 
            T Mon       &   0.048$\pm$0.034 & 0.115$\pm$0.031 & 0.094$\pm$0.015 \\ 
            U Car       &   0.096$\pm$0.024 & 0.110$\pm$0.025 & 0.142$\pm$0.010 \\ 
            \hline
        \end{tabular}
        \begin{tablenotes}
            \item Notes: The IR excess in the $K$, $L$- and $N$-band is derived as the difference between the median magnitude observed in each photometric band and the magnitude of the atmosphere model (see blue bars in Fig.~\ref{fig:ir_excess}). A positive value of $\Delta$ mag represents an excess, and a negative value indicates absorption.
        \end{tablenotes}
    \end{threeparttable}
    
\end{table}

\subsection{Angular diameter at the epoch of observations}
\subsubsection{Pulsation phase at the MATISSE observation epoch}
In order to consistently compare MATISSE observations to the photospheric model of each star it is necessary to derive an accurate pulsation phase corresponding to the epoch of observations. However, pulsation period of Cepheids are subject to period change effect caused by secular evolution or more erratic non-evolutionary effects \citep{turner2006,Csornyei2022,SinghRathour2024}.  For X~Sgr, U~Aql, $\eta$~Aql and U~Car we chose recent pulsation period and reference epoch from \textit{Gaia} DR3 to mitigate the effect of period change with respect to the time of MATISSE observations. This is particularly useful for U~Car because of its large rate of period change \citep{Csornyei2022}. For $\beta$~Dor, $\zeta$~Gem, TT~Aql and T~Mon we adopted the \texttt{SPIPS} values. We summarize our choice in Table \ref{Tab.ephemeris}.

 We assumed secular period evolution for each of our star after inspecting $O-C$ diagrams constructed by \cite{Csornyei2022} and adopting their rate of period change. In the case of $\beta$~Dor, $O-C$ diagram from \cite{Csornyei2022} presents scattered variations, similarly to other Cepheids in the resonance region. We preferred to adopt the period change rate derived by \texttt{SPIPS}. We note that only T~Mon and U~Car are affected by a large rate of period change ($10-100\,$s/yr) that can significantly affect the accuracy of the derivation of the pulsation phase at the MATISSE observation epoch. Additionally, U~Car presents clear wave-like trend in the $O-C$ diagram which is interpreted as perturbation of its orbital motion due to the presence of a companion.

 \subsection{Uniform-Disk angular diameter of the star sample}
 Using the \texttt{SPIPS} model we can now interpolate the limb-darkened (LD) angular diameter at the pulsation phase of each Cepheid corresponding to MATISSE observations. The LD angular diameter is physically equivalent to the bolometric radius of the photosphere entering the Stefan-Boltzmann law. However, interferometric observations in the IR cannot resolve the LD effect of the star but measures only the UD angular diameter. Thus, we converted the LD to the UD angular diameter in the $L$-band following the standard procedure described for example in \cite{Nardetto2020}, and the tables from \cite{claret11}. To this end, we used effective temperature and surface gravity interpolated by \texttt{SPIPS} at the specific date of observation assuming solar metallicity and standard micro-turbulent velocity of 2km/s. We did not correct for the LD effect in the $N$ band because of its negligible effect. Moreover, the measured visibility in the $N$ band does not permit a comparison with the angular diameter of the star as discussed in Sect.~\ref{sect:vis}. The uncertainty of the derived angular diameter is dominated by systematic uncertainties from \texttt{SPIPS} fitting and pulsation phase estimation. For the \texttt{SPIPS} models constrained by complete datasets and good pulsation coverage of the angular diameter (X~Sgr, $\eta$~Aql, $\beta$~Dor and $\zeta$~Gem) we assumed an uncertainty of 2\% on the star angular diameter while we assumed 4\% for the others stars.  We summarize our choice of ephemeris and derived angular diameters at the pulsation phase of MATISSE observations in Table~\ref{Tab.ephemeris}.

\begin{table*}
    \centering
    \begin{threeparttable}
        \caption{\label{Tab.ephemeris} Cepheid ephemeris, fluxes and angular diameters of the star sample.}
        \begin{tabular}{l|c|c|c|c|r|r|r|r}
            \hline
            \hline
            Star & $P_0$ (day)& $T_0$(MJD) & $\dot{P}$ (s/yr) & $\phi_\mathrm{MATISSE}$ & $\theta^\mathrm{SPIPS}_\mathrm{LD}$(mas) & $\theta_\mathrm{UD}(L)$(mas) & $F_L$ (Jy) & $F_N$ (Jy) \\
            \hline
            X Sgr       &7.012   &56926.825  & $+0.29\pm0.01$& 0.04 & 1.35$\pm0.03$  &1.33$\pm0.03$ & $20.1\pm6.9$ & $4.1\pm1.2$ \\ 	
            U Aql       &7.023 &56935.587 & $+0.17\pm0.04$ & 0.88 & 0.73$\pm0.03$ &0.71$\pm0.03$ & $7.7\pm2.7$ & $1.1\pm0.3$ \\
            $\eta$ Aql  & 7.177 &56975.775  & $+0.25\pm0.01$ & 0.58 & 1.80$\pm0.04$ & 1.78$\pm0.04$ & $46.7\pm6.8$ & $7.5\pm1.8$ \\
            $\beta$ Dor &9.842  &50275.770  & $+0.85\pm0.11$ & 0.15 & 1.87$\pm0.04$ & 1.85$\pm0.04$ & $51.9\pm13.1$ & $7.6\pm2.3$ \\
            $\zeta$ Gem & 10.149 &48708.059  & $-3.12\pm0.04$  & 0.60 & 1.61$\pm0.03$ & 1.59$\pm0.03$ & $36.9\pm2.2$ & $5.1\pm0.5$ \\
            TT Aql      &  13.754    & 48310.561     & $-0.70\pm0.11$  &0.03  & 0.76$\pm0.03$ & 0.74$\pm0.03$ & $8.1\pm0.2$ & $1.4\pm0.4$ \\
            T Mon       & 27.029 & 43783.986 & $+14.49\pm0.78$ & 0.03 & 0.87$\pm0.04$ & 0.85$\pm0.04$ & $11.7\pm2.9$ & $2.0\pm0.6$ \\
            U Car       & 38.791 & 56886.190  & $+97.85\pm2.03$ & 0.22 & 0.83$\pm0.03$ & 0.82$\pm0.03$ & $12.8\pm0.5$ & $2.6\pm1.2$ \\ 
            \hline
            \hline
        \end{tabular}
        \begin{tablenotes}
            \item[] \textbf{Notes.} For X~Sgr, U Aql, $\eta$~Aql, U~Car we chose pulsation period $P_0$ and reference epoch at maximum light $T_0$ from \textit{Gaia} DR3. For $\beta$ Dor, $\zeta$~Gem,TT~Aql and T~Mon we adopted the \texttt{SPIPS} values. We used independent rate of period change $\dot{P}$ derived by \cite{Csornyei2022} except for $\beta$~Dor for which we adopted the \texttt{SPIPS} value. These data are used to derive the pulsation phase at the epoch of MATISSE observations $\phi_\mathrm{MATISSE}$, and the corresponding LD diameter $\theta^\mathrm{SPIPS}_\mathrm{LD}$ of each star. The UD diameter in the $L$-band is derived using tables from \cite{claret11}.
        \end{tablenotes}
    \end{threeparttable}
\end{table*}

\section{Measured flux in $L$, $M$ and $N$ bands}\label{sect:flux_cal}
\subsection{Calibration}
We calibrate the total flux of the science targets $F_\mathrm{tot,sci}$  using the known flux of the calibrator $F_\mathrm{tot,cal}$ following

\begin{equation}\label{eq:tot_flux_cal}
F_\mathrm{tot,sci}(\lambda)=\frac{I_\mathrm{tot,sci}(\lambda)}{I_\mathrm{tot,cal}(\lambda)} \times F_\mathrm{tot,cal}(\lambda),
\end{equation}
\noindent where $I_\mathrm{tot,sci}$ and  $I_\mathrm{tot,cal}$ are the observed total raw flux of the science target and the calibrator, respectively. When available we used SED templates of the standard stars from \cite{Cohen1999} to derive $F_\mathrm{tot,cal}$ (see Table~\ref{Tab.cal}). For the following Cepheids we calibrated the total flux owing to accurate calibrator templates : U~Aql, $\eta$~Aql, $\zeta$~Gem, TT Aql and T~Mon. 
However, for three Cepheids (X Sgr, $\beta$ Dor, U Car) none of their calibrators have templates available. In this case, we interpolated the SED from a grid of ATLAS9 models using effective temperature measured by \textit{Gaia} (DR3 or DR2) \citep{GaiaDR2} (see Table \ref{Tab.cal}). We note that \cite{Casagrande2018} warn that effective temperature as measured by \textit{Gaia} are estimated with strong assumptions. However, in the Rayleigh-Jeans domain, the absolute flux level depends mostly on the star angular diameter which is well constrained for all the calibrators. We do not expect strong biases since for two of these calibrators, HD~101162 and HD~59219, the effective temperature is consistent with those obtained from UVES spectra \citep{Alves2015}. Then, we used the UD angular diameter in the $N$ band from JMMC catalog to re-scale the theoretical SED to the specific solid angle of observation. 
We neglect the airmass difference since the airmass between target and calibrators are comparable within less than 10\% except for U~Car (see Table~\ref{tab:log}).   Moreover, while a chromatic correction exists for the $N$ band \citep{Sterzik2005} it is not calibrated (to our knowledge) for the $L$ and $M$ bands. 

In the $N$ band, the total flux of U Aql ($\approx$1$\,$Jy) is at the lower limit of the UTs sensitivity with MATISSE for accurate visibility measurements. Instead, we use the correlated flux, which is the flux contribution from the spatially unresolved structures of the source. We calibrated the correlated flux of the science target $F_\mathrm{corr,sci}$ following :

\begin{equation}\label{eq:corr_flux_cal}
F_\mathrm{corr,sci}=\frac{I_\mathrm{corr,sci}}{I_\mathrm{corr,cal}} \times F_\mathrm{tot,cal} \times V_\mathrm{cal},
\end{equation}

\noindent where $I_\mathrm{corr,sci}$ and  $I_\mathrm{corr,cal}$ are the observed raw correlated fluxes of the science target and the calibrator respectively, and $V_\mathrm{cal}$ the calibrator visibility. For such an absolute calibration of the correlated flux, we had a robust interferometric calibrator with an atmospheric template given by \cite{Cohen1999}.

\begin{figure*}[!htbp]
\centering
\begin{subfigure}[b]{.35\textwidth}
  \includegraphics[width=\linewidth]{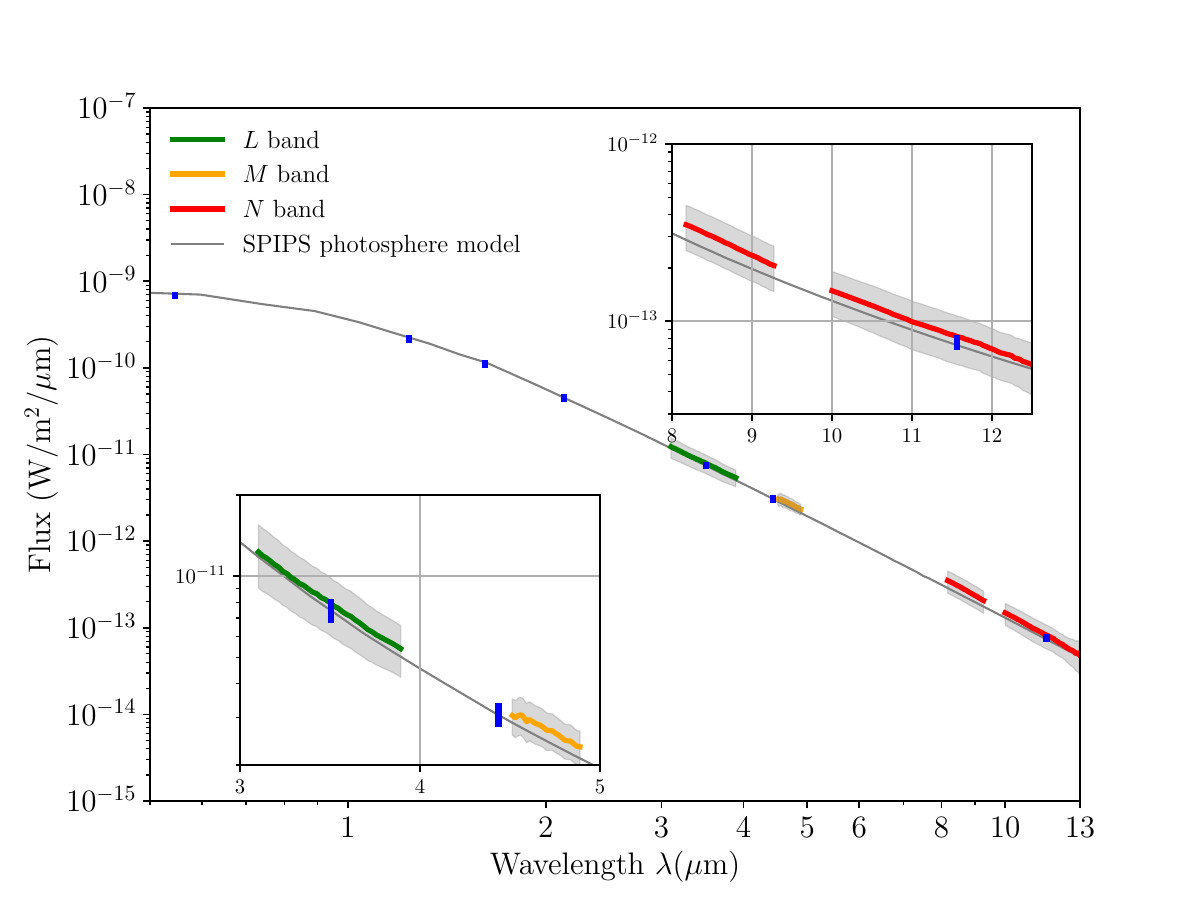}
  \caption{X Sgr, 7.01$\,$d}
  \label{fig:x_sgr_flux}
\end{subfigure}%
\begin{subfigure}[b]{.35\textwidth}
  \includegraphics[width=\linewidth]{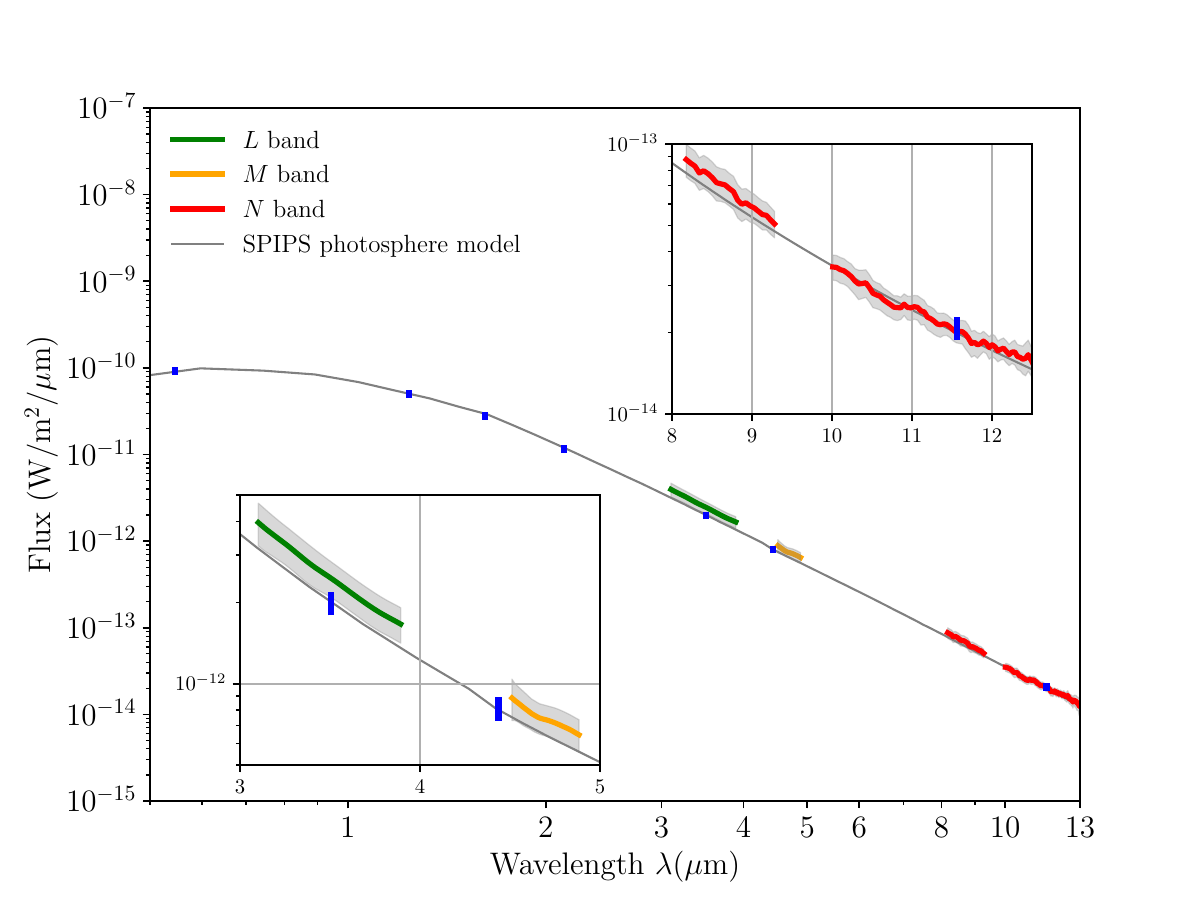}
  \caption{U Aql, 7.02$\,$d}
  \label{fig:u_aql_flux}
\end{subfigure}%
\begin{subfigure}[b]{.35\textwidth}
  \includegraphics[width=\linewidth]{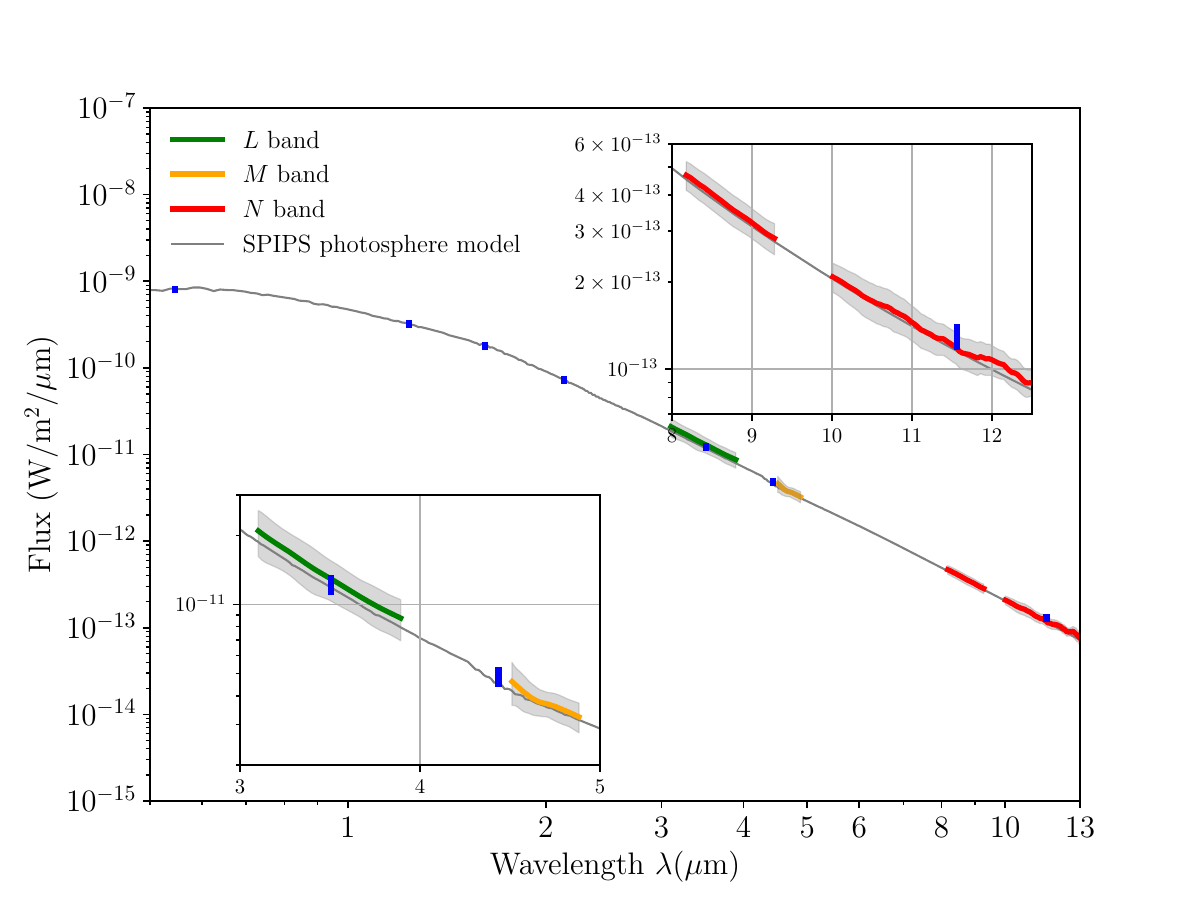}
  \caption{$\eta$ Aql, 7.17$\,$d}
  \label{fig:eta_aql_flux}
\end{subfigure}\vskip\baselineskip
\begin{subfigure}[b]{.35\textwidth}
  \includegraphics[width=\linewidth]{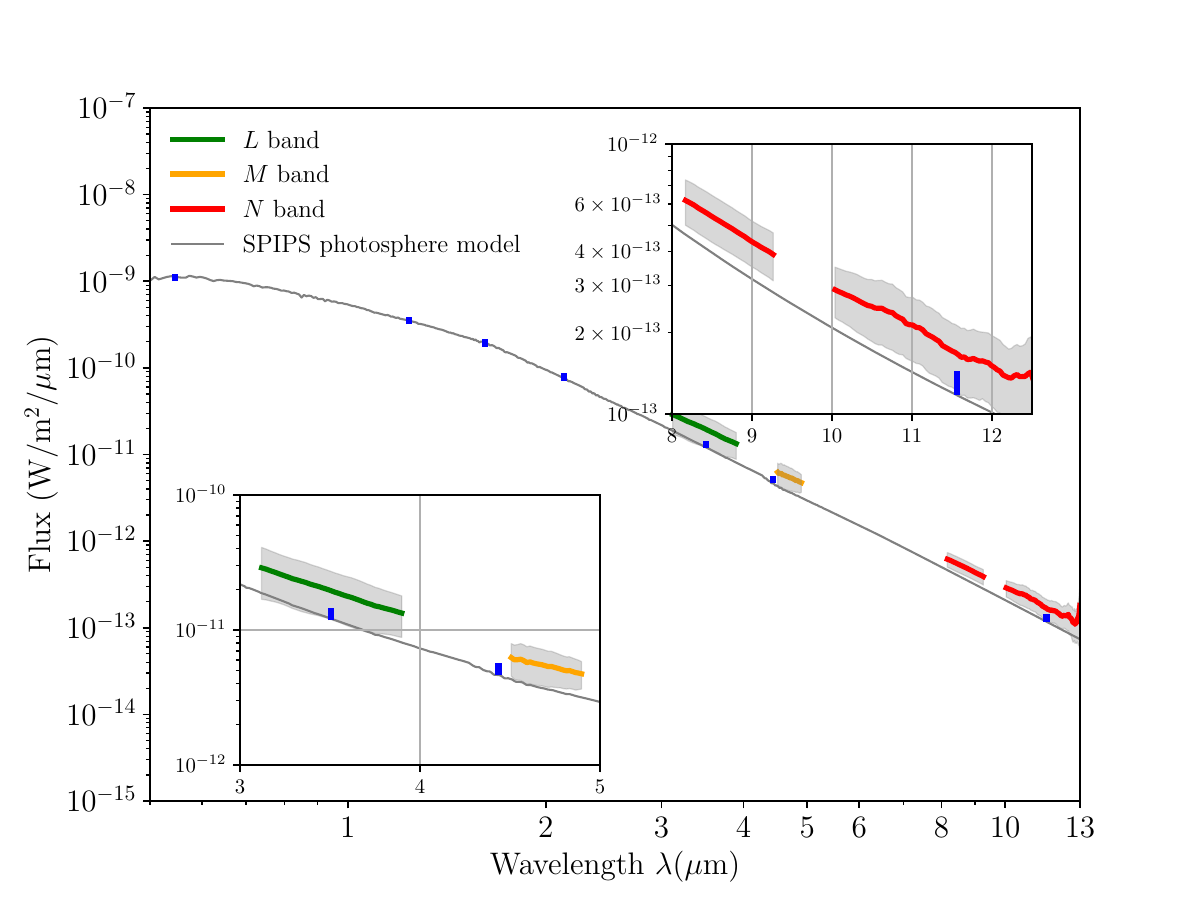}
  \caption{$\beta$ Dor, 9.84$\,$d}
  \label{fig:beta_dor_flux}
\end{subfigure}%
\begin{subfigure}[b]{.35\textwidth}
  \includegraphics[width=\linewidth]{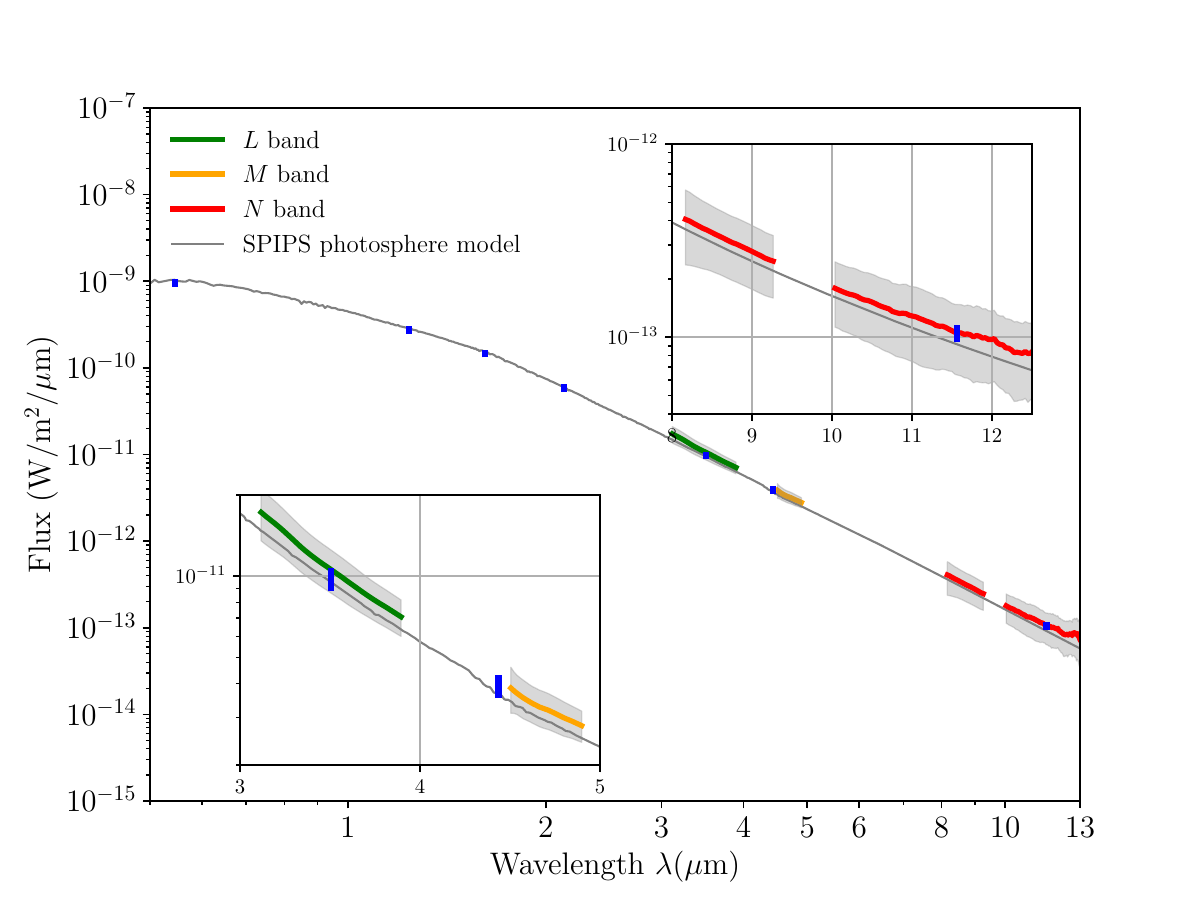}
  \caption{$\zeta$ Gem, 10.15$\,$d}
  \label{fig:zeta_gem_flux}
\end{subfigure}%
\begin{subfigure}[b]{.35\textwidth}
  \includegraphics[width=\linewidth]{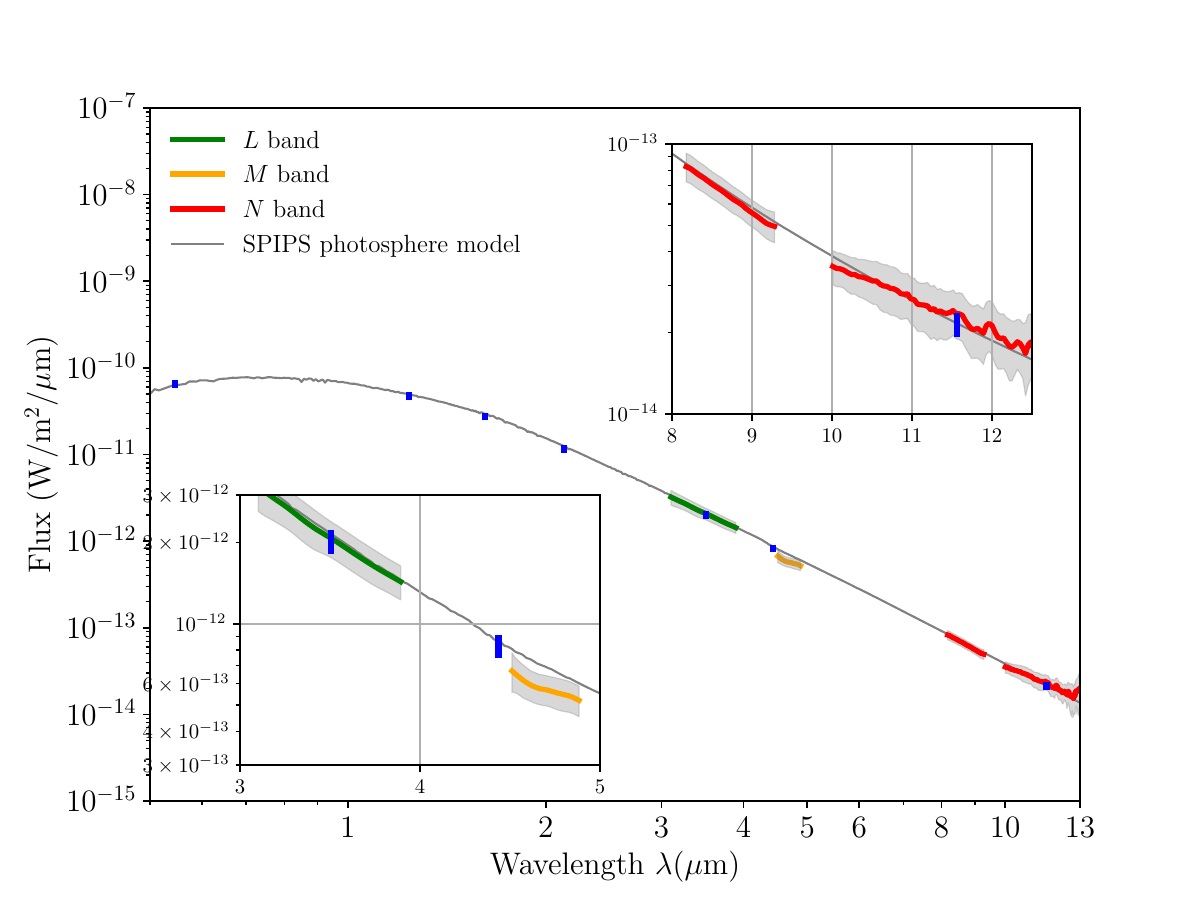}
  \caption{TT Aql, 13.75$\,$d}
  \label{fig:tt_aql_flux}
\end{subfigure}\vskip\baselineskip
\begin{subfigure}[b]{.35\textwidth}
  \includegraphics[width=\linewidth]{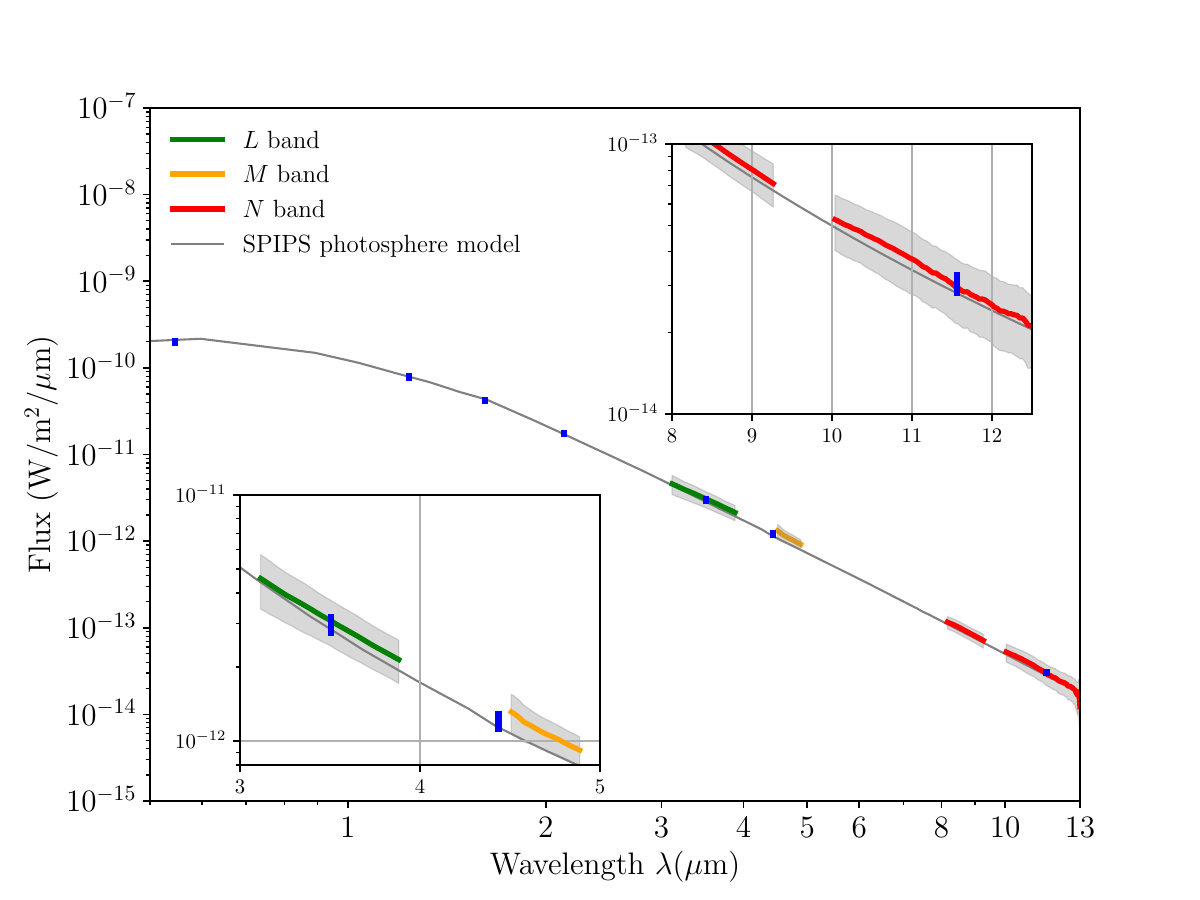}
  \caption{T Mon, 27.03$\,$d}
  \label{fig:t_mon_flux}
\end{subfigure}%
\begin{subfigure}[b]{.35\textwidth}
  \includegraphics[width=\linewidth]{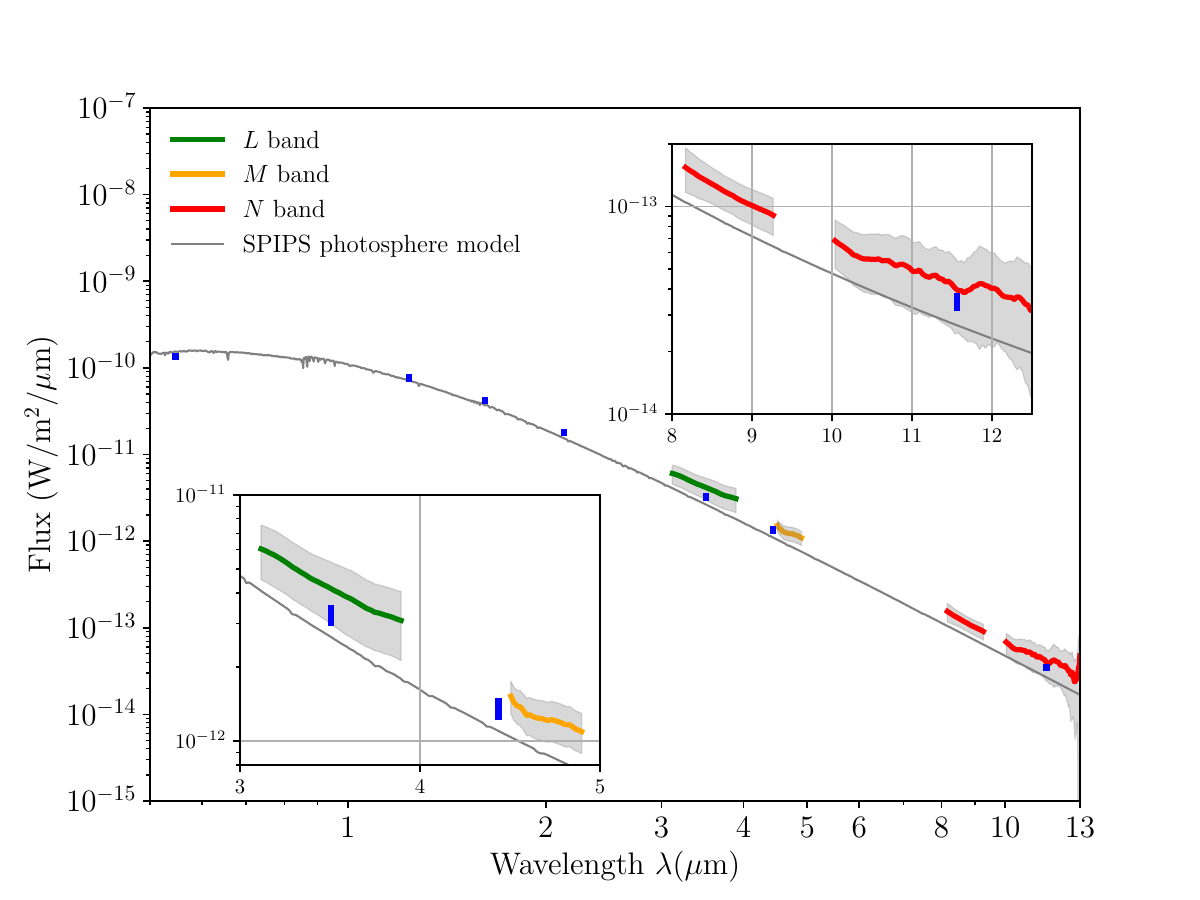}
  \caption{U Car, 38.87$\,$d}
  \label{fig:u_car_flux}
\end{subfigure}%
\caption{MATISSE calibrated flux in $L$, $M$ and $N$ bands together with ATLAS9 atmospheric model interpolated for each star by \texttt{SPIPS} at the specific pulsation phase of observations. The  blue bars represent the photometry interpolated at the phase of MATISSE observations by \texttt{SPIPS}. In the case of U Aql, the $N$-band flux is the correlated flux. See Section \ref{sect:flux_cal} for details of the calibration.}
\label{fig:total_flux}
\end{figure*}

The calibrated total fluxes in $L$, $M$ bands and $N$-band are presented in Fig.~\ref{fig:total_flux}. We discarded the spectral region between 9.3 and 10$\,\mu$m, due to the presence of a telluric ozone absorption feature around 9.6$\,\mu$m that strongly impacts the data quality. We also overplot the \texttt{SPIPS} SED of each Cepheid at the specific phase of the MATISSE observations as derived in Sect.~\ref{sect:spips} (see grey curves in Fig.~\ref{fig:total_flux}). 

\begin{figure*}[]
\centering

\begin{subfigure}[b]{0.24\textwidth}
    \includegraphics[width=\linewidth]{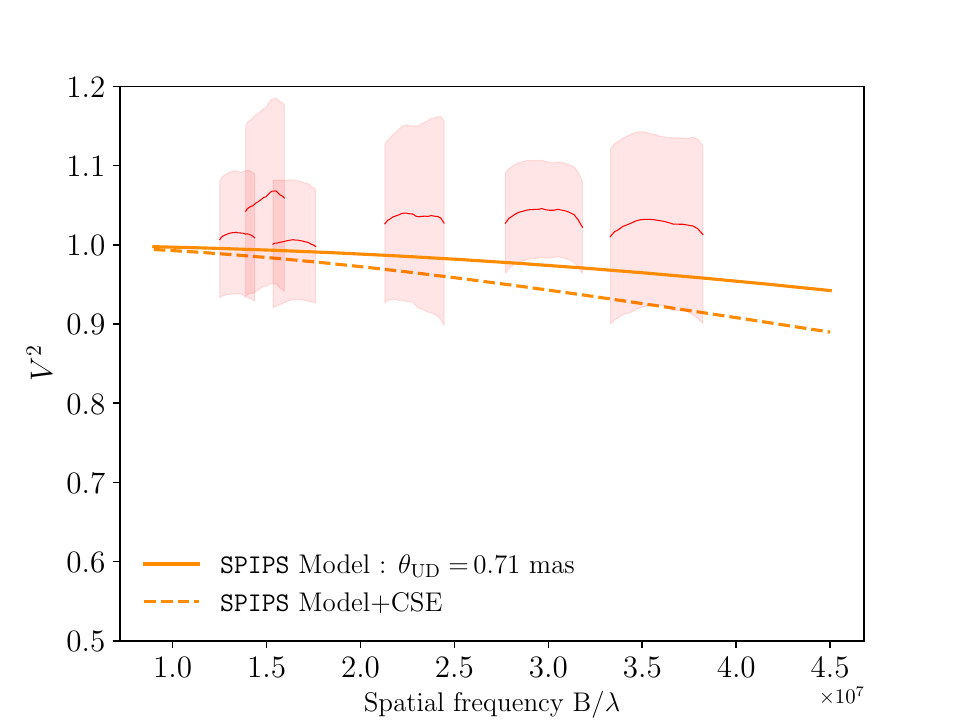}
    \caption{U Aql, 7.02$\,$d}
    \label{fig:u_aql_LM}
\end{subfigure}
\begin{subfigure}[b]{0.24\textwidth}
    \includegraphics[width=\linewidth]{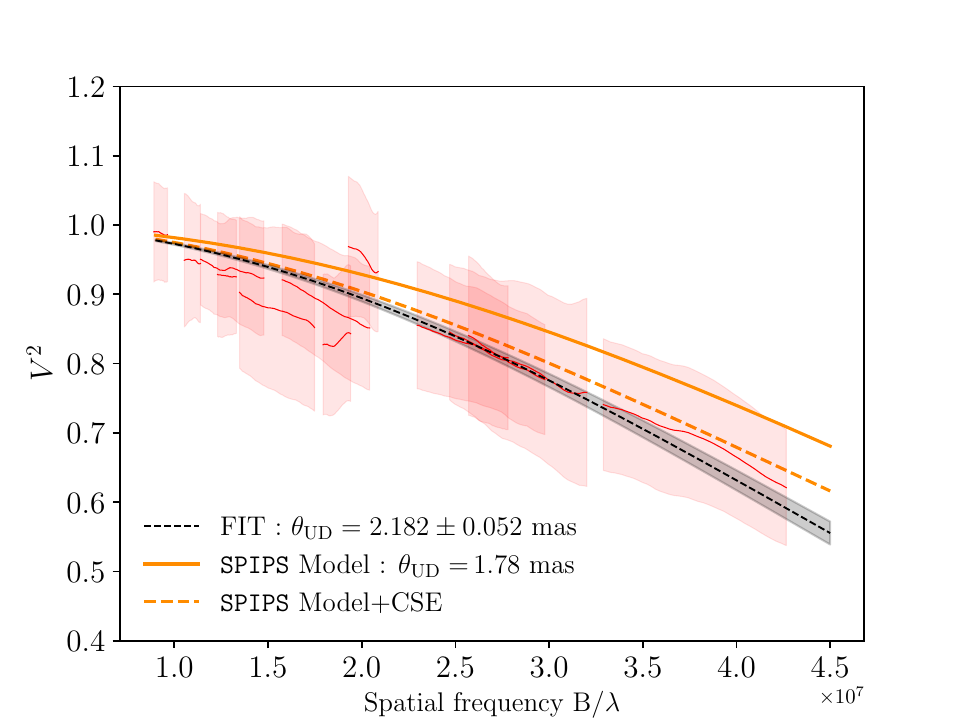}
    \caption{$\eta$ Aql, 7.17$\,$d}
    \label{fig:eta_aql_LM}
\end{subfigure}
\begin{subfigure}[b]{0.24\textwidth}
    \includegraphics[width=\linewidth]{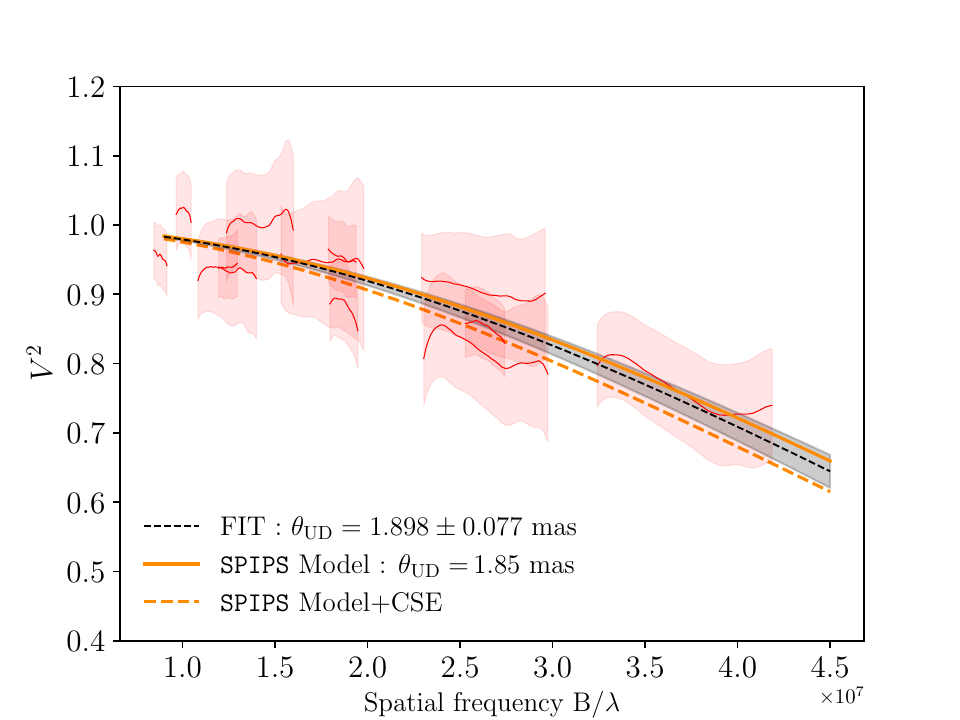}
    \caption{$\beta$ Dor, 9.84$\,$d}
    \label{fig:beta_dor_LM}
\end{subfigure}

\vskip\baselineskip
\begin{subfigure}[b]{0.24\textwidth}
    \includegraphics[width=\linewidth]{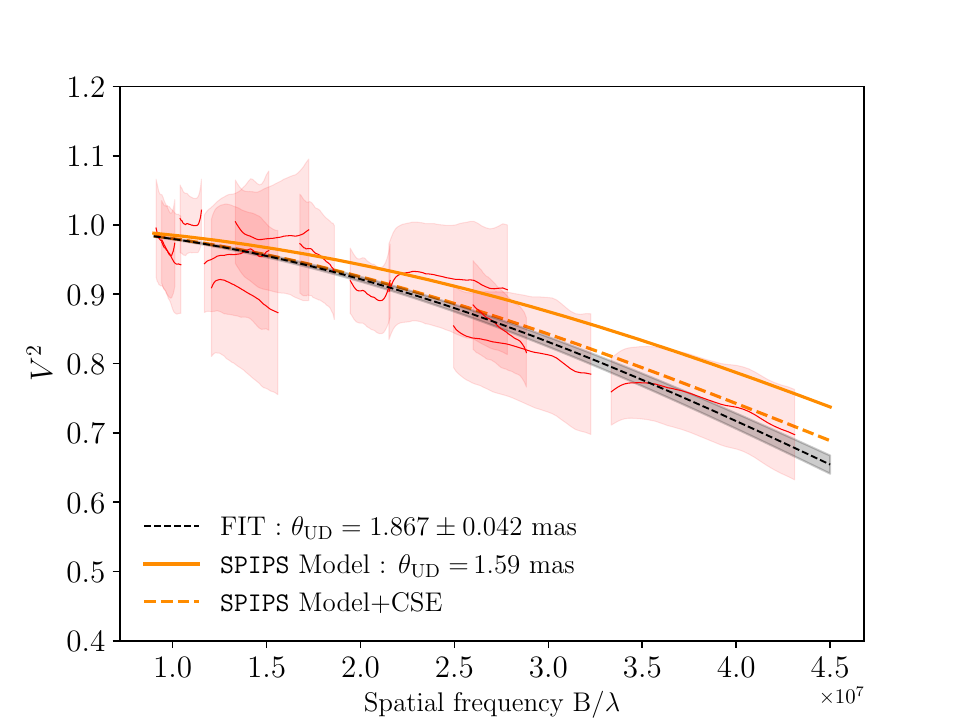}
    \caption{$\zeta$ Gem, 10.15$\,$d}
    \label{fig:zet_gem_LM}
\end{subfigure}
\begin{subfigure}[b]{0.24\textwidth}
    \includegraphics[width=\linewidth]{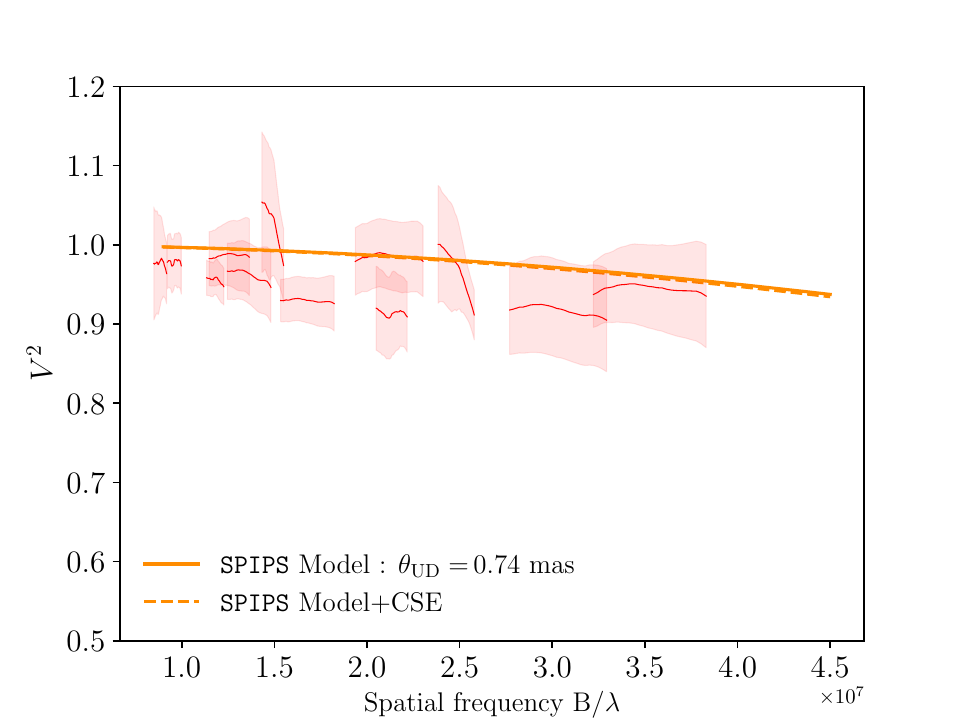}
    \caption{TT Aql, 13.75$\,$d}
    \label{fig:tt_aql_LM}
\end{subfigure}
\begin{subfigure}[b]{0.24\textwidth}
    \includegraphics[width=\linewidth]{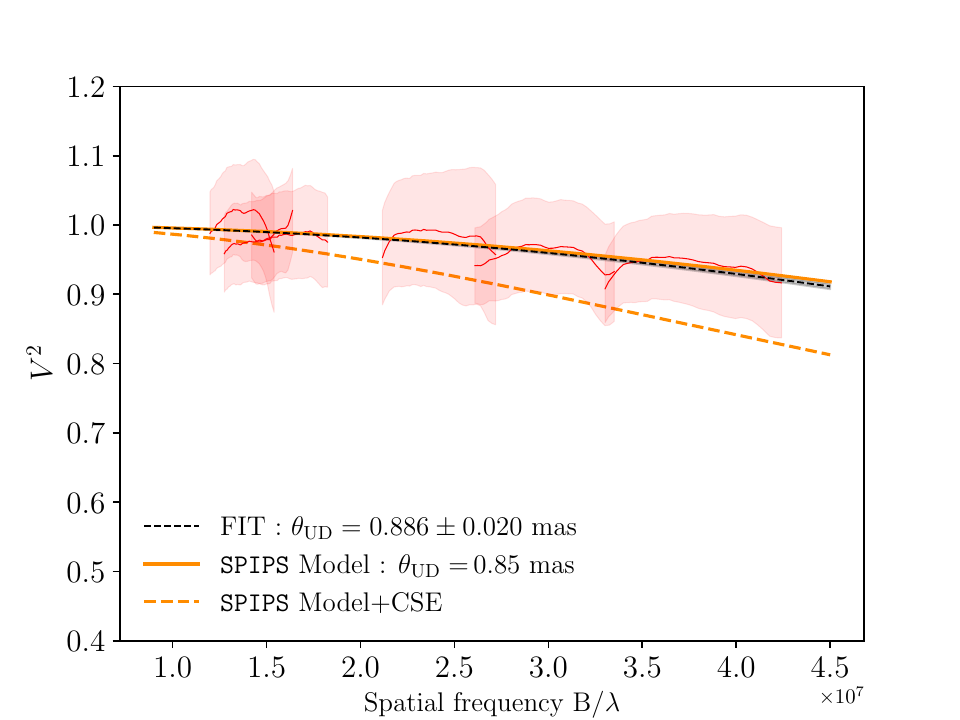}
    \caption{T Mon, 27.03$\,$d}
    \label{fig:t_mon_LM}
\end{subfigure}
\begin{subfigure}[b]{0.24\textwidth}
    \includegraphics[width=\linewidth]{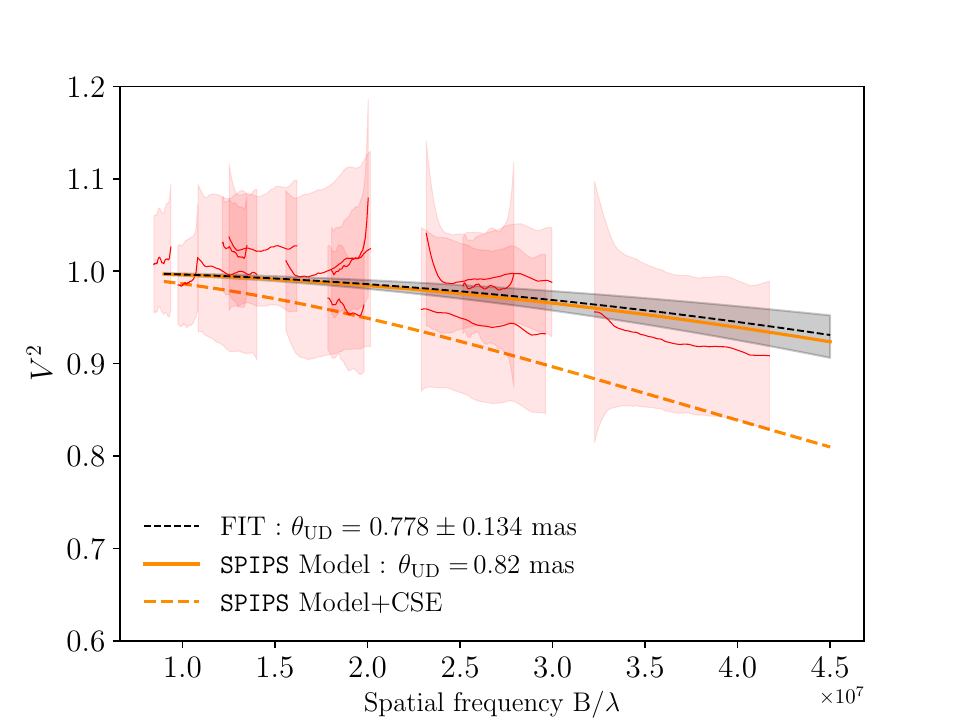}
    \caption{U Car, 38.87$\,$d}
    \label{fig:u_car_LM}
\end{subfigure}
\caption{MATISSE calibrated squared visibilities in $L$ and $M$ bands. The orange thick curve is the visibility derived from the UD diameter of the star by \texttt{SPIPS} at the specific phase of MATISSE observation. The orange dashed line represents the visibility derived for the star plus the CSE model by \texttt{SPIPS} (see Sect.~\ref{sect:ir}). The dashed black curve is the fit of the UD angular diameter with error derived from the bootstrap method (see Table \ref{tab:fit}). No model adjustments (labelled FIT) are presented for U Aql and TT Aql because these stars are mostly unresolved and are more affected by uncertainties in the absolute visibility scale. X Sgr data are impacted by a too short coherence time and thus are not presented.}
 \label{fig:vis_L}
\end{figure*}

\begin{figure*}[]
\begin{subfigure}[b]{.35\textwidth}
  \includegraphics[width=\linewidth]{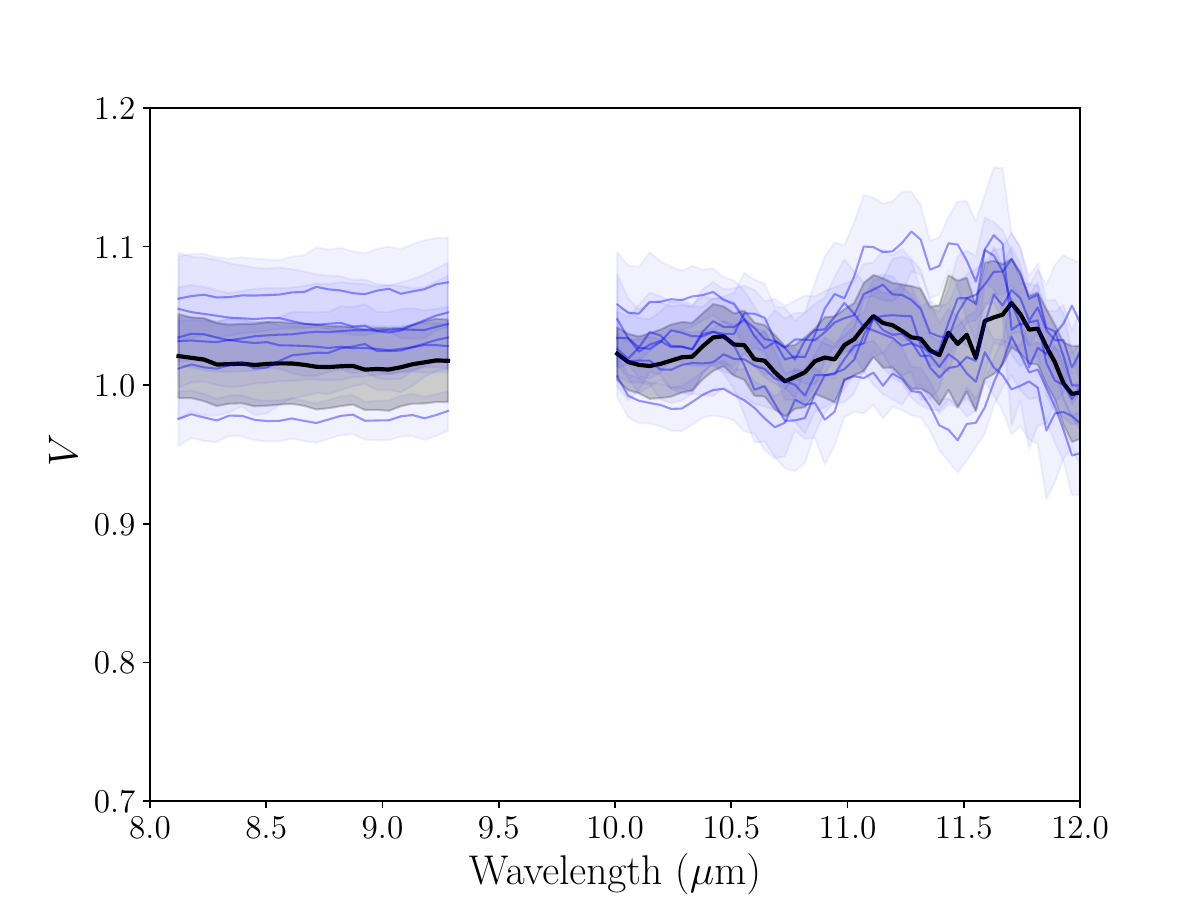}
  \caption{$\eta$ Aql, 7.17$\,$d}
  \label{fig:1}
\end{subfigure} 
\begin{subfigure}[b]{.35\textwidth}
  \includegraphics[width=\linewidth]{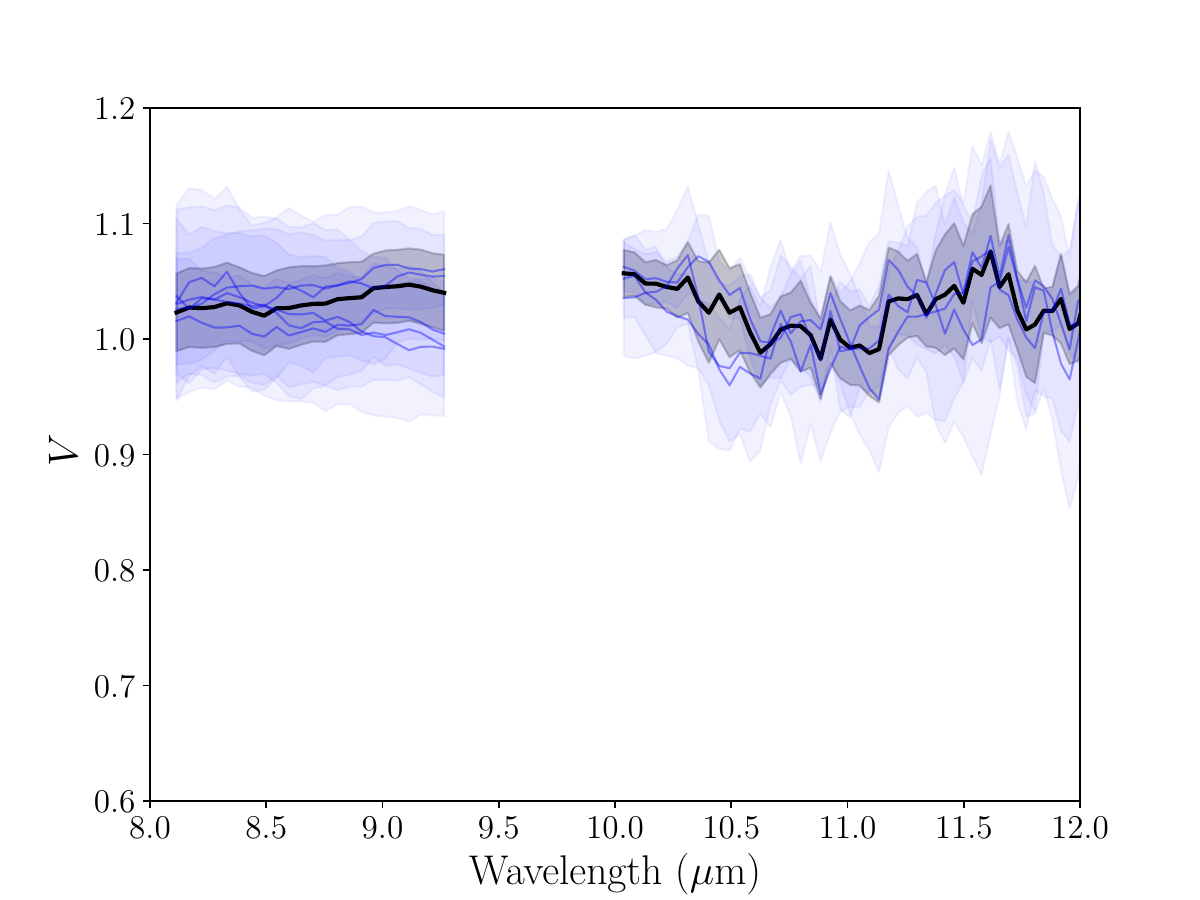}
  \caption{$\beta$ Dor, 9.84$\,$d}
  \label{fig:4}
\end{subfigure}
\begin{subfigure}[b]{.35\textwidth}
  \includegraphics[width=\linewidth]{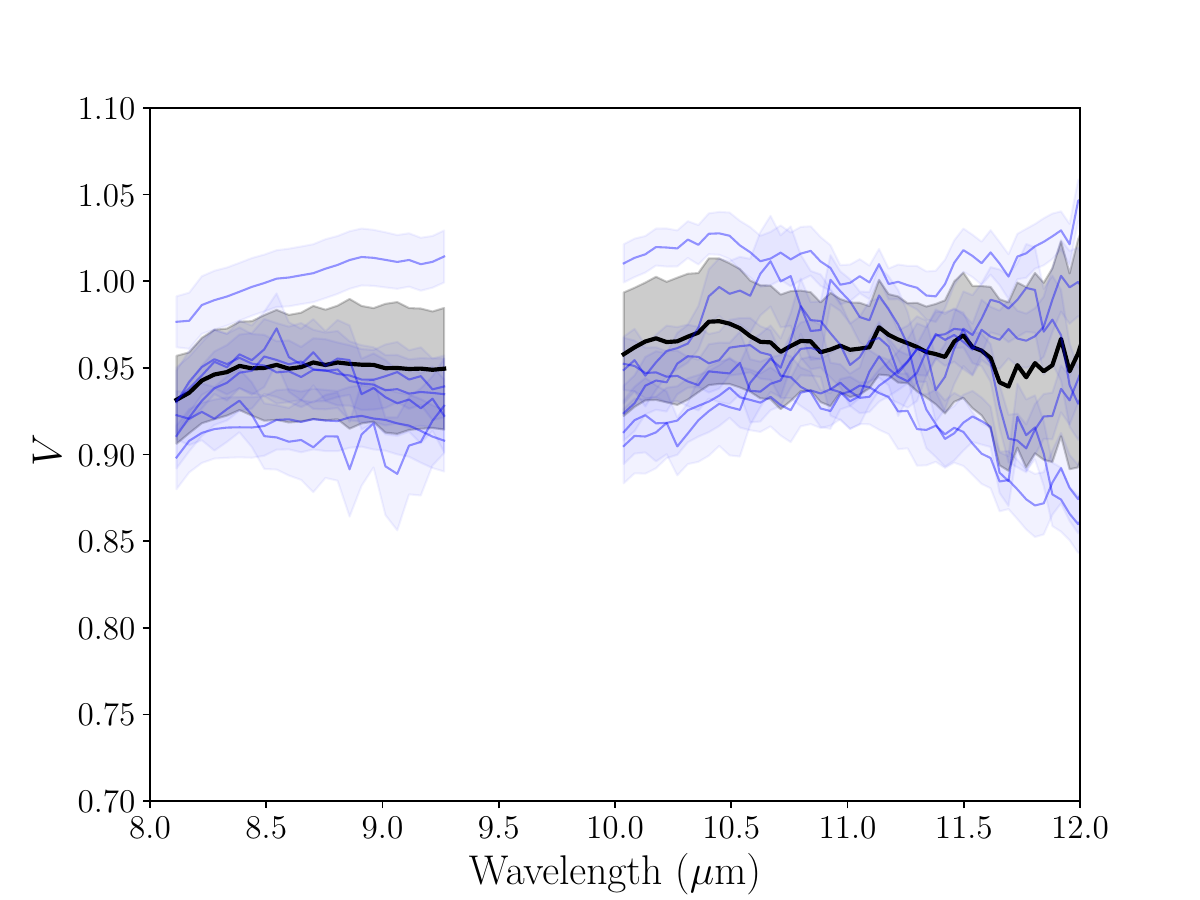}
  \caption{$\zeta$ Gem, 10.15$\,$d}
  \label{fig:1}
\end{subfigure} 
\begin{subfigure}[b]{.35\textwidth}
  \includegraphics[width=\linewidth]{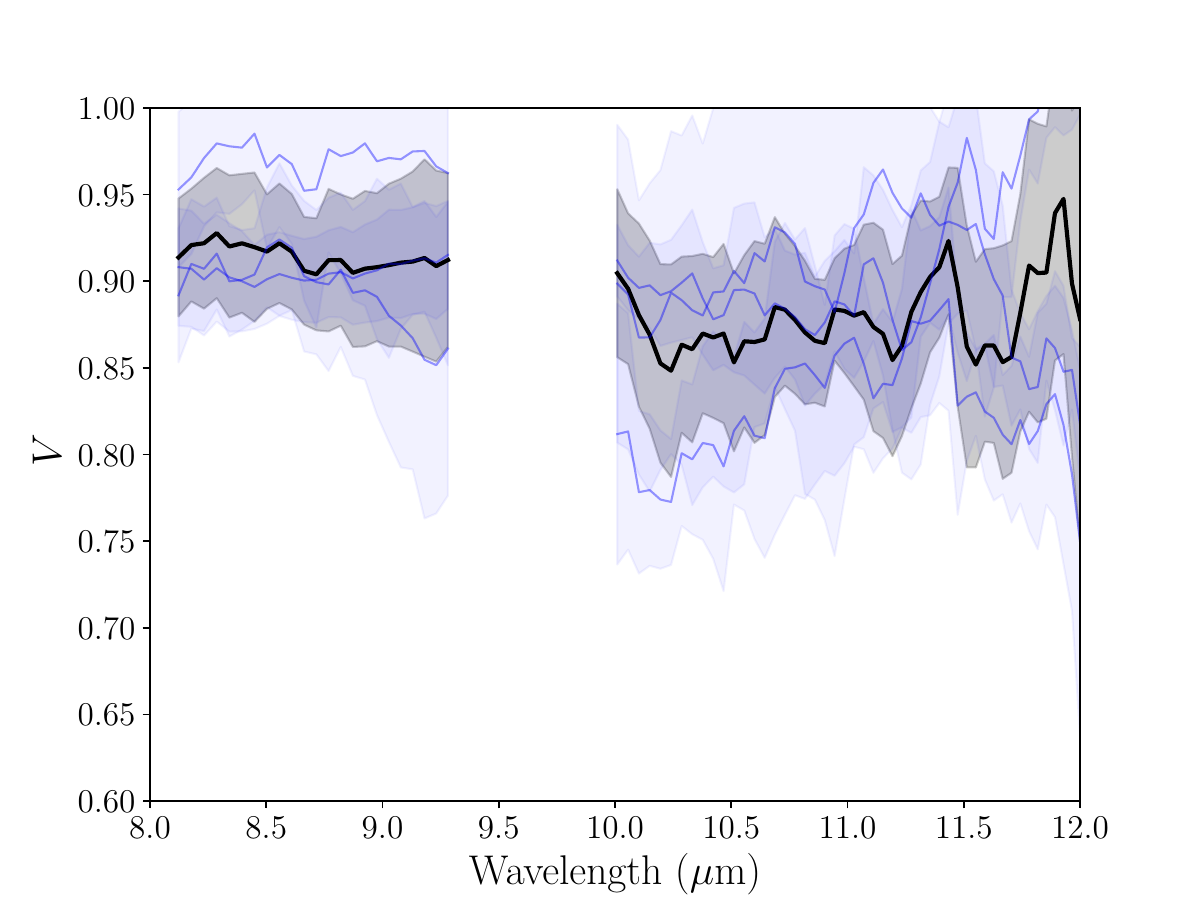}
  \caption{TT Aql, 13.75$\,$d}
  \label{fig:4}
\end{subfigure}
\begin{subfigure}[b]{.35\textwidth}
  \includegraphics[width=\linewidth]
  {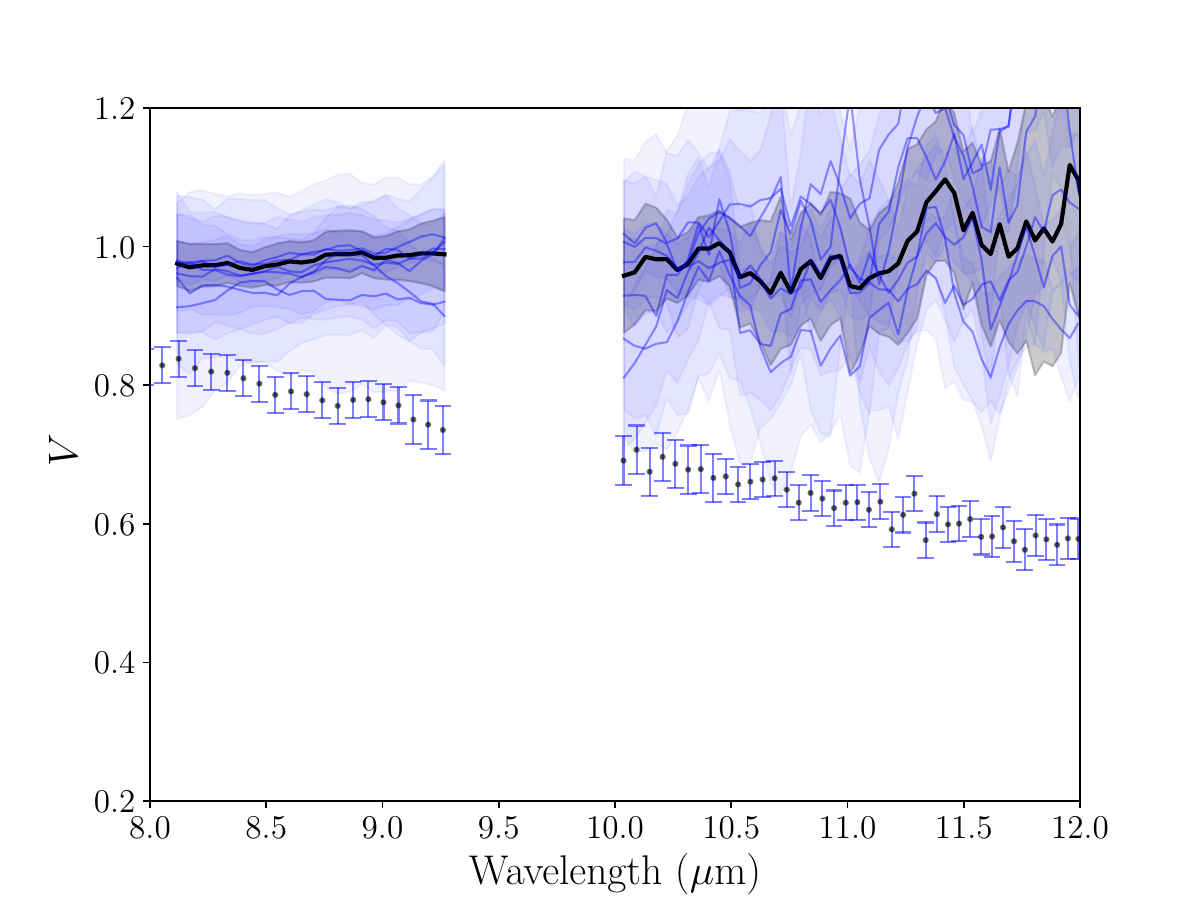}
  \caption{T Mon, 27.03$\,$d}
  \label{fig:t_mon_N}
\end{subfigure}
\begin{subfigure}[b]{.35\textwidth}
  \includegraphics[width=\linewidth]{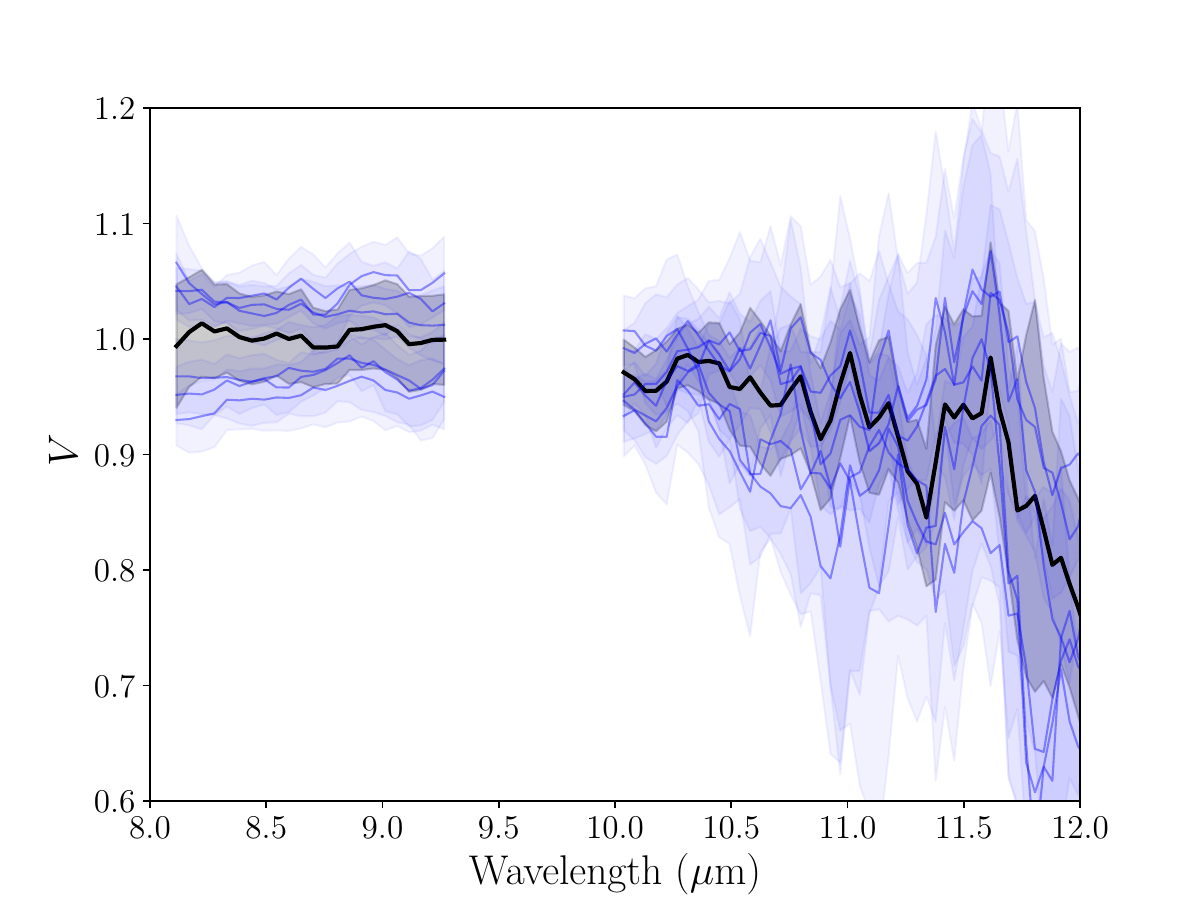}
  \caption{U Car, 38.87$\,$d}
  \label{fig:4}
\end{subfigure}
\caption{Visibility in $N$-band plotted against the wavelength. The black curve represents the weighted mean of the observations. U~Aql is not presented since we measured only the correlated flux and X~Sgr photometry in the $N$-band was not usable. Spectral band between 9.3 and 10$\,$ micron is removed because the atmosphere is not transmissive. In the case of T~Mon, we overplot the weighted mean of MIDI/VLTI observation as given by \cite{gallenne13b} for comparison.}
 \label{fig:vis_N}
\end{figure*}


\subsection{Analysis}
We draw here several important preliminary conclusions. First of all, we can see that the $LM$ and $N$-band flux are aligned with a pure Rayleigh-Jeans slope. In particular, the $N$ band spectra indicates the absence of oxygen-rich dust emission for all Cepheids of our sample, such as silicates which exhibit a prominent feature between 9 and 12 micron \citep{DraineLee1984}. In the case of $\eta$~Aql, $\zeta$~Gem and $\ell$~Car \textit{Spitzer} low-resolution spectra already revealed the absence of dust emission \citep{Hocde2020a,Hocde2021}. Interestingly, \cite{gallenne13b} suggested silicate dust features based on MIDI/VLTI observations for X~Sgr and T~Mon while our observations yield an opposite result for these two stars, as we do not observe any peculiar trends in the $N$-band flux. We discuss further this difference in the analysis of the MATISSE visibility in Sect.~\ref{sect:vis}.

Secondly, the SEDs as derived by \texttt{SPIPS} at the specific phase of MATISSE observations are in agreement with the calibrated flux within the uncertainties, except for $\beta$ Dor for which we derive an excess up to about 50\% of the MATISSE flux in $L$,$M$ and $N$ bands. We did not find the origin of this discrepancy since the photometric band from \texttt{SPIPS} are well adjusted, and on the other hand, the calibration seems reliable with good observing conditions, similar airmass, and consistent template of the calibrator HD 59219. However it is not relevant to derive the IR excess from MATISSE flux observation that could be due to CSEs because the measurements have rather large uncertainties ($\geq$ 10\%$\approx$0.10$\,$mag) which is not suited for deriving subtle IR excesses of the order of 5\% of the total flux.

\section{Closure phase}\label{sect:T3}
MATISSE also provides closure phase measurements, which contain information about the spatial centro-(a)symmetry of the brightness distribution of the source. Given the phase measurements of a telescope triplet $ijk$, the closure phase is the combination $\phi_{ijk}=\phi_{ij}+\phi_{jk}-\phi_{ik}$. One of the advantages of the closure equation is that it cancels out any atmospheric or baseline-dependent phase error \citep{Jennison1958}. For all the  closure phase measurements, we find an average of about 0$^\circ$ in the $L$, $M$ and the $N$ band (see Fig.~\ref{fig:T3}). In $N$-band, the spectral window beyond 10$\,\mu$m is too noisy to obtain meaningful conclusion for U~Aql, TT Aql and T Mon. Overall, down to a sub-degree level, our closure phase measurements are consistent with the absence of significant brightness spatial asymmetries in the environment for all Cepheids in $L$, $M$ bands and $N$ bands.

\section{Visibility in $LM$ and $N$ bands}\label{sect:vis}
\subsection{Calibration}
As discussed in Sect.~\ref{sect:cal_sci} the targets were calibrated using specific calibrators for the LM-band and the $N$-band. Hence, we performed by default a CAL-SCI sequence of calibration separately for both $L$ and $N$-bands. In some cases, we obtained a better calibration in the $L$-band using the brighter $N$-band calibrator instead, as it is the case for example of $\eta$ Aql. For X~Sgr, $\beta$~Dor and $\zeta$~Gem we could only observe one calibrator. In this case the same calibrator was used for the calibration in $L$ and $N$-bands. Moreover, each target required a special attention because of peculiar atmospheric conditions or technical problems during the observations that could affect the visibility curves. Hence, we scrutinized the raw visibilities to ensure their overall quality for both target and calibrators. When we observed an obvious discrepant exposure for a corresponding baseline --generally deviating by more than 2$\sigma$ compared to other exposures of adjacent baselines-- we simply excluded this latter from the analysis. For U Aql we note that the calibrated visibilities are slightly above 1. 
The results of the calibration for the squared visibility $V^2$ in $LM$-band and the visibility $V$ in the $N$-band are presented in Figs.~\ref{fig:vis_L} and \ref{fig:vis_N} respectively; the error bars represent the standard deviation between the visibilities of the different exposures, within one observing sequence. The visibility curve associated with the expected UD angular diameter of the star as derived from the \texttt{SPIPS} analysis is shown for comparison (see orange curves in Figs.~\ref{fig:vis_L} and \ref{fig:vis_N}). We also displayed the visibility of the latter model taking into account the CSE model derived by \texttt{SPIPS} in each case (see dashed curves in Figs.~\ref{fig:vis_L} and \ref{fig:vis_N}).

\subsection{Results in the $LM$-bands}\label{sect:resuls_LM}
For all Cepheids of our sample, the MATISSE observations in the $L$ and $M$ bands are in agreement with the \texttt{SPIPS} UD angular diameter. Moreover, we derived the UD angular diameter from the MATISSE observations in $LM$ band by fitting the squared visibility $V^2_\mathrm{UD}(f)$ following:
\begin{equation}\label{eq:v_ud}
    V_\mathrm{UD}^2(f)= \left( 2\frac{J_1(\pi \theta_\mathrm{UD}f)}{\pi \theta_\mathrm{UD}f}\right)^2,
\end{equation}
where $J_1$ is the Bessel function of the first order and where $f$=$B_p$/$\lambda$ is the spatial frequency ($B_p$ the length of the projected baseline). We used the \texttt{PMOIRED}\footnote{Code available at \url{https://github.com/amerand/PMOIRED}}
code \citep{PMOIRED2022} to perform the fit of the UD angular diameter and estimate the uncertainties via the bootstrap method. We provide the UD angular diameters in Table \ref{tab:fit}. From these results it is not possible to resolve a CSE emission of the order of 0.05 to 0.10$\,$mag in the $L$-band. As we can see from Fig.~\ref{fig:vis_L}, the \texttt{SPIPS} CSE model (dashed line) cannot be resolved by MATISSE. We provide upper limits for both the size and the CSE flux contribution in Sect.~\ref{sect:upper_limit}.

\begin{table}[t]
\centering
\begin{threeparttable}
\caption{Fits of UD angular diameter.}\label{tab:fit}
\begin{tabular}{l|c|c}
\hline
\hline
Star    & Date (MJD)     &  $\theta_\mathrm{UD}$ (mas)  \\
\hline
$\eta$ Aql         &    59420.19          & 2.182$\pm$0.052\\
$\beta$ Dor        &  59244.03            & 1.898$\pm$0.077 \\ 
$\zeta$ Gem        &  59300.06            & 1.867$\pm$0.042\\
T Mon              &   58923.06           & 0.886$\pm$0.020 \\
U Car              &   59300.12           & 0.778$\pm$0.134 \\
\hline  
\end{tabular}
\begin{tablenotes}
    \item \textbf{Notes.} Uncertainties are derived using the boostrap method implemented in \texttt{PMOIRED} \citep{PMOIRED2022}.
\end{tablenotes}
\end{threeparttable}
\end{table}

\subsection{Results in the $N$-band}
In the $N$-band, the visibilities are more affected by the thermal background which impact the absolute visibility level and the signal to noise ratio. However, it is clear from the Fig.~\ref{fig:vis_N} that we do not resolve any environment for the six stars presented since we obtain rather horizontal visibility level without any significant trend. This observation motivates us to perform a weighted mean of the visibility at different baseline which highlight the horizontal visibility trend (see black curve in Fig.~\ref{fig:vis_N}). This result is in agreement with the SED in the $N$-band presented in Sect.~\ref{sect:flux_cal} which does not present silicate emission features which should produce a large dip of the visibility centered on 9.7$\mu$m. As for the SED analysis, the measured visibilities with MATISSE indicates that there is not resolved CSE around T~Mon which disagrees with the results found by \cite{gallenne13b} with MIDI/VLTI (see Fig.~\ref{fig:t_mon_N}. This discrepancy cannot be attributed to a difference of calibrator, as we chose the same calibrator as in G13, namely 30~Gem and 18~Mon. We note that \cite{Gro2020} also found contradictory results with the G13 SED of T~Mon. However, G13 reported that the excess of T~Mon might suffer from sky background contamination which might be at the origin of this difference. In the case of X~Sgr, the MATISSE $N$-band visibility is also strongly affected by the poor atmospheric conditions and is not discussed. Thus, we cannot compare with the results from G13 who resolved a CSE in the $N$-band with MIDI/VLTI with a relative CSE contribution of $f= 13.3\pm0.7$\%. However, the total flux of X~Sgr in the $N$-band as measured by MATISSE suggests that there is an absence of silicate features (see Figs.~\ref{fig:x_sgr_flux}). This result disagrees with the observations from \cite{Gallenne2012,gallenne13b} who detected a trend in the SED, modeled with a radiative transfer of CSE with dust mixture.

In the case of U~Car, \cite{Gallenne2012} detected a spatially extended emission with a significant residual flux of $\Delta F = 16.3\pm1.4$\% in the $N$-band. They also proposed a model of silicate dust mixture to explain the SED. While the IR excess in the $N$-band as derived by \texttt{SPIPS} is in close agreement with $\Delta$mag$=0.13\,$mag (see Fig.~\ref{fig:u_car_ir} and Table~\ref{tab:spips_mag}), we do not find silicate feature in the MATISSE $N$-band spectrum and visibility, which suggest another physical source of the IR emission. \cite{Gallenne2012} note however that their observations could be affected by the interstellar-cirrus background emission in the region of this star also observed from IRAC 8$\,\mu$m observations \citep{Barmby2011}.

\subsection{Upper limits on the CSE in the $LM$ band}\label{sect:upper_limit}

\begin{figure}[]
\begin{subfigure}{0.5\textwidth}
\includegraphics[width=\linewidth]{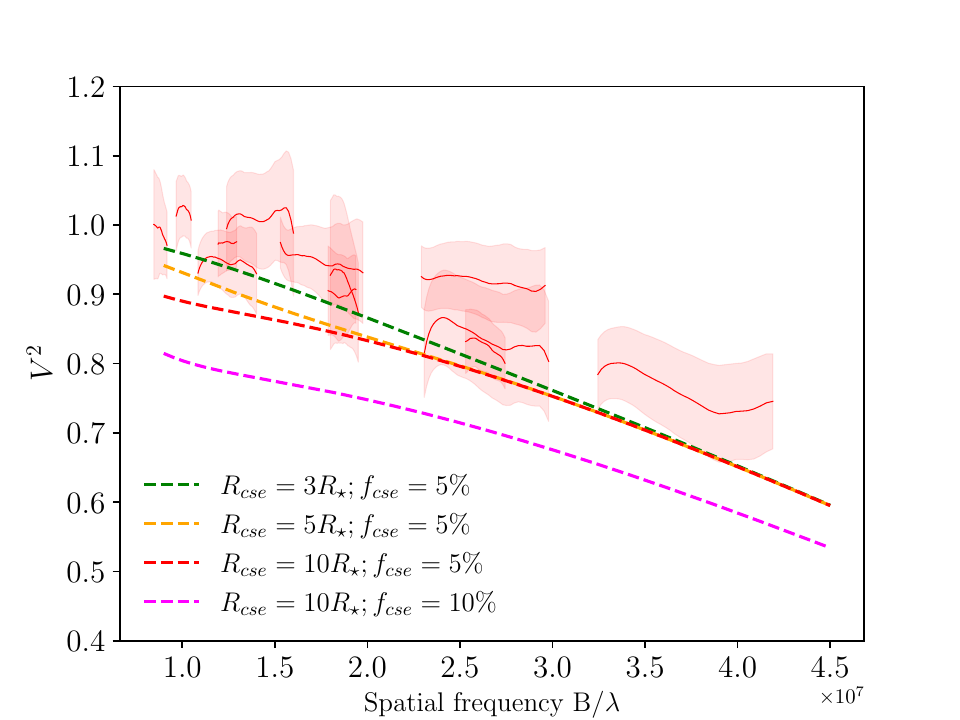}
\caption{}  
\end{subfigure}\hspace*{\fill}

\begin{subfigure}{0.5\textwidth}
\includegraphics[width=\linewidth]{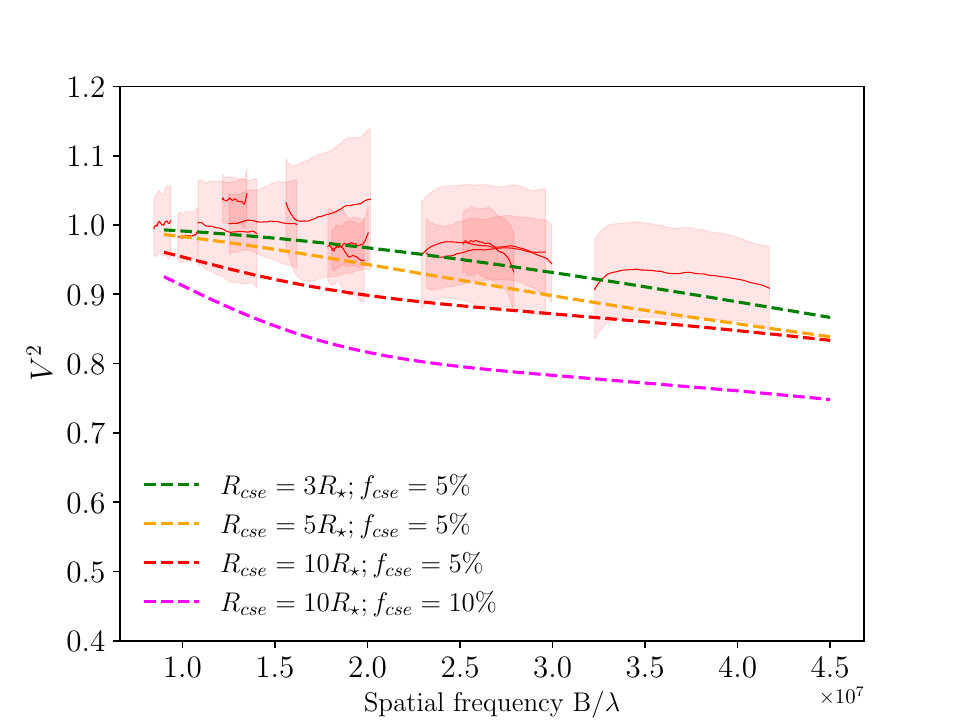}
\caption{}  
\end{subfigure}
\caption{Examples of CSE models with a Gaussian brightness intensity distribution in the case of (a) $\beta$~Dor (1.86$\,$mas) and (b) U~Car (0.80$\,$mas) which illustrate the difficulty to resolve a compact CSE with MATISSE/VLTI. For a CSE flux contribution of 5\% of the total flux, compact CSEs between 5 and 10$\,R_\star$ cannot be resolved (see green, orange and red models). The large and bright model shown as a magenta dashed line is however excluded in both stars. Upper limits for different CSE models are given in Table~\ref{tab:star_radii}.} \label{fig:upper}
\end{figure}

\begin{figure}[!htbp]
\centering

\begin{subfigure}[b]{.24\textwidth}
  \includegraphics[width=\linewidth]{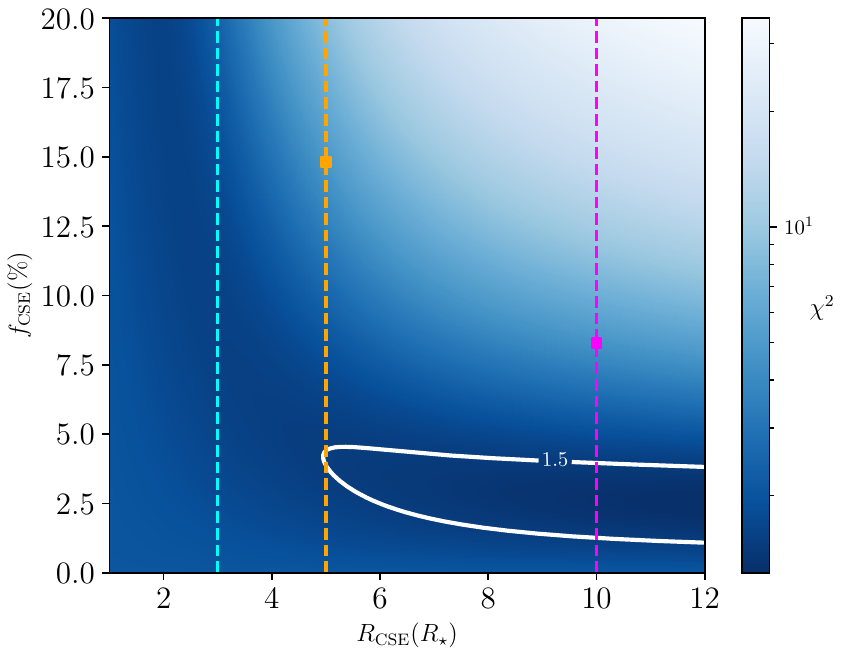}
  \caption{U Aql, 7.02$\,$d}
\end{subfigure}%
\begin{subfigure}[b]{.24\textwidth}
  \includegraphics[width=\linewidth]{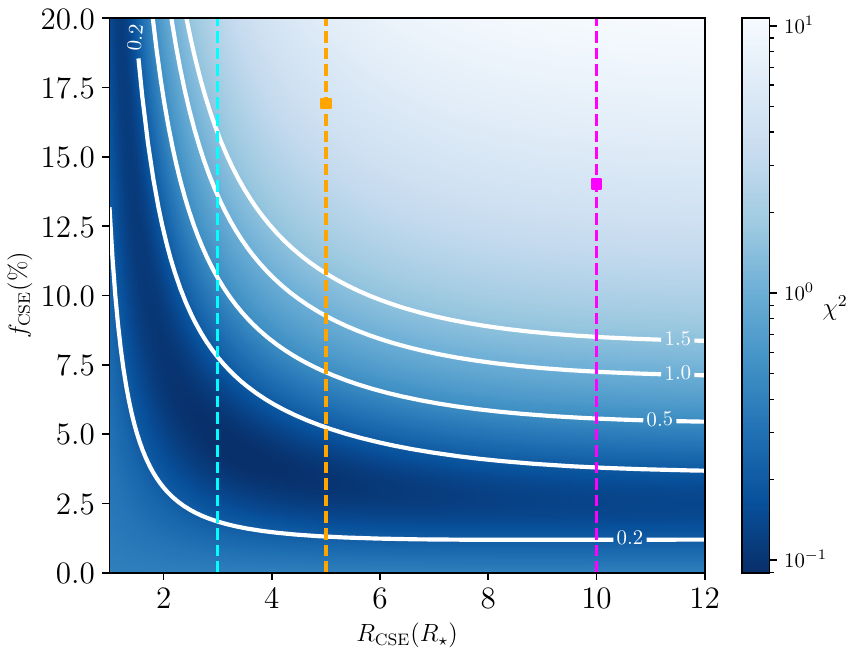}
  \caption{$\eta$ Aql, 7.17$\,$d}
\end{subfigure}\vskip\baselineskip
\begin{subfigure}[b]{.24\textwidth}
  \includegraphics[width=\linewidth]{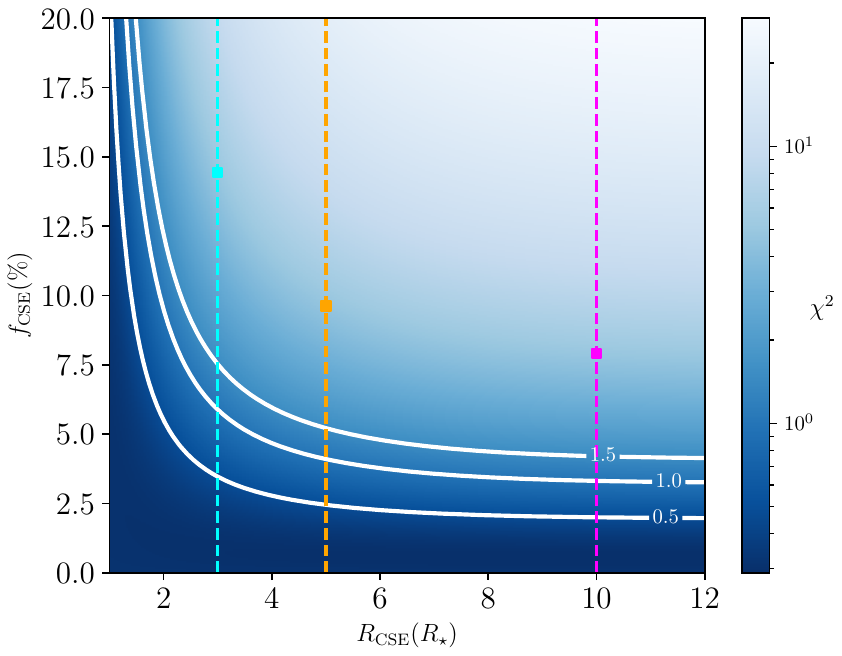}
  \caption{$\beta$ Dor, 9.84$\,$d}
\end{subfigure}%
\begin{subfigure}[b]{.24\textwidth}
  \includegraphics[width=\linewidth]{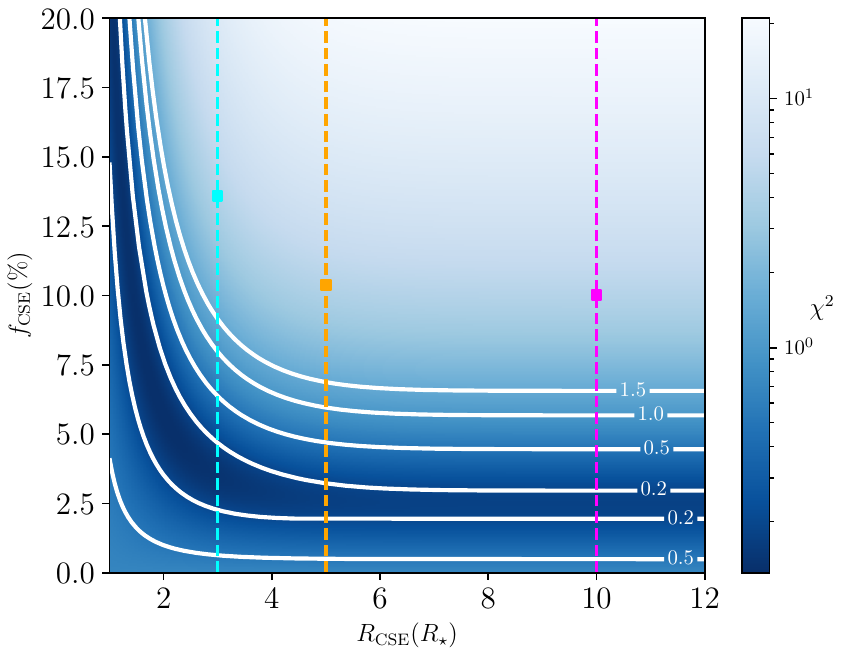}
  \caption{$\zeta$ Gem, 10.15$\,$d}
\end{subfigure}\vskip\baselineskip
\begin{subfigure}[b]{.24\textwidth}
  \includegraphics[width=\linewidth]{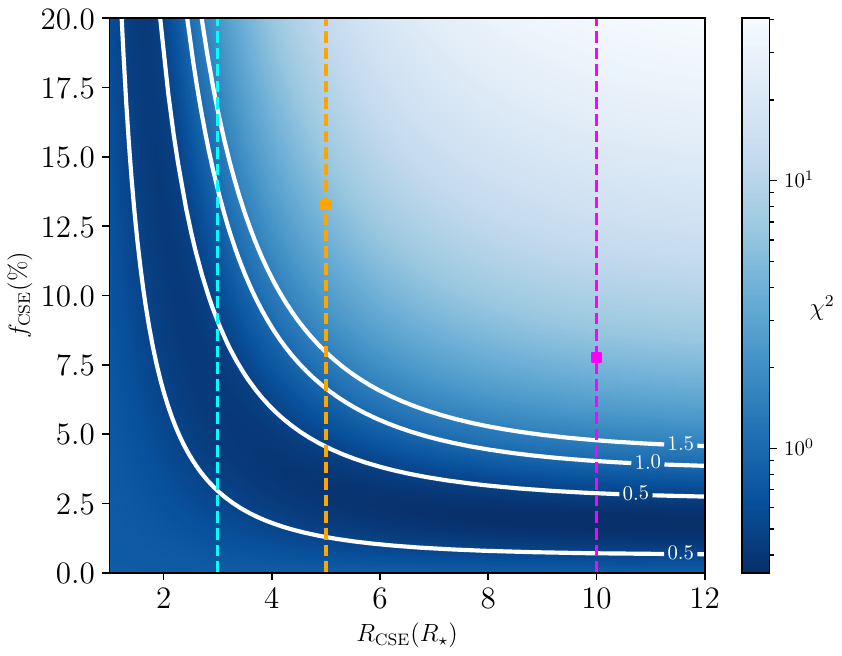}
  \caption{TT Aql, 13.75$\,$d}
\end{subfigure}
\begin{subfigure}[b]{.24\textwidth}
  \includegraphics[width=\linewidth]{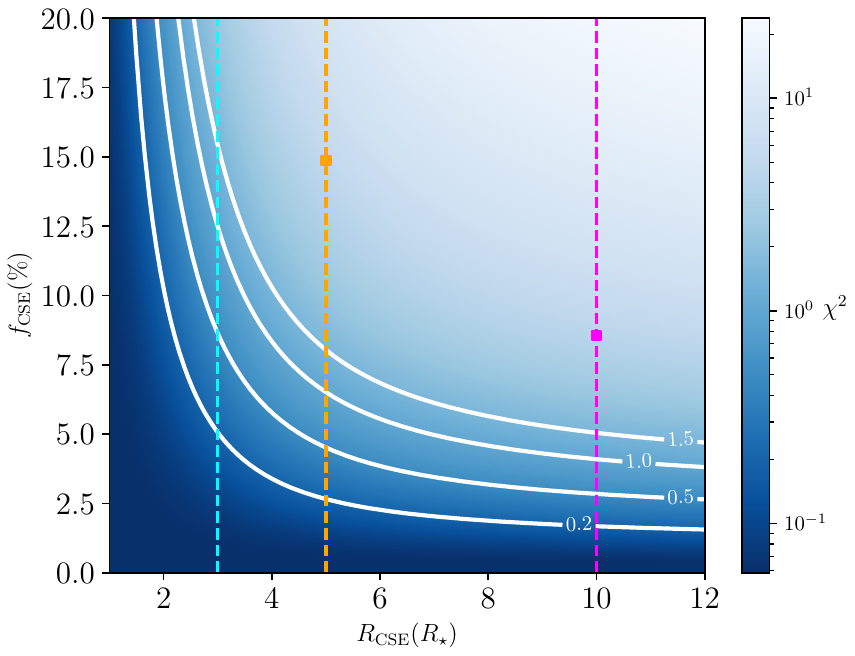}
  \caption{T Mon, 27.03$\,$d}
\end{subfigure}\vskip\baselineskip
\begin{subfigure}[b]{.24\textwidth}
  \includegraphics[width=\linewidth]{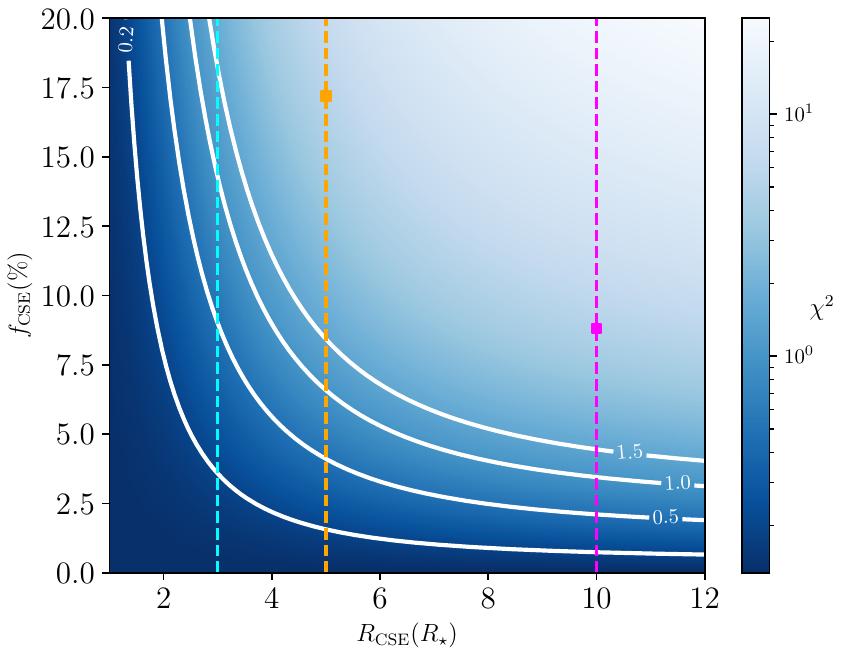}
  \caption{U Car, 38.87$\,$d}
\end{subfigure}%
\caption{Reduced $\chi^2$ map as a function of flux contribution $f_\mathrm{CSE}$ and FWHM ($R_\mathrm{CSE}$) of the Gaussian CSE model in $L$ and $M$ bands. Cyan, orange and purple dashed lines represent different CSE radius (FWHM) for which we provided upper limit on the CSE flux contribution displayed as a square (see Table \ref{tab:star_radii}). A weak flux contribution (<5\%) does not permit to constrain the CSE dimensions. Conversely, the flux of a compact envelope (<3-4$\,R_\star$) cannot be determined. Large and bright models ((R>$10\,R_\star$;$f>10\%$, square purples) can be however excluded.}
\label{fig:chi2}
\end{figure}

Our results in $LM$ and $N$-bands suggest that we do not resolve any significant circumstellar emission for all the star of our sample. However, the large error bars prevent the detection of a compact and/or small emission around these stars. For example, we can see that the \texttt{SPIPS} CSE model is also consistent with the MATISSE visibilities (see orange dashed lines in Fig.~\ref{fig:vis_L}). To better illustrate the difficulty to resolve compact CSE on the $L$-band visibility curve, we present different CSE models in Fig.~\ref{fig:upper}. The shape of the first visibility lobe provides crucial information regarding the size and flux of the CSE. However, it is difficult to consistently define upper limits for each star as the flux and size of the CSE are fully degenerate within the uncertainty. In order to provide upper limits to the extension and the emission of the CSE, we applied a geometrical model on the measured $LM$ band visibilities. Each Cepheid is modeled with a UD angular diameter $\theta_\mathrm{UD}$ which is fixed to the one at the specific phase of the observation derived by \texttt{SPIPS} in the $L$ band from Table \ref{Tab.ephemeris}. The visibility $V_\mathrm{UD}(f)$ is derived according to Eq. \ref{eq:v_ud} presented in Sect.~\ref{sect:resuls_LM}. We superimposed to this star model a CSE modeled with a Gaussian intensity distribution with a full-width at half maximum (FWHM) $\theta_\mathrm{CSE}$. The visibility of the envelope $V_\mathrm{CSE}(f)$ is derived following the Van Cittert-Zernicke theorem \citep{Berger2007}:
\begin{equation}
    V_\mathrm{CSE}(f)=\exp \Big[ -\frac{(\pi\theta_\mathrm{CSE}f)^2}{4 \mathrm{ln}2}\Big].
\end{equation}
We note that other models were used to for the CSE of Cepheids such as a shell or optically thin envelope  \citep{merand06,Gallenne2012}. The choice of a Gaussian intensity distribution for the envelope is used for simplicity, as we do not have physical justifications at this time. This model is however consistent with a centro-symmetric structure given by a zero closure-phase in Sect.~\ref{sect:T3}. Thus, we computed the total squared visibility of the star plus CSE model:
\begin{equation}\label{eq:vis2}
    V_{\mathrm{tot}}^2(f)=\Big(\lvert  F_\star \, V_\mathrm{UD}(f) + F_\mathrm{CSE} \, V_\mathrm{CSE}(f)  \rvert \Big)^2
\end{equation}
 where $F_\star$ and  $F_\mathrm{CSE}$ are the normalized stellar and CSE flux contribution to the total flux respectively, normalized to unity $F_\star$ + $F_\mathrm{CSE}$ = 1. In this process, we compute $V_{\mathrm{tot}}^2(f)$ for all combination of CSE flux contribution and size. We then derive the reduced $\chi^2$ map by comparing with MATISSE visibilities (see Fig.~\ref{fig:chi2}).

As shown by the $\chi^2$ map for all stars of our sample, assuming a compact envelope below 3-4$\,R_\star$ it is not possible to constrain the flux contribution from these observations. Conversely a faint circumstellar emission is also consistent for different CSE size (see green dashed line in Fig.~\ref{fig:chi2}). In order to provide quantitative limits on the CSE characteristics, we calculated the residuals between the observations and models, excluding models with residuals above 2$\,\sigma$. We provide upper limits of the CSE flux contribution for different CSE radius in Table \ref{tab:star_radii} which can also be vizualized in Fig.~\ref{fig:chi2}. These upper limits are overall similar from one star to another. In particular, we can securely rule out envelopes that are simultaneously large and bright (R>$10\,R_\star$;$f>10\%$) for all the star sample.

\begin{table}[h!]
    \centering
        \caption{Upper limit at 2$\sigma$ for the flux of the CSE.}
    \begin{tabular}{c|c|c|c}
        \hline
        \hline
        Star & $3\,R_\star$ & $5\,R_\star$ & $10\,R_\star$ \\
        \hline
        U Aql &31.9 &14.8 &8.3 \\
        $\eta$ Aql & 25.4 & 16.9& 14.0\\
        $\beta$ Dor &14.4 & 9.6&7.9 \\
        $\zeta$ Gem & 13.6 & 10.4&10.0 \\
        TT Aql &28.9 &13.3 &7.7 \\
        T Mon &30.8 &14.8 &8.6 \\
        U Car &38.1 &17.2 &8.8 \\
        \hline
    \end{tabular}
    \begin{tablenotes}
        \item Notes: The flux contribution of the envelope $F_\mathrm{CSE}$  is given in percent for three different radius (FWHM) of the envelope model. These limits are also displayed in Fig.~\ref{fig:chi2}.
    \end{tablenotes}

    \label{tab:star_radii}
\end{table}

\subsection{Radiative transfer models of dusty CSE}\label{sect:dusty_model}
\begin{figure*}[]
\begin{subfigure}[b]{.35\textwidth}
  \includegraphics[width=\linewidth]{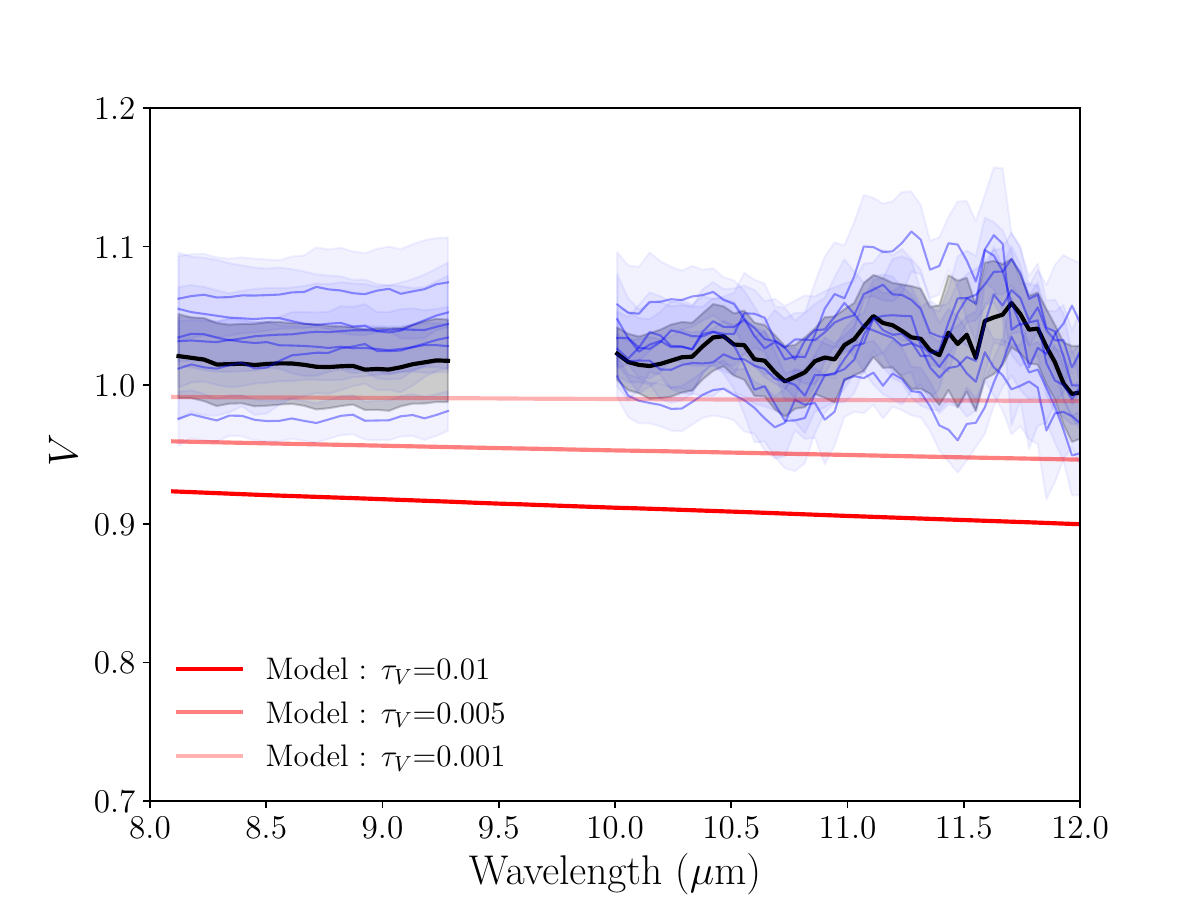}
  \caption{}
  \label{fig:1}
\end{subfigure} 
\begin{subfigure}[b]{.35\textwidth}
  \includegraphics[width=\linewidth]{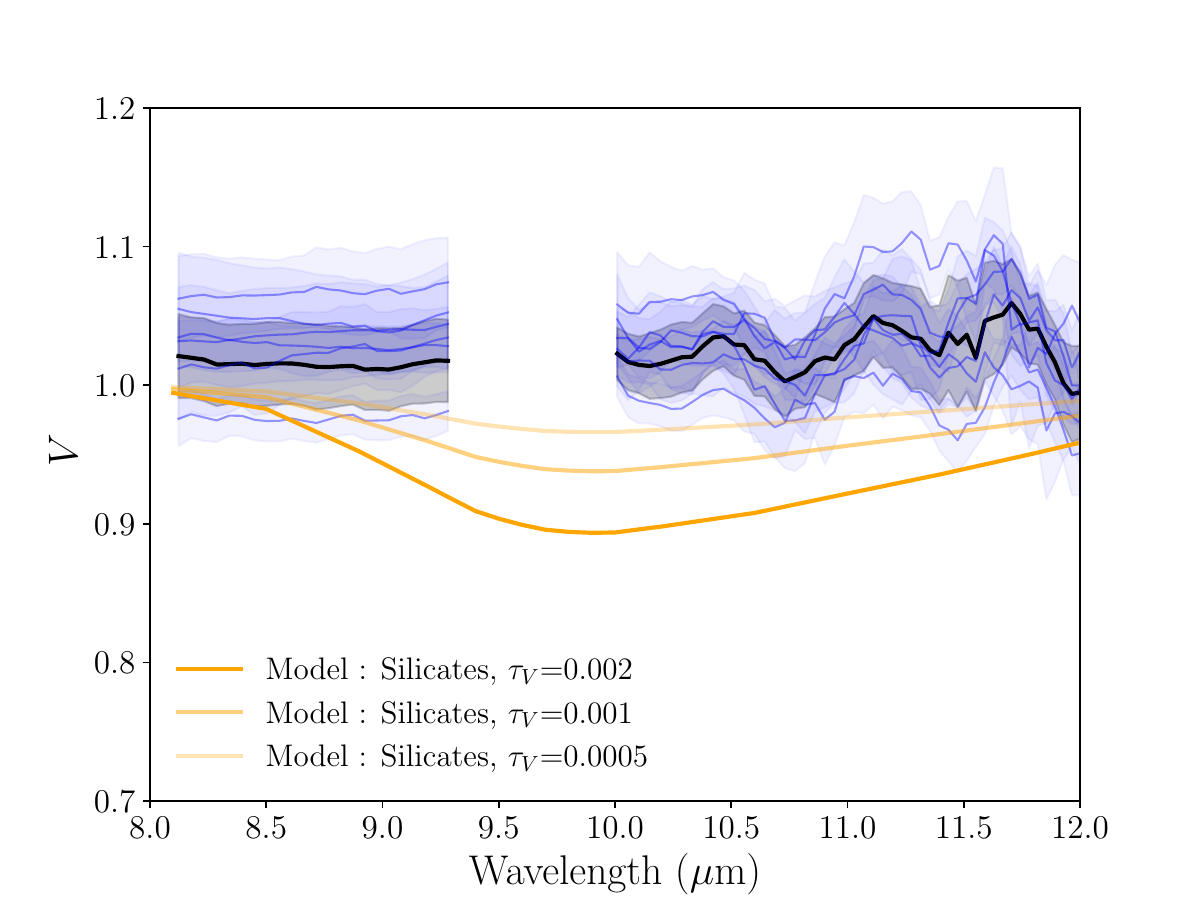}
  \caption{}
  \label{fig:4}
\end{subfigure}
\begin{subfigure}[b]{.35\textwidth}
  \includegraphics[width=\linewidth]{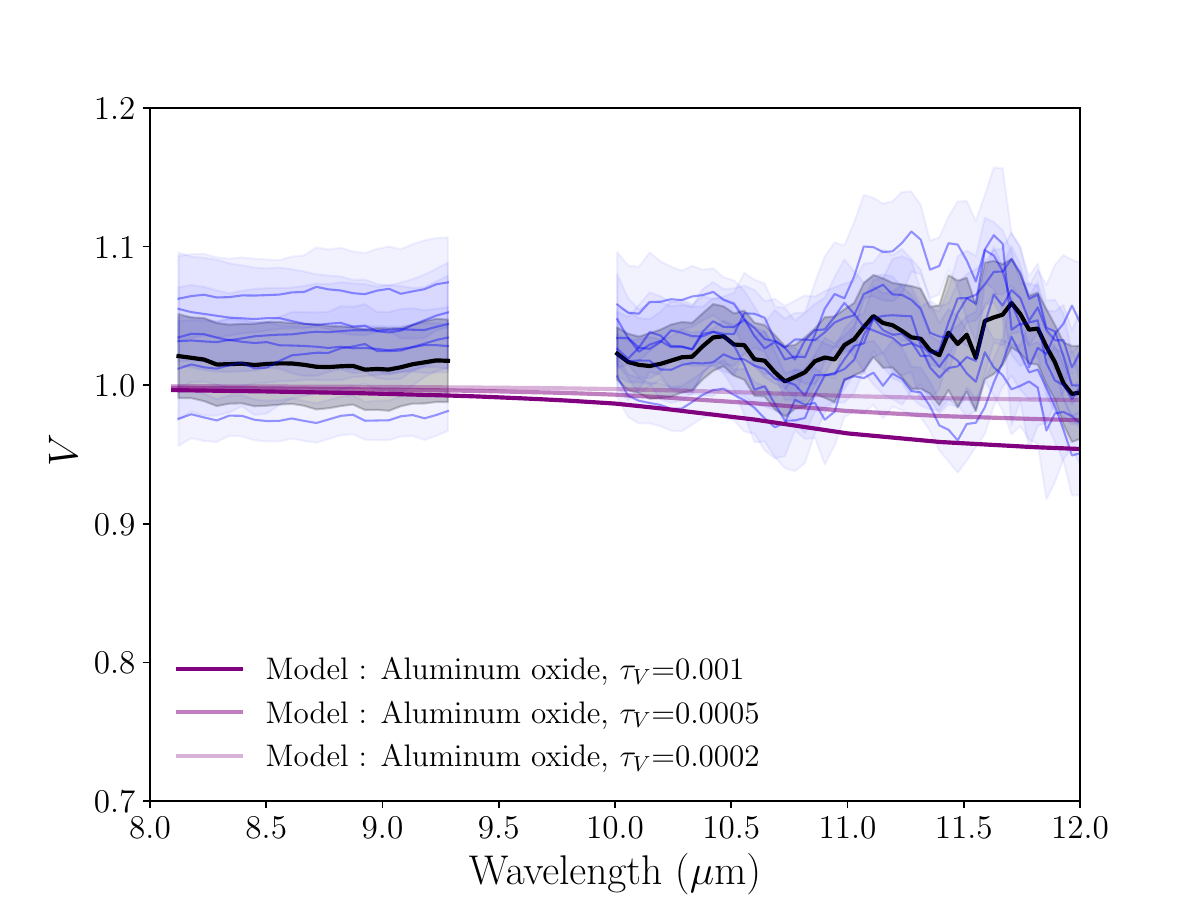}
  \caption{}
  \label{fig:1}
\end{subfigure} 
\begin{subfigure}[b]{.35\textwidth}
  \includegraphics[width=\linewidth]{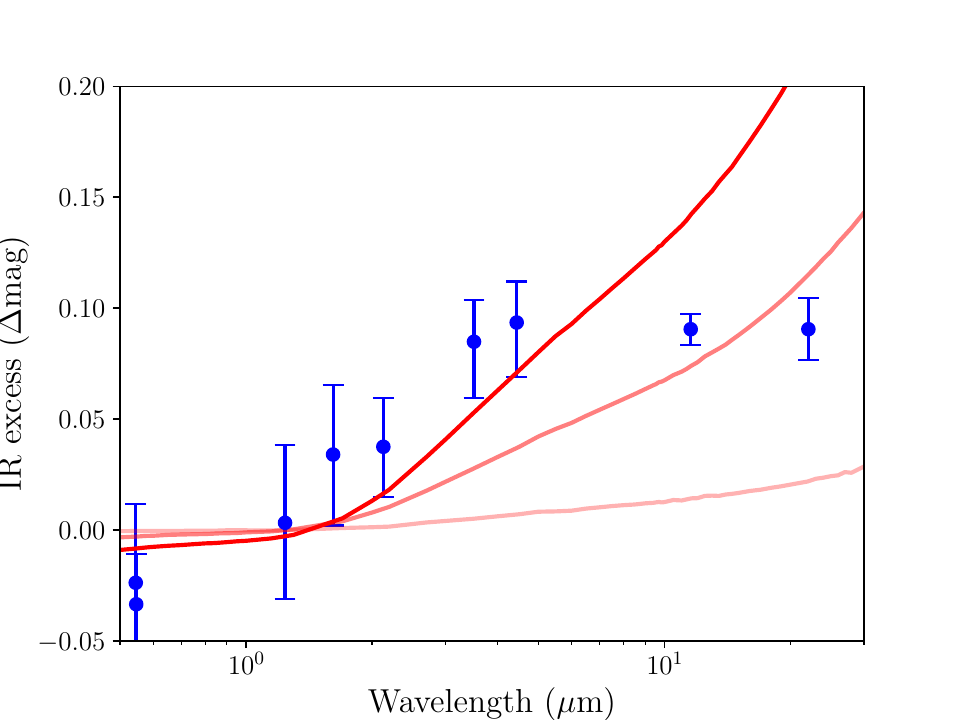}
  \caption{}
  \label{fig:4}
\end{subfigure}
\begin{subfigure}[b]{.35\textwidth}
  \includegraphics[width=\linewidth]{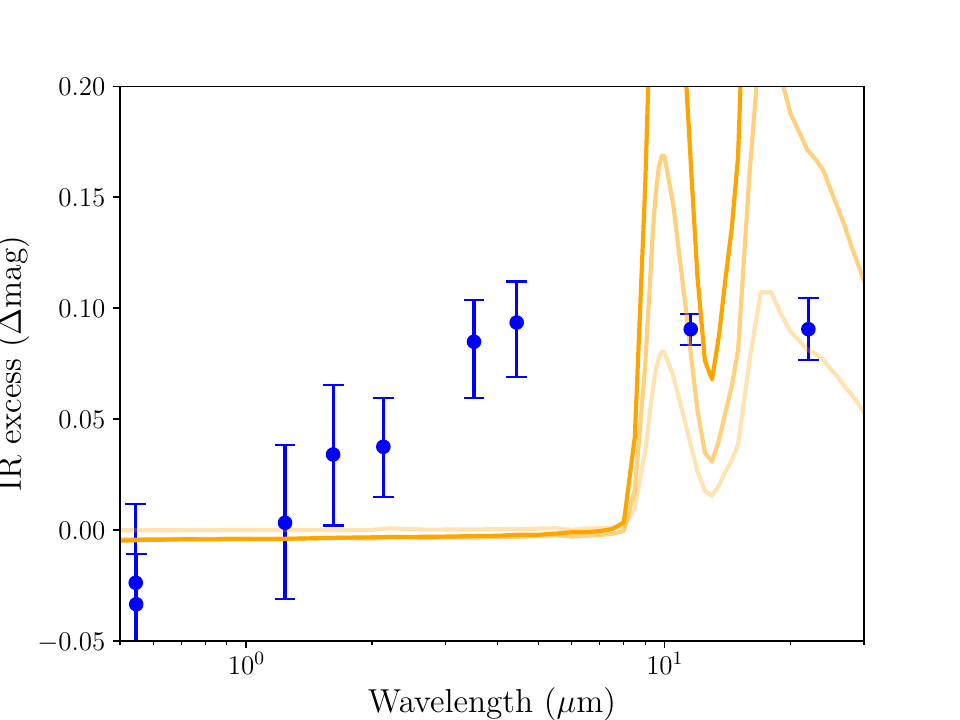}
  \caption{}
  \label{fig:t_mon_N}
\end{subfigure}
\begin{subfigure}[b]{.35\textwidth}
  \includegraphics[width=\linewidth]{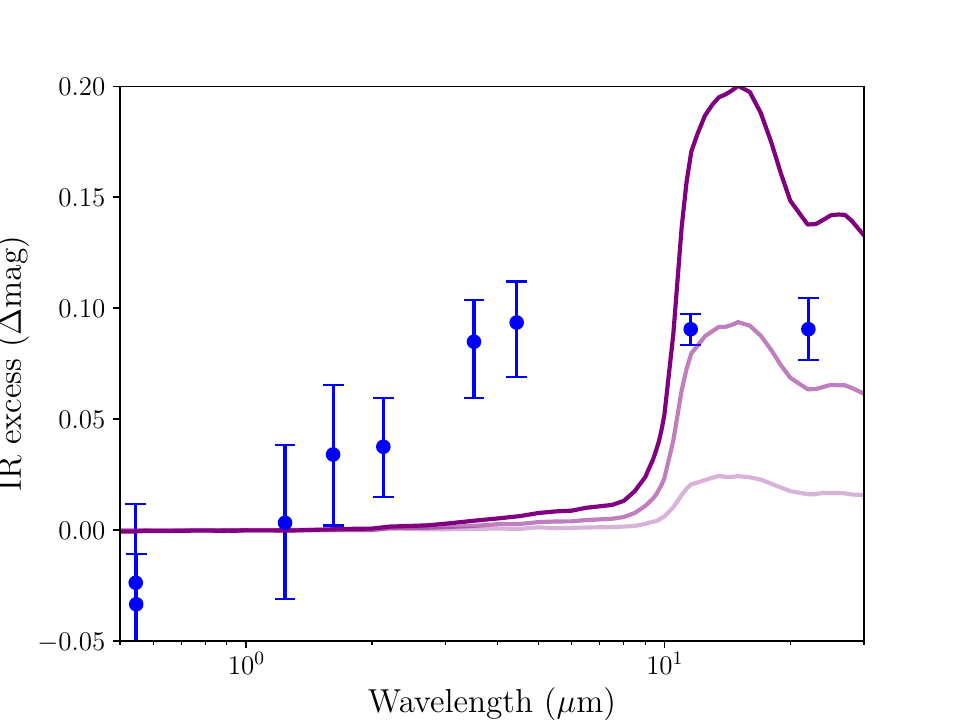}
  \caption{}
  \label{fig:4}
\end{subfigure}
\caption{Visibility in the $N$-band and IR excess from CSE models of dust computed with \texttt{DUSTY} for the Cepheid $\eta$ Aql. From left to right the three cases correspond to iron, silicates and aluminum oxide in red, yellow and purple respectively. For each CSE models we used different optical depth as indicated in the legends.}
 \label{fig:DUSTY_MODELS_etaaql}
\end{figure*}

From the IR excess and the $N$-band observations it is possible to constrain the CSE models based on dust grains.
To this end, we computed dusty CSE models with the radiative transfer code \texttt{DUSTY} \citep{DUSTY} in order to show the theoretical visibility and IR excess in those cases. While fitting CSE models by fine-tuning dust composition and other characteristics is feasible, we opted for a simpler approach, selecting representative test cases to illustrate these models. We chose $\eta$ Aql and T Mon as test cases, representing stars with differing angular diameters ($\approx$1.8 and 0.8 mas, respectively). For each star, we modeled a black-body as a central source by adopting the stellar luminosity, temperature, and radius from \texttt{SPIPS} at the phase corresponding to the MATISSE observations. To explore characteristic mineralogy of oxygen-rich stars, we modeled three types of dust compositions: iron, silicates, and alumina using optical constants from \cite{henning1996}, \cite{Ossenkopf1992} and \cite{begemann1997} respectively. For each model, the condensation temperature was arbitrary fixed at 1200\,K, slightly higher than condensation models from \cite{Gail1999} at relatively low gas pressure. A standard grain size distribution \citep{MRN} and dust density ($n_d\propto r^{-2}$) were used, while optical depth was varied for each dust type. From these models, we computed the $N$-band visibility for a baseline of 130 m, matching the maximum VLTI baseline, and we derived the corresponding IR excess for comparison with \texttt{SPIPS} results. The results are presented in Figs.~\ref{fig:DUSTY_MODELS_etaaql} and \ref{fig:DUSTY_MODELS_tmon}.

These results allow to constrain the optical depth for each dust type. We find that for both $\eta$~Aql and T~Mon, an optical depth of the order of $\tau_v = 0.001$ or lower aligns well with MATISSE observations, indicating that, if dust exists around Cepheids, it is likely too faint to be resolved. However, CSEs with optical depths of at least $\tau_V = 0.01$ would be clearly resolved for every dust type. For comparison, \cite{gallenne13b} fitted $\tau_V = 0.151\pm0.042$ for explaining MIDI/VLTI visibility of T Mon, with 80\% of iron content, which is excluded from our observations in Fig.~\ref{fig:DUSTY_MODELS_tmon}. On the other hand, \cite{gallenne13b} derived $\tau_V =0.008\pm0.002$ for X Sgr, with visibility range of 0.9-1.0 measured by MIDI/VLTI. Given the relatively small angular diameter of X Sgr ($\approx$1.35\,mas), we do not expect to resolve such faint envelope with MATISSE. Our simple tests are also in agreement with \cite{Gro2020} who fitted the IR excess of Galactic Cepheid with iron as the main dust component, with optical depth from about $10^{-4}$ to $10^{-3}$. In particular, the optical depth derived by \cite{Gro2020} in the case of the CSE of $\eta$ Aql and $\zeta$ Gem (with iron content above 90\%) would yield a visibility of $V\approx0.95$ in agreement with our MATISSE observations. As we can see from Figs~\ref{fig:DUSTY_MODELS_etaaql} and \ref{fig:DUSTY_MODELS_tmon}, silicate and alumina emits predominantly in the $N$-band, a level of $\approx$0.10\,mag in the $N$-band gives no IR excess in $K$ and $L$-band. The only exception is iron component which is able to produce noticeable near-IR emission. Indeed, for both $\eta$ Aql and T Mon, an iron optical depth of $\tau_V = 0.01$ would result in IR excess closely matching the \texttt{SPIPS} estimates, with corresponding values of $\Delta K = 0.02\,$mag, $\Delta L = 0.05\,$mag, and $\Delta N = 0.13\,$mag for $\eta$~Aql, and $\Delta K = 0.03\,$mag, $\Delta L = 0.07\,$mag, and $\Delta N = 0.15\,$mag for T~Mon. However, we consider it unlikely that almost only iron would form in these environment. As discussed in the introduction, a compact envelope of about 2-4$\,$R$_\star$ might be too hot for dust condensation. A more plausible explanation for the observed near-IR excess is a compact, ionized gas envelope that behaves similarly to a chromosphere and emits continuous free-free emission \citep{Hocde2020a}.

\section{Conclusion}\label{sect:conclusion}
\begin{figure}[]
\centering
\includegraphics[scale=0.72]{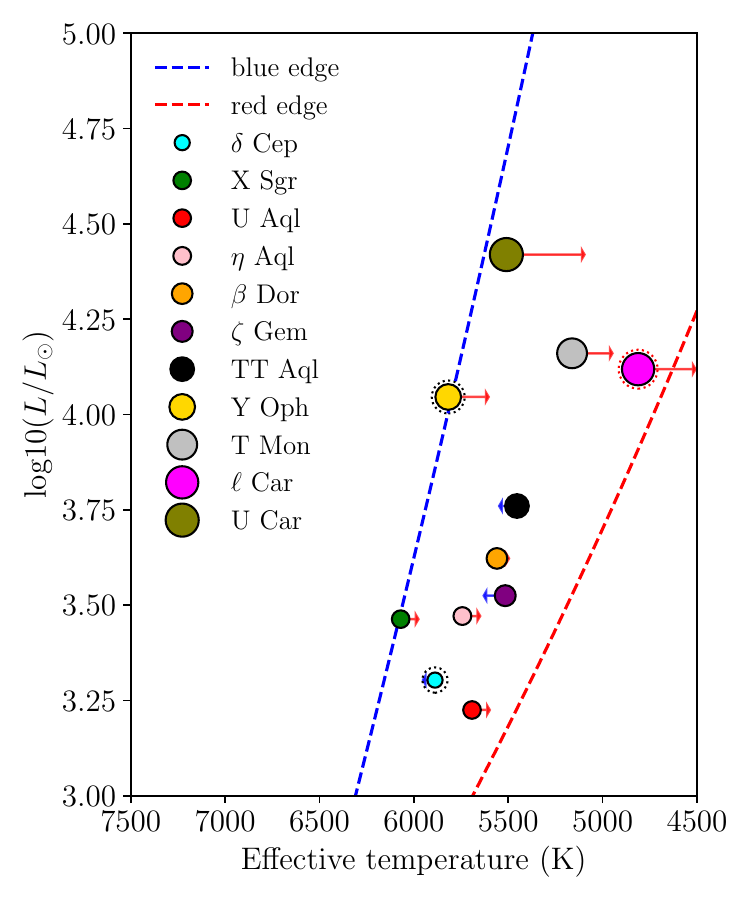}
\caption{\small Cepheid sample observed with MATISSE/VLTI \citep[This work, and ][]{Hocde2021} completed with $\delta$ Cep and Y Oph observed with CHARA \citep{merand06,merand07,nardetto16}.  The mean luminosity and effective temperature in the instability strip are derived by \texttt{SPIPS}. Blueward or redward crossing are indicated by arrows depending on the sign of the period change. Black dotted circles indicate stars where CSEs are suspected to be resolved by interferometry based on previous studies. The red dotted circle represents a star with a confirmed CSE around $\ell$ Car.
\label{fig:IS}}
\end{figure}

In this paper we explored the circumstellar emission for a group of Cepheids from pulsation period from 7 up to about 40$\,$days and distributed at different location within the instability strip as shown in Fig.~\ref{fig:IS}. From the results of MATISSE observations alone, we have shown that the closure phase for all Cepheids are consistent with central-symmetry in $L$, $M$ ad $N$ bands. We have also found the absence of dust emission for a relatively large set of Cepheids with different pulsation period and stellar parameters. It is however not excluded that a circumstellar dust envelope can form for Cepheids of larger pulsation period \citep[see, for example,][]{Kovtyukh2024}.  In this context, it is important to enlarge the sample of Cepheids with interferometric observations in the $N$-band to definitely conclude on the absence of dusty envelopes. However, an important caveat is that the current MATISSE observations cannot effectively constrain the size of the CSE, especially when its flux contribution is weak, around 5\%. While a CSE simultaneously bright ($\approx$10\%) and large ($\approx10\,R_\star$) can be ruled out, the observations are compatible with the presence of a compact CSE within the uncertainties, as modeled in previous studies. We note that our analysis is based on a single snapshot for each star, thus these observations are exposed to calibration bias. However, the homogeneity of these results based on eight different Cepheids provide a robust picture of the CSE model across the instability strip.

On the other hand, interferometric detection of CSEs are biased toward larger angular diameter such as $\ell$~Car, hence it is not possible to definitely conclude on the characteristics of the envelope through the instability strip with MATISSE/VLTI. Given the few observations up to date, it is also difficult to observe any relation as shown in Fig.~\ref{fig:IS}, that would depend on the evolution's stage of the star. Among the resolved CSE emission, $\ell$~Car is likely the most significant since it was observed with MIDI, VISIR and MATISSE. All others CSEs resolved up to now such as around $\delta$~Cep, Y~Oph and Polaris must be confirmed with additional observations from the northern hemisphere.

In future studies, follow-up observations using GRAV4MAT in higher angular resolution could significantly reduce errors and provide more precise constraints, especially for bright targets in the mid-IR. On the other hand, observations at higher spatial frequency in the near-IR is necessary to resolve circumstellar emission of smaller angular diameter stars. Further observations of Cepheids using GRAVITY/VLTI in $K$-band \citep{GRAVITY2017} and CHARA/SPICA in the visible domain \citep{Mourard2022} will be also essential to disentangle the star and the CSE contribution.

\section{Data availability}
MATISSE reduced data are made available at ESO science archive facility.
Figures of Appendix \ref{app:spips} are made available at Zenodo \href{url}{https://zenodo.org/records/14505155}.
\begin{acknowledgements}
We thank the referee for the careful reading which helped to improve the paper. The research leading to these results has received funding from the European Research Council (ERC) under the European Union’s Horizon 2020 research and innovation programme (grant agreements No 695099 and No 951549). This work received the funding from the Polish-French Marie Skłodowska-Curie and Pierre Curie Science Prize awarded by the Foundation for Polish Science. This work also received funding from the Polish Ministry of Science and Higher Education grant agreement 2024/WK/02. The authors acknowledge the support of the French Agence Nationale de la Recherche (ANR) under grant
ANR-23-CE31-0009-01 (Unlock-pfactor). RS is supported by the National Science Center, Poland, Sonata BIS project 2018/30/E/ST9/00598. This research made use of the SIMBAD and VIZIER (\url{http://cdsweb.u- strasbg.fr/}) databases at CDS, Strasbourg (France) and the electronic bibliography maintained by the NASA/ADS system. Based on observations made with ESO telescopes at Paranal observatory under program IDs: 104.D-0554(B), 106.21RL.001, 106.21RL.002 and 106.21RL.003.  This research has benefited from the help of SUV, the VLTI user support  service of the Jean-Marie Mariotti Center (\url{http://www.jmmc.fr/suv.htm}). This research has also made use of the Jean-Marie Mariotti Center \texttt{Aspro}
service \url{http://www.jmmc.fr/aspro}.
This research also made use of Astropy, a community-developed corePython package for Astronomy \citep{astropy2018}. 
\end{acknowledgements}
\bibliographystyle{aa}  
\bibliography{bibtex_vh} 

\begin{appendix}

\section{Additional material}
\begin{table}[]
\caption{\label{tab:log} \small Journal of the observations.}
\begin{center}
\begin{tabular}{l|c|c|c|c|c}
\hline
\hline
 \textbf{Target} \&	&	Date 	& seeing 	&	$\tau_0$  &AM & $\theta_\mathrm{fit}$-$\theta_\mathrm{JSDC}$\\
	Calibrator & (MJD)		&		($^{\prime \prime}$)		&	(ms)& 	&(\%)\\
\hline
30 Gem         &	 58923.04          &	0.94	         &		3.1       & 	1.33 &       +4.0\\
\textbf{T Mon}	&	58923.06	&	0.90 & 4.0	 &	 1.35 &-\\
18 Mon	         &	  58923.08        &		1.05         &		 3.6      & 	1.31 &   +5.3\\
\hline
\textbf{$\beta$ Dor}	&	59244.03	&	0.67       & 7.1    &   	 1.28 &-\\
HD59219	                &	 59244.06           &		   0.60      &	7.1      &1.23 	&$+1.8$\\
\hline		
87 Gem		           &	59300.04	            &	  0.60	 &	8.8   & 1.69&$+0.1$\\
\textbf{$\zeta$ Gem}	&	59300.06	&	0.75    & 5      &  1.66 &-\\
\hline
HD101162	   &	 59300.11            &		     0.44  & 7.6	& 	1.41     &$+0.9$\\
\textbf{U Car}	&	59300.12	&	0.50  & 7.7	&	 1.23& -\\
$\alpha$ Char	&	59300.15            &		     0.51  & 8.8	& 	1.82     &$-0.7$\\
\hline
HD189533	       &	59420.17             &	0.47      	    & 4.0  & 1.14	&$+2.6$\\
\textbf{$\eta$ Aql}	&	59420.19	&	0.54    &4.0	&	1.11 &-\\
HD189695	   &	   59420.21           &		     0.65    & 4.1	 & 1.21	&$+1.2$\\
\hline
HD188390	           &	59484.01          &		0.50   & 4.7	 & 	1.05     &      $+4.7$\\
\textbf{TT Aql(1)}	   &	59484.03  &	0.60       & 3.5	 &	 1.18&-\\
HD189695	           &	59484.05          &	   0.50    & 5.5     & 	1.23  &$+1.4$\\
\hline
HD189695	&	    59484.06         &		     	0.70     &3.3	      & 	1.27 &$+3.8$\\
\textbf{U Aql}	&	59484.08	&	0.52  & 6.5	   &	 1.23&-\\
HD184574	   &	59484.09          &		         	0.46 &	 7.1     &1.27 	  & $+4.9$\\
\hline
\textbf{X Sgr}	&	59809.09	&	1.07  & 2.2	&	 1.07&-\\
HD183925	   &	59809.12    &	0.83	  &	2.4 & 1.01	&+1.7\\
\hline
HD188390	&	 59837.10&	0.40	&	5.0	&	1.11& 2.0\\
\textbf{TT Aql(2)}	& 59837.12&		0.36 &	5.4  &1.49 &-\\
HD189695	&	59837.14 &	0.38	&		5.5	&1.49& 1.4\\
\hline
\end{tabular}
\normalsize
\end{center}
\begin{tablenotes}
    \item \textbf{Notes.} The Modified Julian Date (MJD), the seeing (in arcsecond), the coherence time $\tau_0$ (in millisecond)  and the air mass (AM) are indicated. The last column lists the difference of the MATISSE diameter measurements of the calibrator $\theta_\mathrm{fit}$ in comparison to JSDC catalogue $\theta_\mathrm{JSDC}$ following the method from \cite{Robbe2022} (see Sect.~\ref{sect:cal_sci}). T~Mon is part of GTO program 104.D-0554(B). TT Aql(1) was used for the total flux and visibility in the $N$-band and TT Aql(2) for the calibration in the $LM$-band. For U Aql, only correlated flux was acquired due to its low brightness.
\end{tablenotes}
\end{table}

\begin{figure*}[]
\begin{subfigure}[b]{.35\textwidth}
  \includegraphics[width=\linewidth]{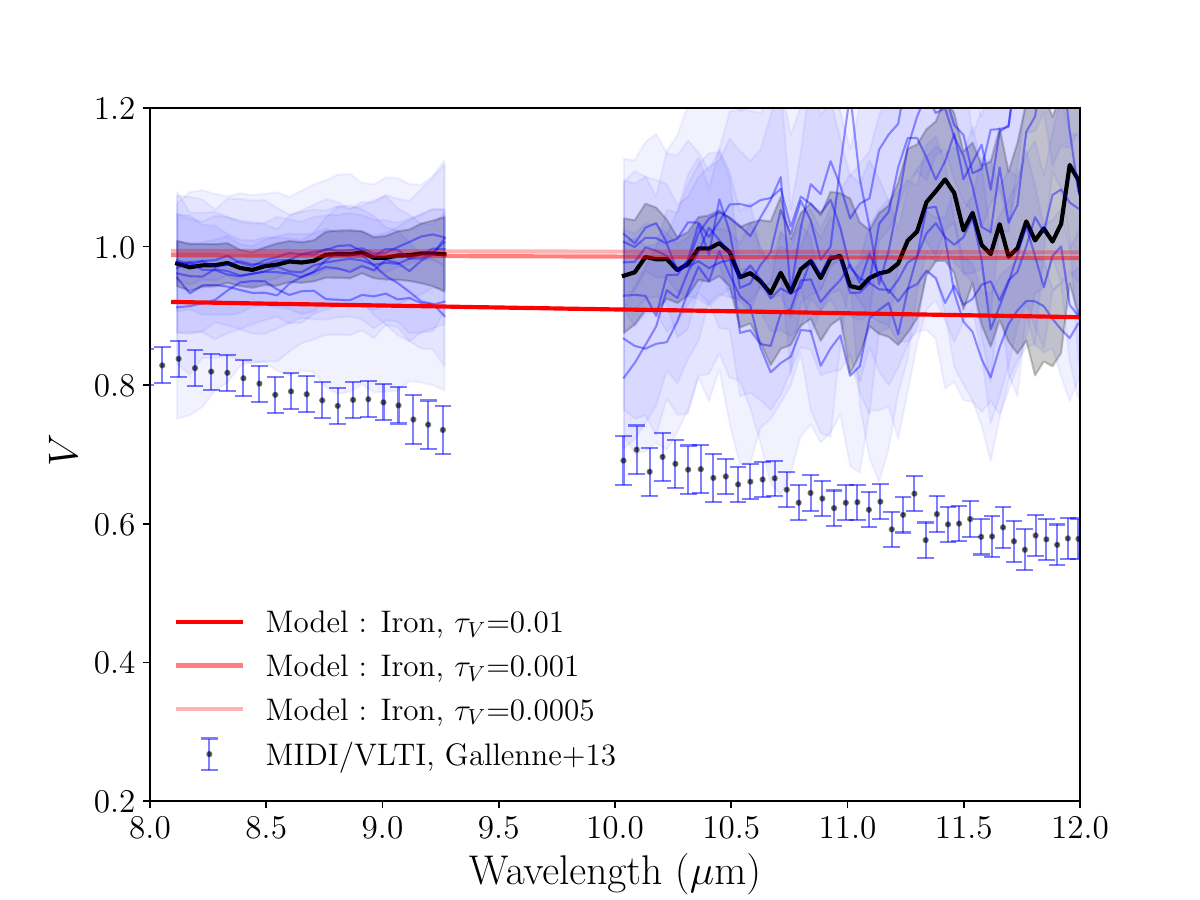}
  \caption{}
  \label{fig:1}
\end{subfigure} 
\begin{subfigure}[b]{.35\textwidth}
  \includegraphics[width=\linewidth]{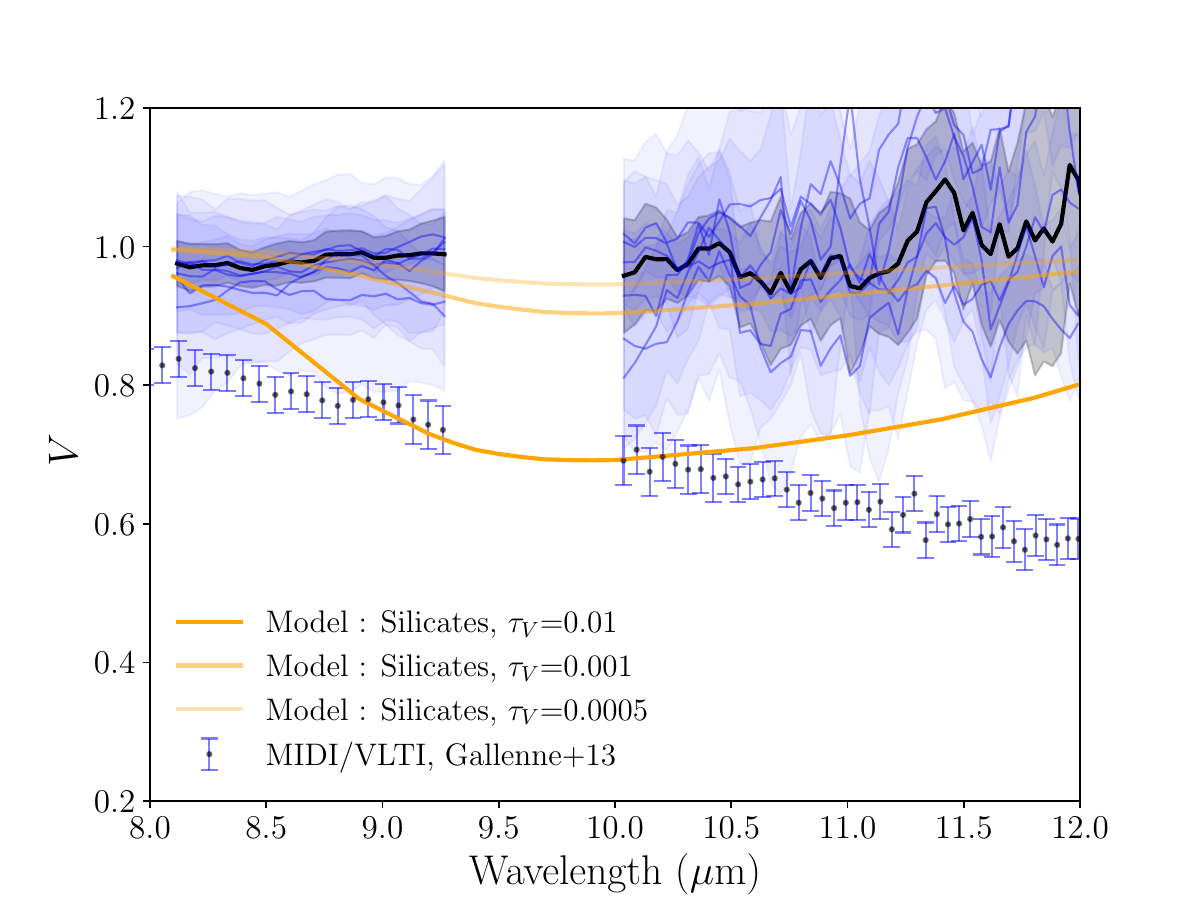}
  \caption{}
  \label{fig:4}
\end{subfigure}
\begin{subfigure}[b]{.35\textwidth}
  \includegraphics[width=\linewidth]{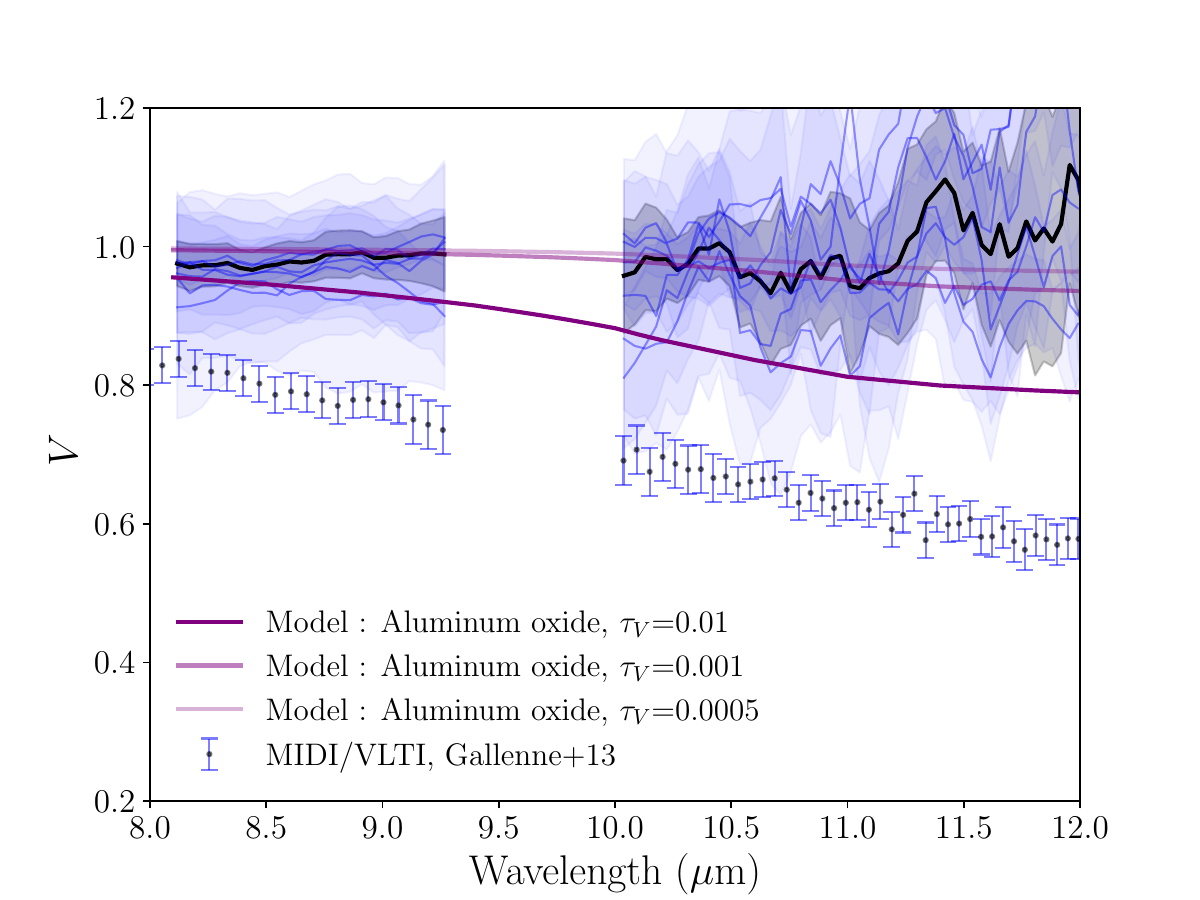}
  \caption{}
  \label{fig:1}
\end{subfigure} 
\begin{subfigure}[b]{.35\textwidth}
  \includegraphics[width=\linewidth]{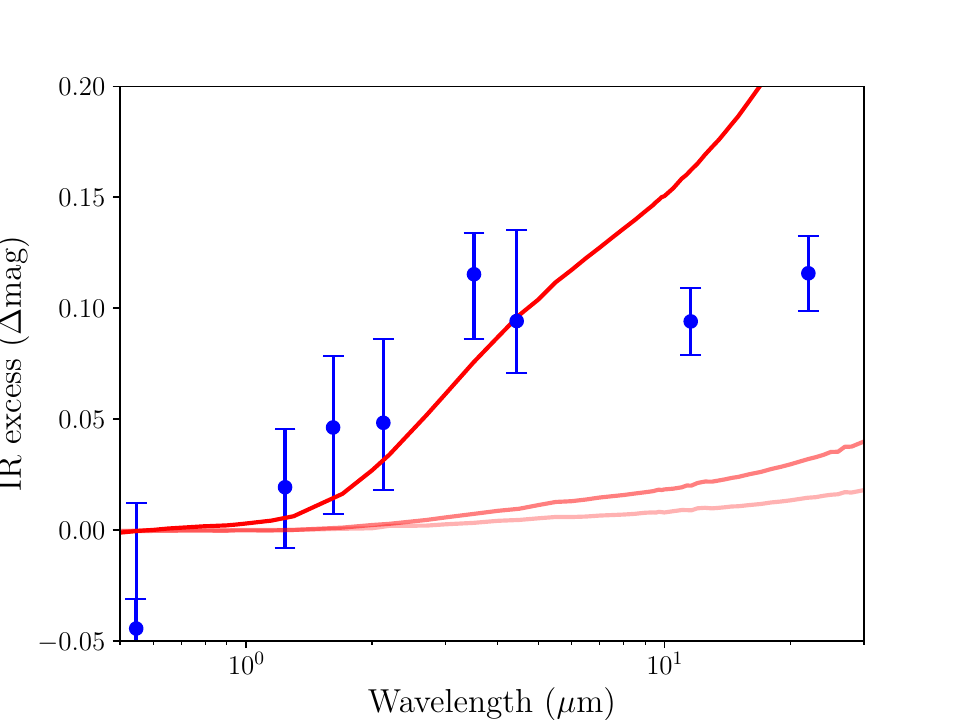}
  \caption{}
  \label{fig:4}
\end{subfigure}
\begin{subfigure}[b]{.35\textwidth}
  \includegraphics[width=\linewidth]{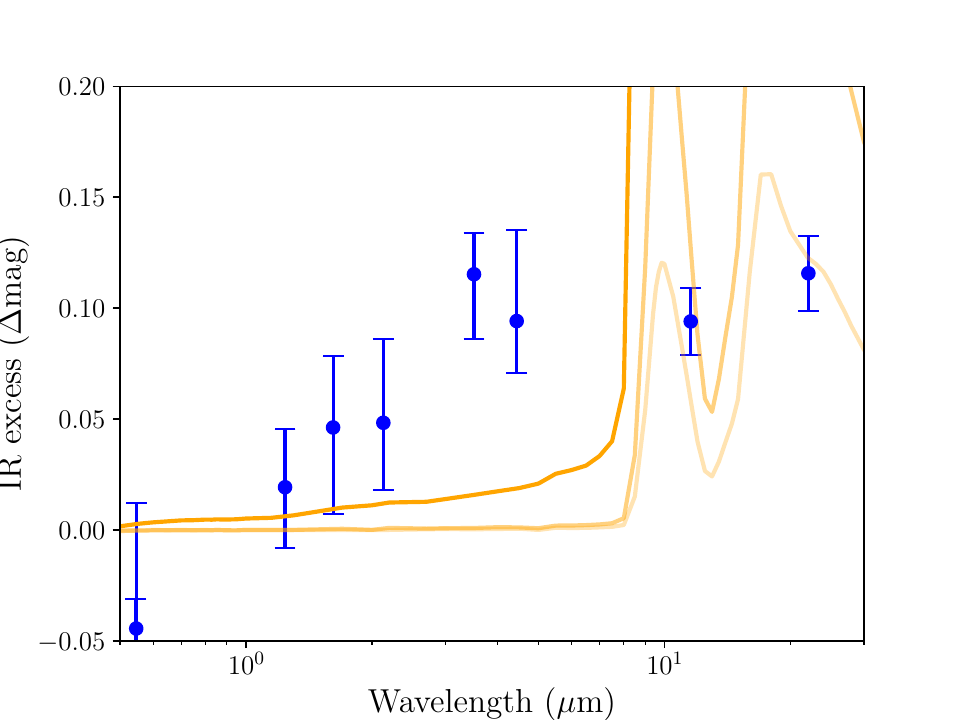}
  \caption{}
  \label{fig:t_mon_N}
\end{subfigure}
\begin{subfigure}[b]{.35\textwidth}
  \includegraphics[width=\linewidth]{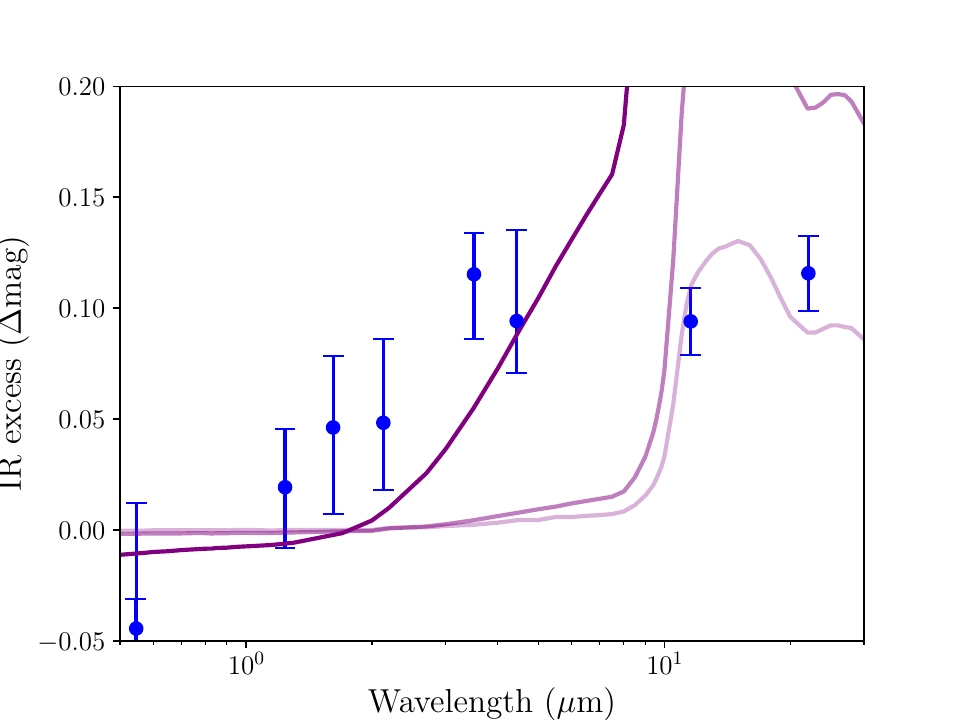}
  \caption{}
  \label{fig:4}
\end{subfigure}
\caption{Same as Fig.~\ref{fig:DUSTY_MODELS_etaaql} but in the case of T Mon.}
 \label{fig:DUSTY_MODELS_tmon}
\end{figure*}

\section{SPIPS data set and fitted pulsational model of the star sample.\label{app:spips}}\label{tab:comparison}

\begin{table*}[h!]
\centering
\caption{Comparison of effective temperature, color excess and LD angular diameter derived by \texttt{SPIPS} from \cite{Trahin2021} and \cite{Gallenne2021}.}

\begin{tabular}{l|ccc|ccc|ccc}
\hline
\hline
Star & \multicolumn{3}{c}{$T_\mathrm{eff}$ (K)} & \multicolumn{3}{c}{$E(B-V)$} & \multicolumn{3}{c}{<$\theta_\mathrm{LD}$> (mas)} \\
\hline
 & This work & T21 & G21  & This work & T21& G21 & This work & T21 & G21 \\
\hline
X Sgr & 6068$_{7}$ & 6047$_{80}$ & 6080$_{60}$ &0.271$_{0.004}$ & 0.277$_{0.021}$ & 0.305$_{0.002}$& 1.347$_{0.004}$  & 1.345$_{0.027}$&1.382$_{0.001}$ \\
U Aql & 5717$_{13}$ & 5735$_{80}$ & 5733$_{21}$ &  0.418$_{0.011}$ & 0.417$_{0.021}$ & 0.424$_{0.002}$& 0.773$_{0.004}$  & 0.778$_{0.016}$ & 0.7862$_{0.0006}$ \\
$\eta$ Aql & 5719$_{11}$ & 5790$_{80}$ &5747$_{19}$ &0.150$_{0.004}$ & 0.162$_{0.021}$ & 0.166$_{0.003}$ & 1.753$_{0.002}$ & 1.728$_{0.035}$ &1.760$_{0.004}$ \\
$\beta$ Dor & 5550$_{8}$ & 5581$_{80}$ & 5564$_{66}$  & 0.047$_{0.003}$& 0.063$_{0.021}$ & 0.078$_{0.004}$& 1.771$_{0.003}$  & 1.763$_{0.035}$& 1.811$_{0.003}$ \\
$\zeta$ Gem & 5513$_{10}$& 5520$_{80}$ & 5536$_{34}$ &0.040$_{0.004}$  &  0.040$_{0.021}$ & 0.020$_{0.009}$  & 1.652$_{0.002}$  & 1.637$_{0.033}$ & 1.58$_{0.02}$ \\
TT Aql & 5557$_9$ & 5470$_{80}$ & 5504$_{17}$ & 0.547$_{0.003}$ & 0.551$_{0.021}$ & 0.564$_{0.002}$  & 0.757$_{0.001}$ & 0.760$_{0.015}$ & 0.7617$_{0.0008}$\\
T Mon & 5012$_{4}$ &5172$_{80}$ & 5206$_{22}$  &  0.235$_{0.021}$ & 0.271$_{0.021}$ & 0.299$_{0.005}$& 0.919$_{0.003}$& 0.944$_{0.019}$& 0.963$_{0.002}$\\
U Car & 5707$_{34}$ & -& 5397$_{24}$ & 0.447$_{0.012}$ & - & 0.382$_{0.007}$ & 0.893$_{0.003}$ &-& 0.917$_{0.003}$ \\
\hline
\end{tabular}
\end{table*}

\begin{table*}[h]
\centering
\caption{References of observations used for the \texttt{SPIPS} fitting}\label{Tab.SPIPS}
\begin{tabular}{l|l|l|l|l}
\hline
\hline
Star & Photometry & Radial Velocity & Effective Temperature& Angular diameter \\ 
\hline
X Sgr & 1,2,3,4,5,6  & 42 &24 & 30,31\\ 

U Aql &5,6,7,17 & 38,39 &24 & 31\\ 

$\eta$ Aql &3,5,6,8,9,10,11,12 & 23,40,41 &25,26,27 & 30,32,33\\ 

$\beta$ Dor &1,3,5,6,7,14 & 23 & 24,27& 30,34 \\ 

$\zeta$ Gem &1,2,4,5,6 & 23 &27 & 30,32,35 \\ 

TT Aql &1,2,5,6,8,10,11,12,15,16,17 & 23& 24,28&35,36 \\ 

T Mon &2,3,6,8,12,14,15,17,18,19 & 23 &28 & 31,35\\ 

U Car & 5,6,8,14,20& 23 & 24,29& 31, 35\\ 

$\ell$ Car &1,2,6,14,21,22 & 23 &24 & 30,37\\ 
\hline
\end{tabular}
\begin{tablenotes}
    \item \textbf{Notes : Photometry:} 1: \cite{tycho2006}: 2: \cite{Berdnikov2008}, 3: \cite{Pel1976}, 4: \cite{Feast2008}, 5: \cite{Monson2012}, 6: \cite{WISE2010}, 7: \cite{berdnikov02}, 8: \cite{Welch1984}, 9: \cite{hipparcos1997}, 10: \cite{barnes97}, 11: \cite{kiss98}, 12: \cite{moffett84}, 14: \cite{laney1992}, 15: \cite{Coulson1985}, 16: \cite{barnes97}, 17: \cite{Monson2011}, 18: \cite{Szabados1981}, 19: \cite{AAVSO2003}, 20: \cite{Walraven1964}, 21: \cite{Madore1975}, 22: \cite{Marengo2010Spitzer}. \textbf{Radial Velocity:} 23: \cite{Anderson2024}, 38: \cite{Eaton2020}, 39: \cite{Borgniet2019}, 40: \cite{storm04}, 41: \cite{barnes05}, \cite{storm11b}. \textbf{Effective Temperature:} 24: \cite{Luck2018}, 25: \cite{LuckAndrievsky2004}, 26: \cite{Kovtyukh2010}, 27: \cite{Proxauf2018}, 28: \cite{Kovtyukh2005}, 29: \cite{Usenko2011}. \textbf{Angular diameter:} 30: VINCI/VLTI, \cite{Kervella2004VINCI}, 31: MIRC/CHARA or PIONIER/VLTI, \cite{Gallenne2019a}, 32: 
PTI, \cite{Lane2002}, 33: FLUOR/CHARA, \cite{Trahin2021}, 34: SUSI, \cite{Davis2008}, 35: PIONIER, \cite{Trahin2021}, 36: CLASSIC, \cite{Trahin2021}, 37: \cite{Anderson2016}.
    \end{tablenotes}
\label{table:star_data}
\end{table*}

\begin{figure*}[h]
\begin{center}
\includegraphics[scale=0.55]{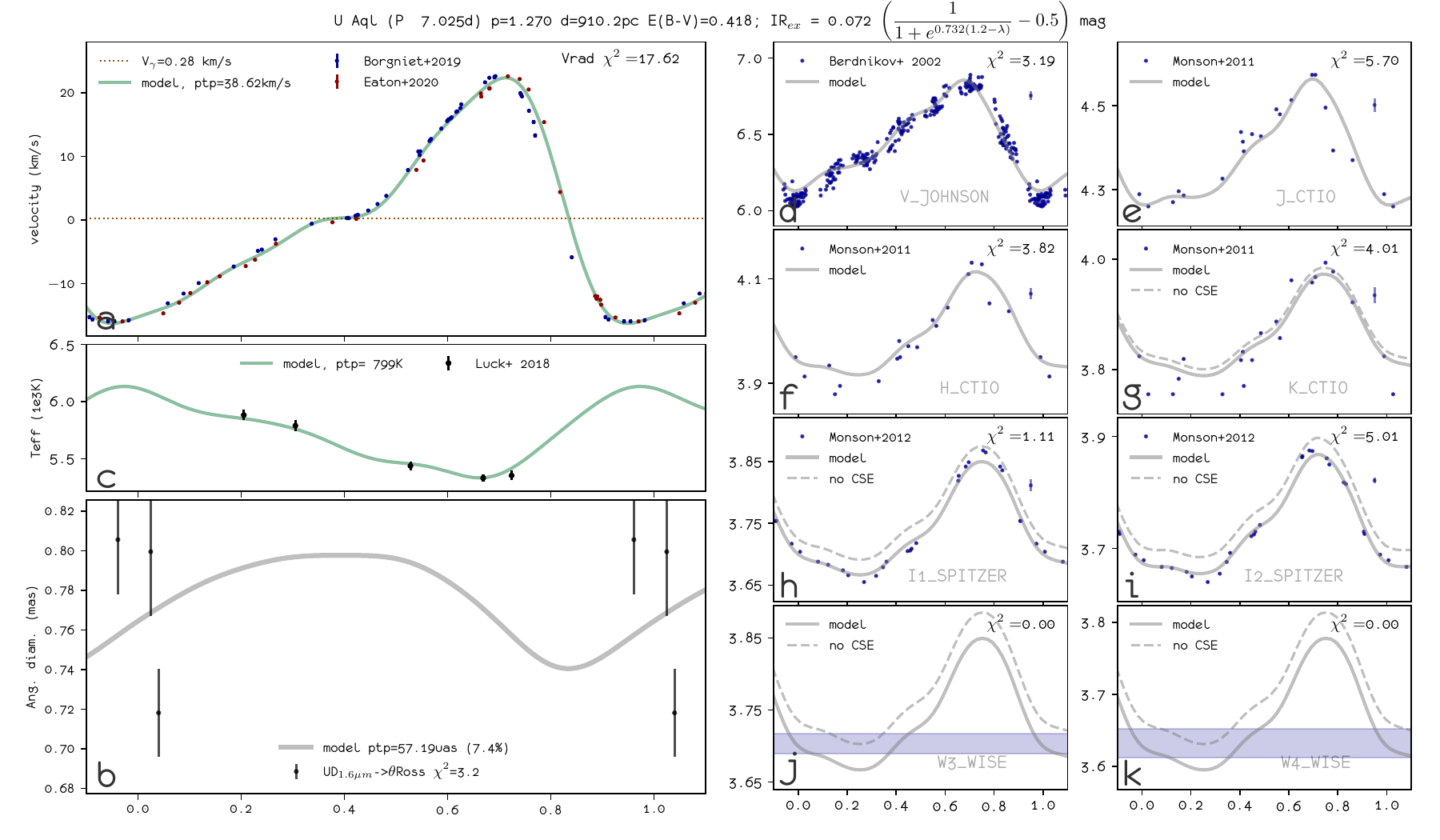}
\caption{\label{u_aql_spips}\small The SPIPS results of U Aql.}
\end{center}
\end{figure*}

\begin{figure*}[h]
\begin{center}
\includegraphics[scale=0.55]{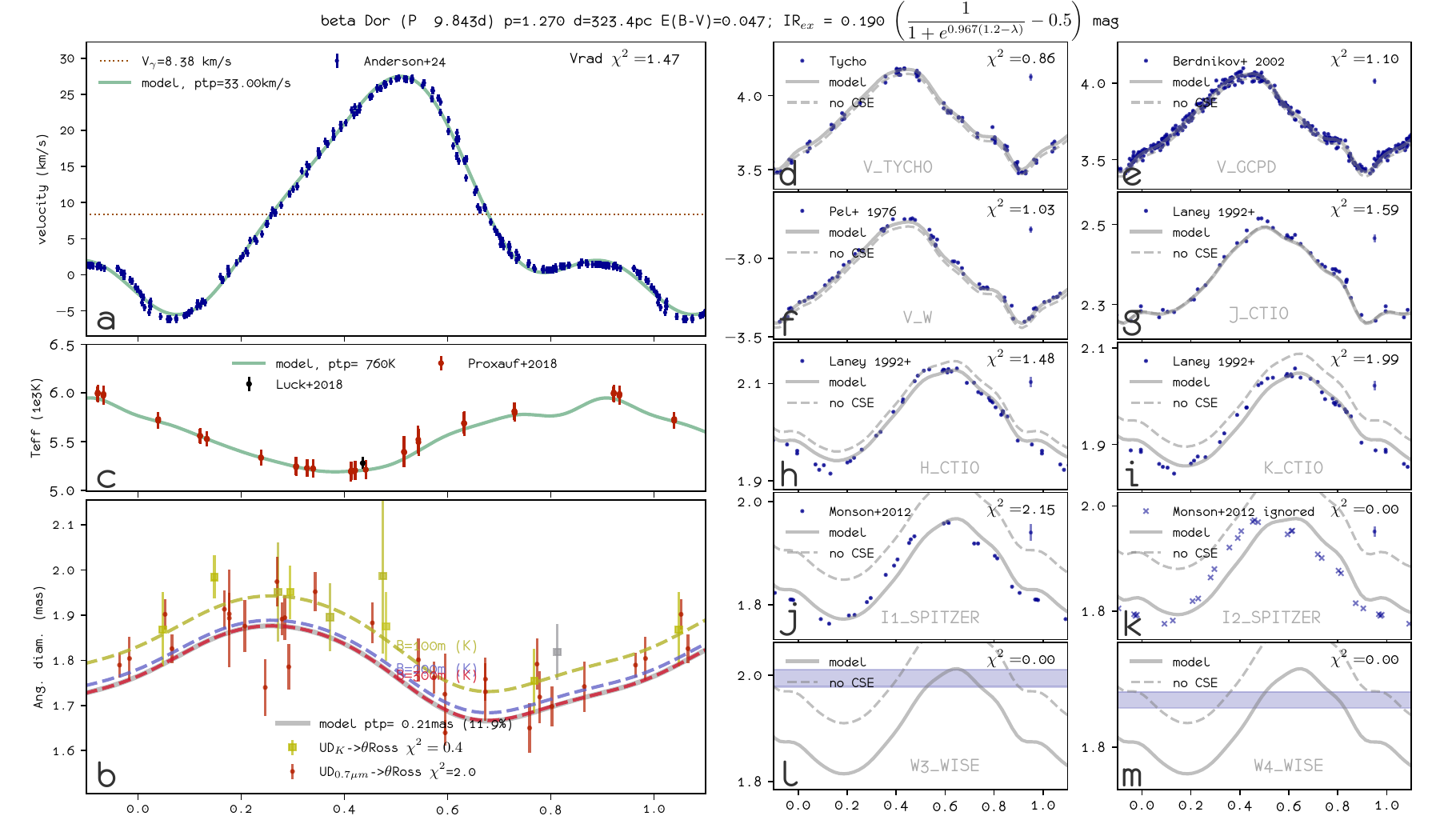}
\caption{\label{fig:spips_beta_dor}\small The SPIPS results of $\beta$ Dor.}
\end{center}
\end{figure*}

\begin{figure*}[h]
\begin{center}
\includegraphics[scale=0.55]{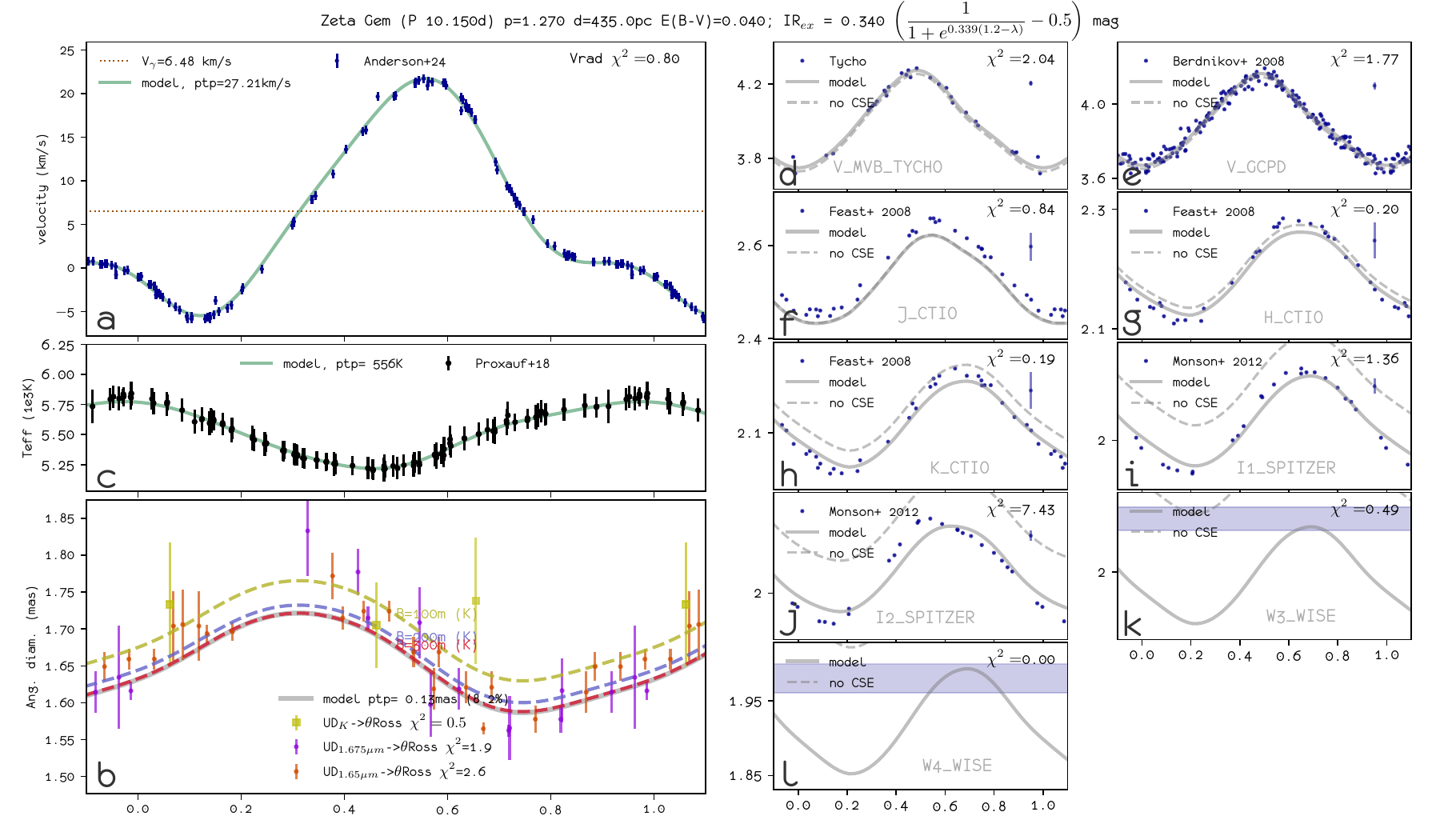}
\caption{\label{fig:zeta_gem_spips}\small The SPIPS results of $\zeta$ Gem.}
\end{center}
\end{figure*}

\begin{figure*}[h]
\begin{center}
\includegraphics[scale=0.55]{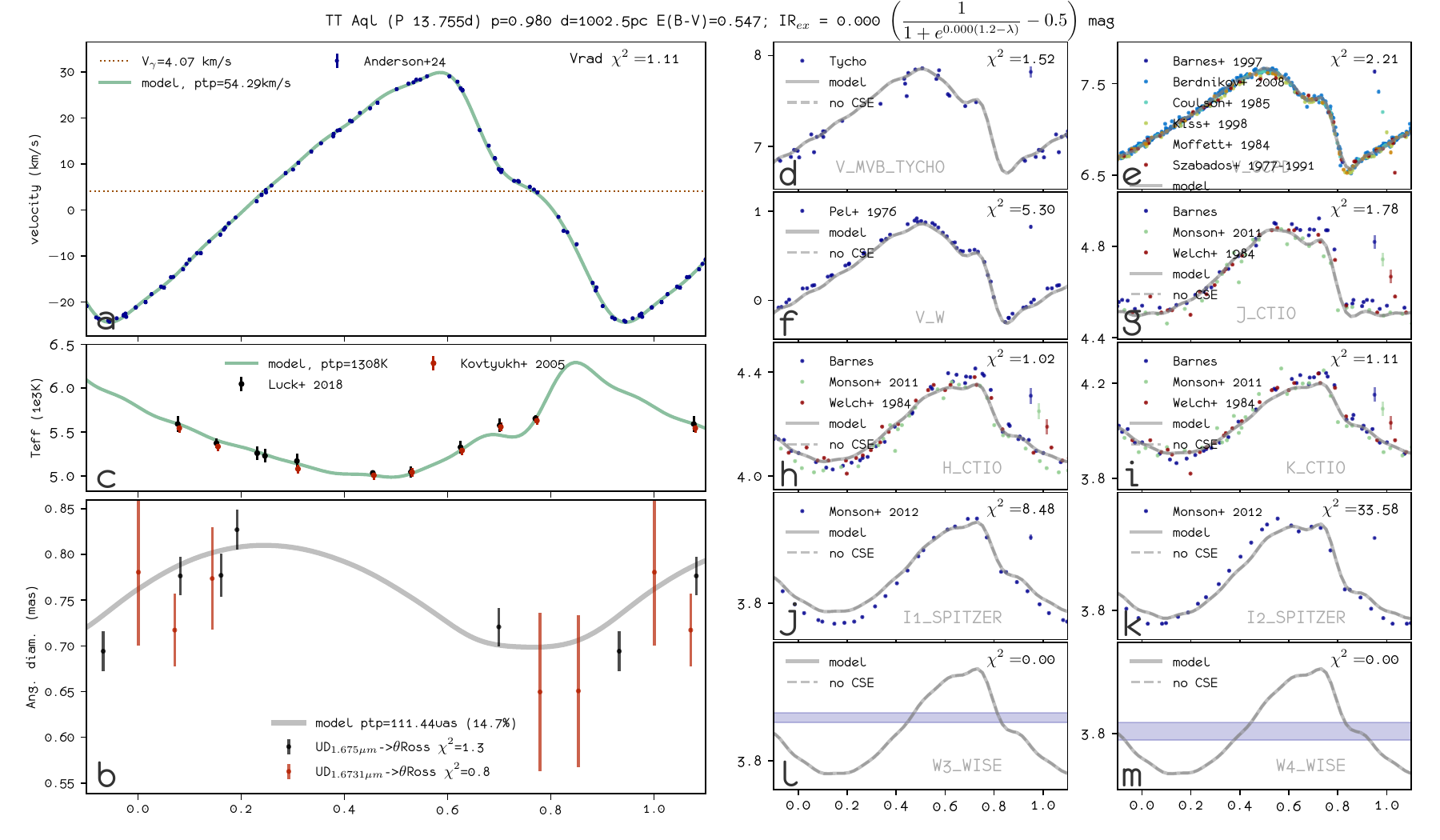}
\caption{\label{fig:tt_aql_spips}\small The SPIPS results of TT Aql.}
\end{center}
\end{figure*}

\begin{figure*}[h]
\begin{center}
\includegraphics[scale=0.55]{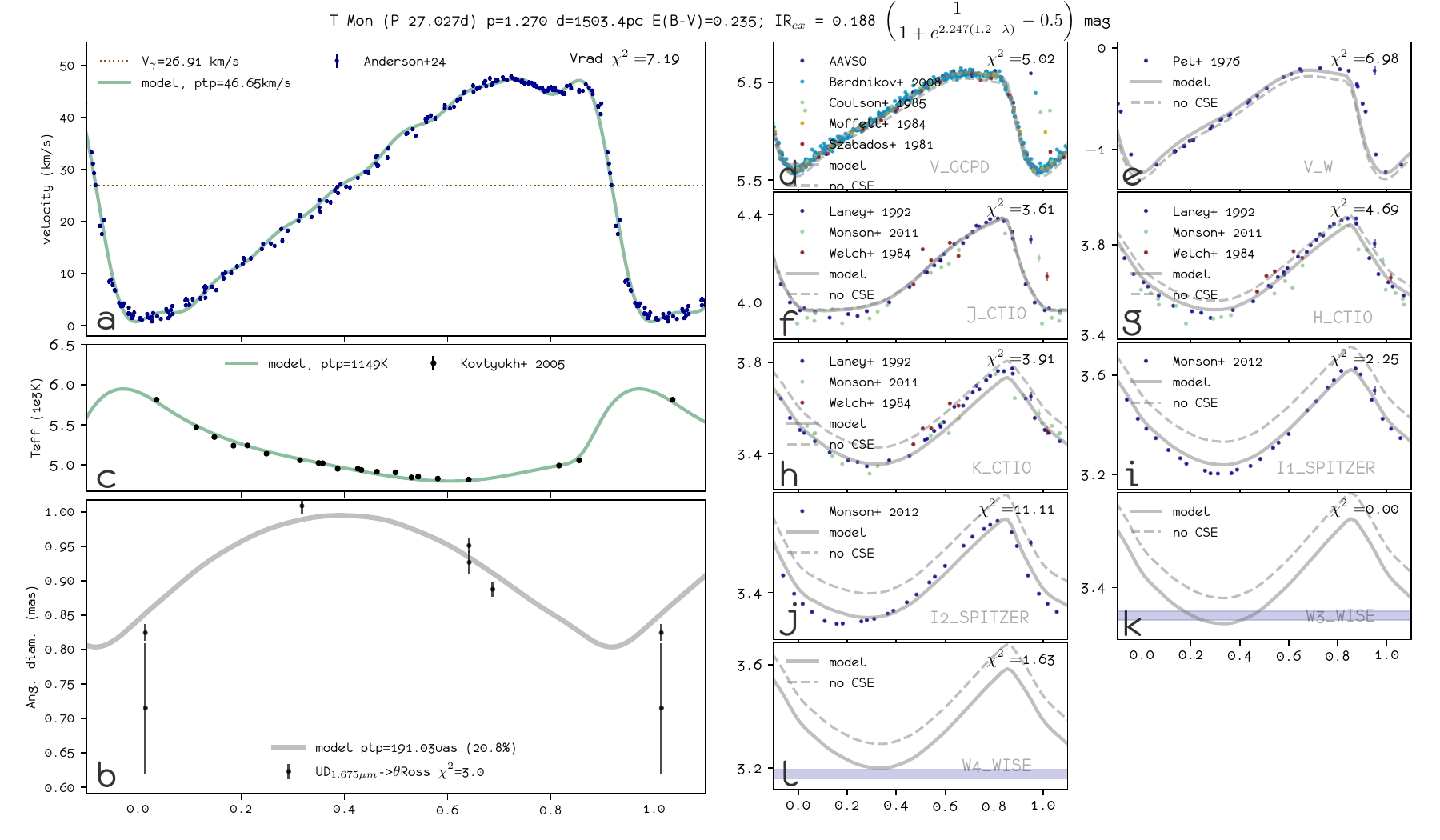}
\caption{\label{t_mon_spips}\small The SPIPS results of T Mon.}
\end{center}
\end{figure*}

\begin{figure*}[h]
\begin{center}
\includegraphics[scale=0.55]{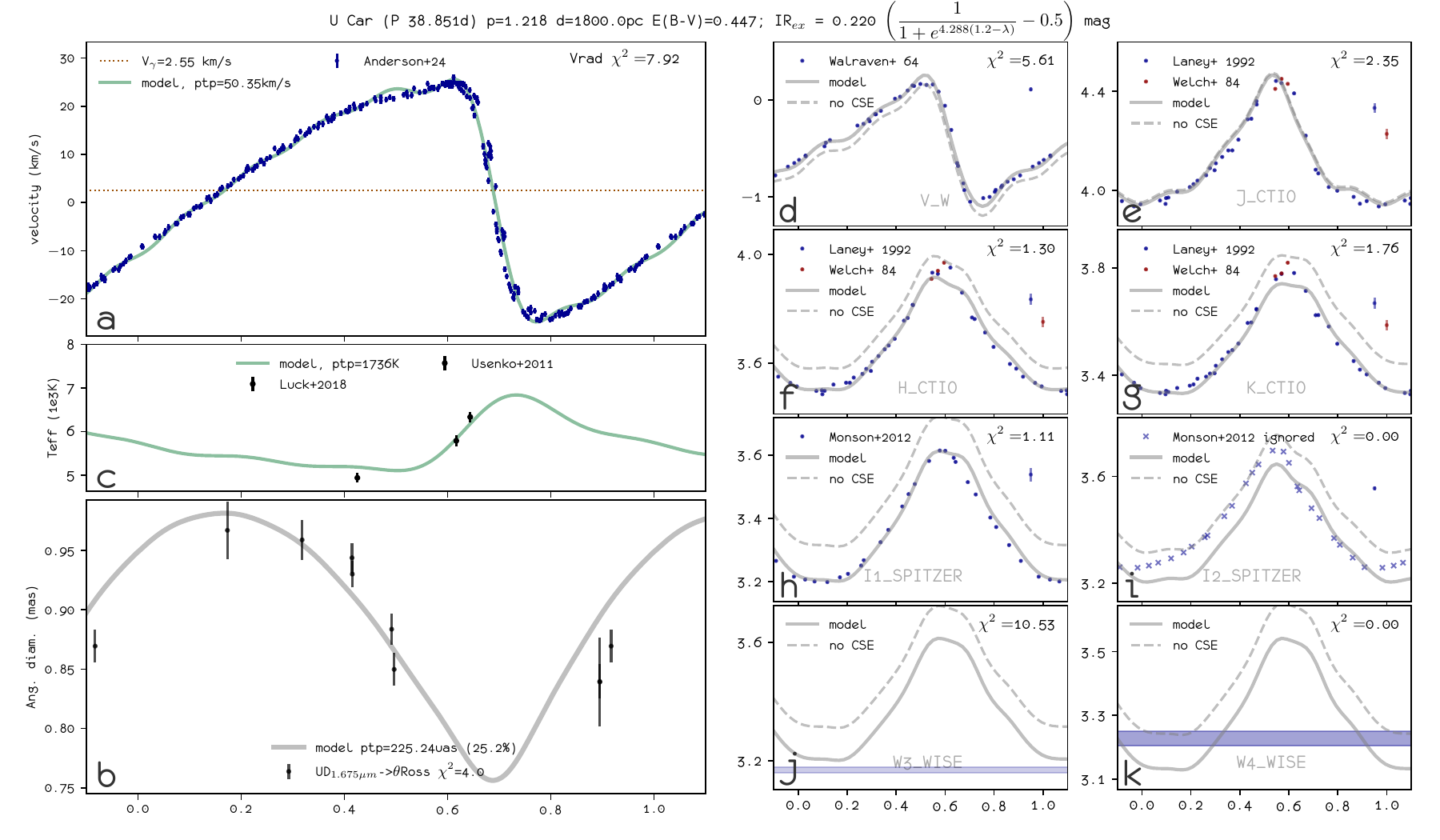}
\caption{\label{u_car_spips}\small The SPIPS results of U Car.}
\end{center}
\end{figure*}

\section{Closure phase observations}
\begin{figure*}[htb]
\centering 
\begin{subfigure}[b]{.24\textwidth}
  \includegraphics[width=\linewidth]{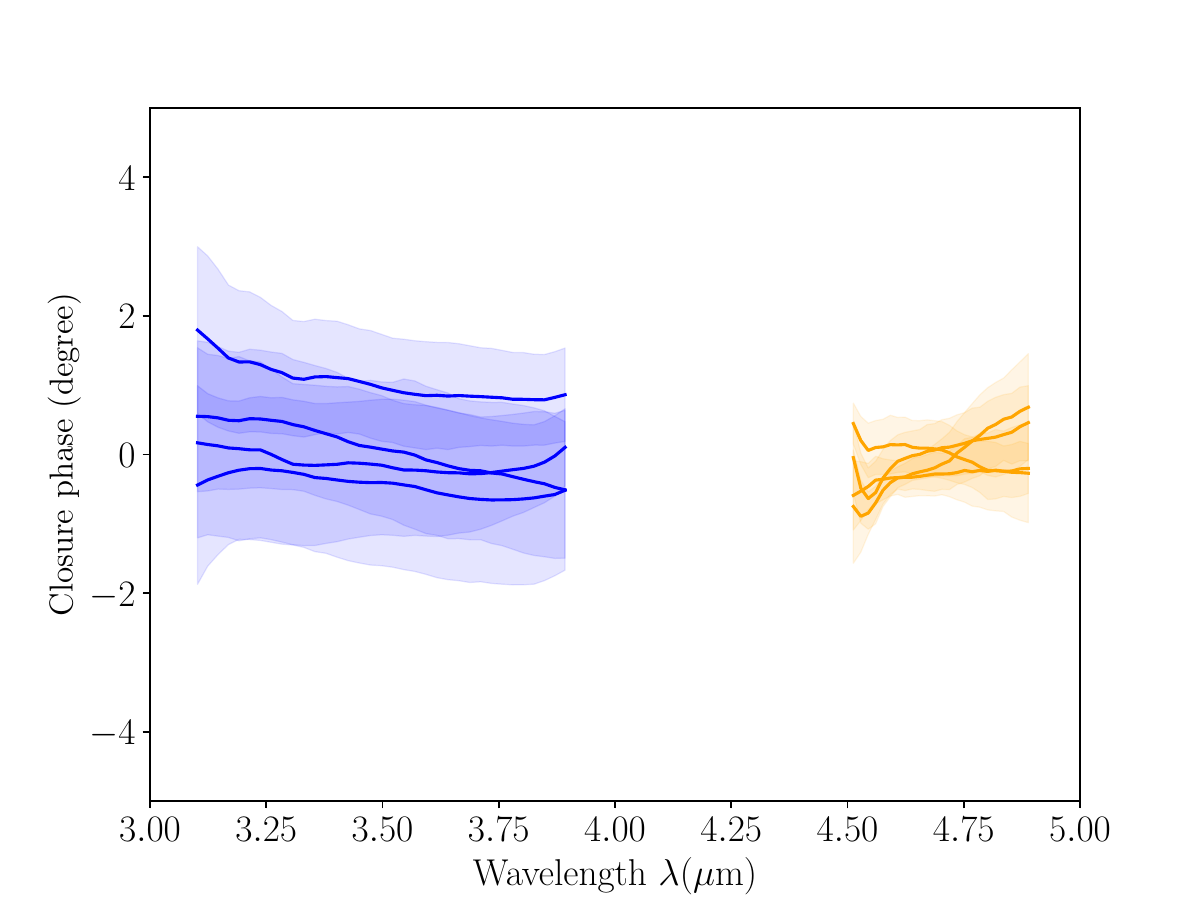}
  \caption{X Sgr, $LM$}
  \label{fig:4}
\end{subfigure}
\begin{subfigure}[b]{.24\textwidth}
  \includegraphics[width=\linewidth]{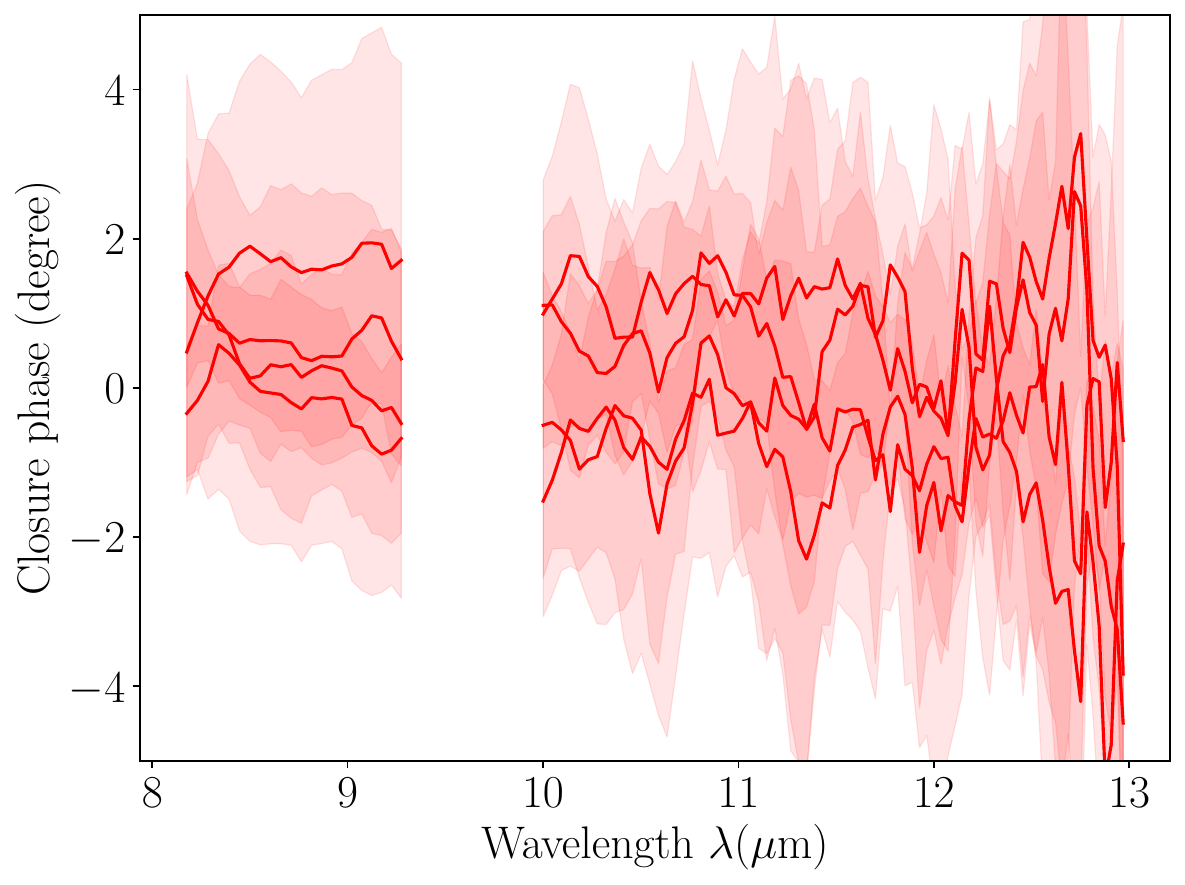}
  \caption{X Sgr, $N$}
  \label{fig:4}
\end{subfigure}
\begin{subfigure}[b]{.24\textwidth}
  \includegraphics[width=\linewidth]{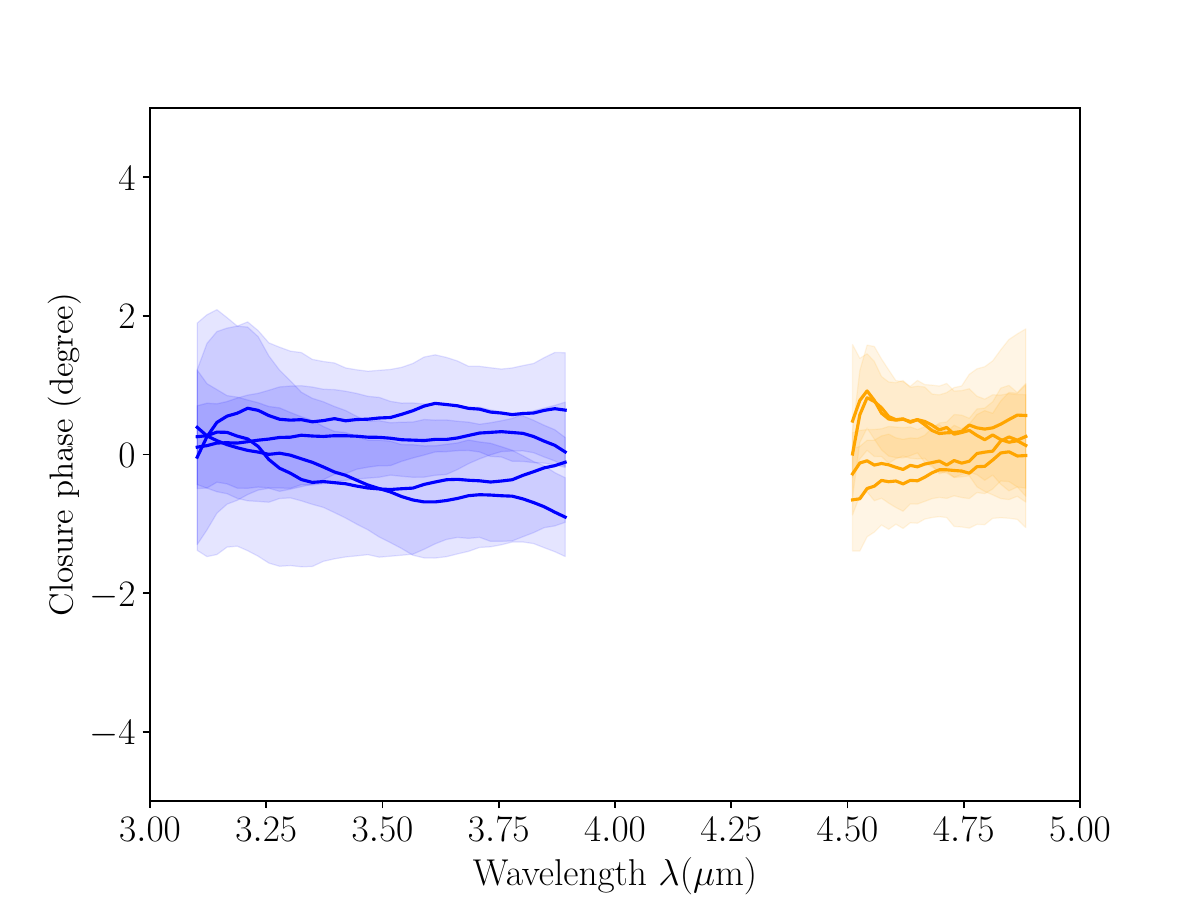}
  \caption{U Aql, $LM$}
  \label{fig:1}
\end{subfigure} 
\begin{subfigure}[b]{.24\textwidth}
  \includegraphics[width=\linewidth]{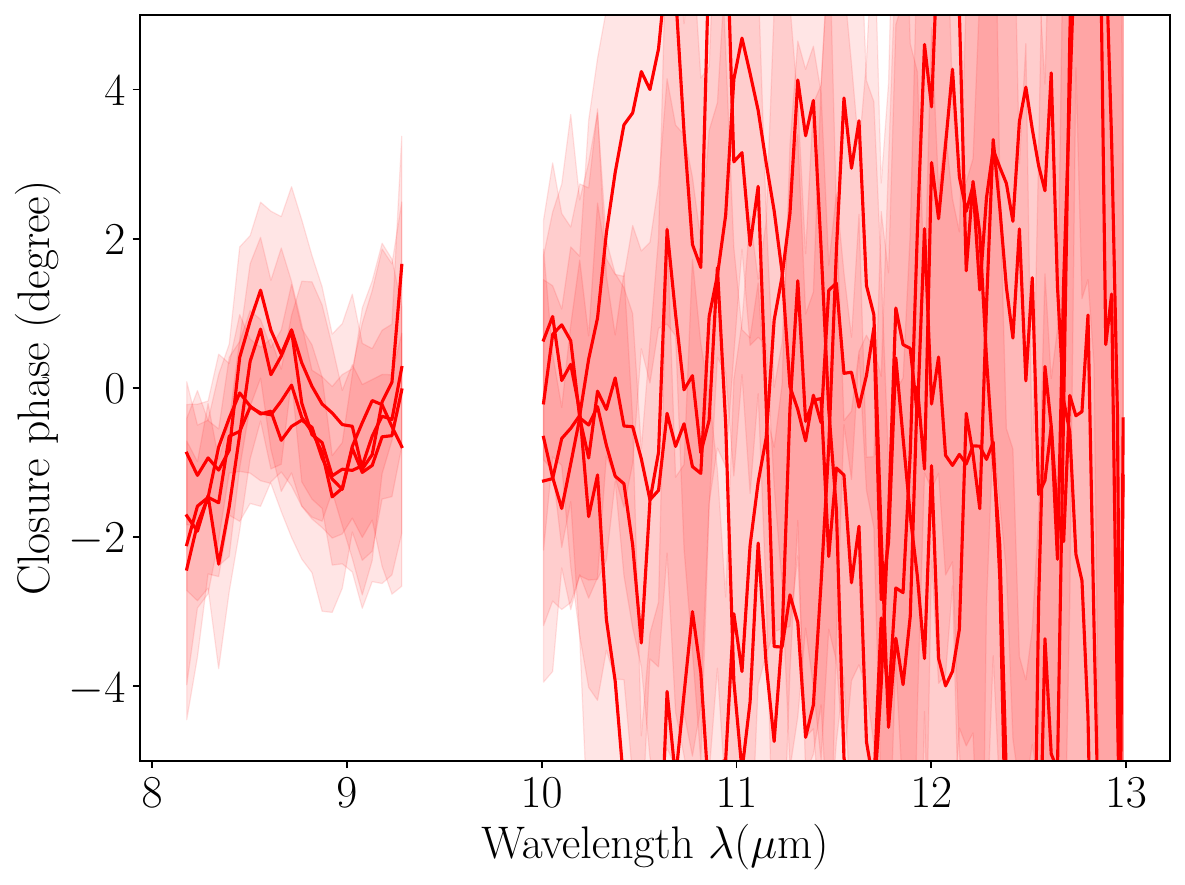}
  \caption{U Aql, $N$}
  \label{fig:4}
\end{subfigure}

\begin{subfigure}[b]{.24\textwidth}
  \includegraphics[width=\linewidth]{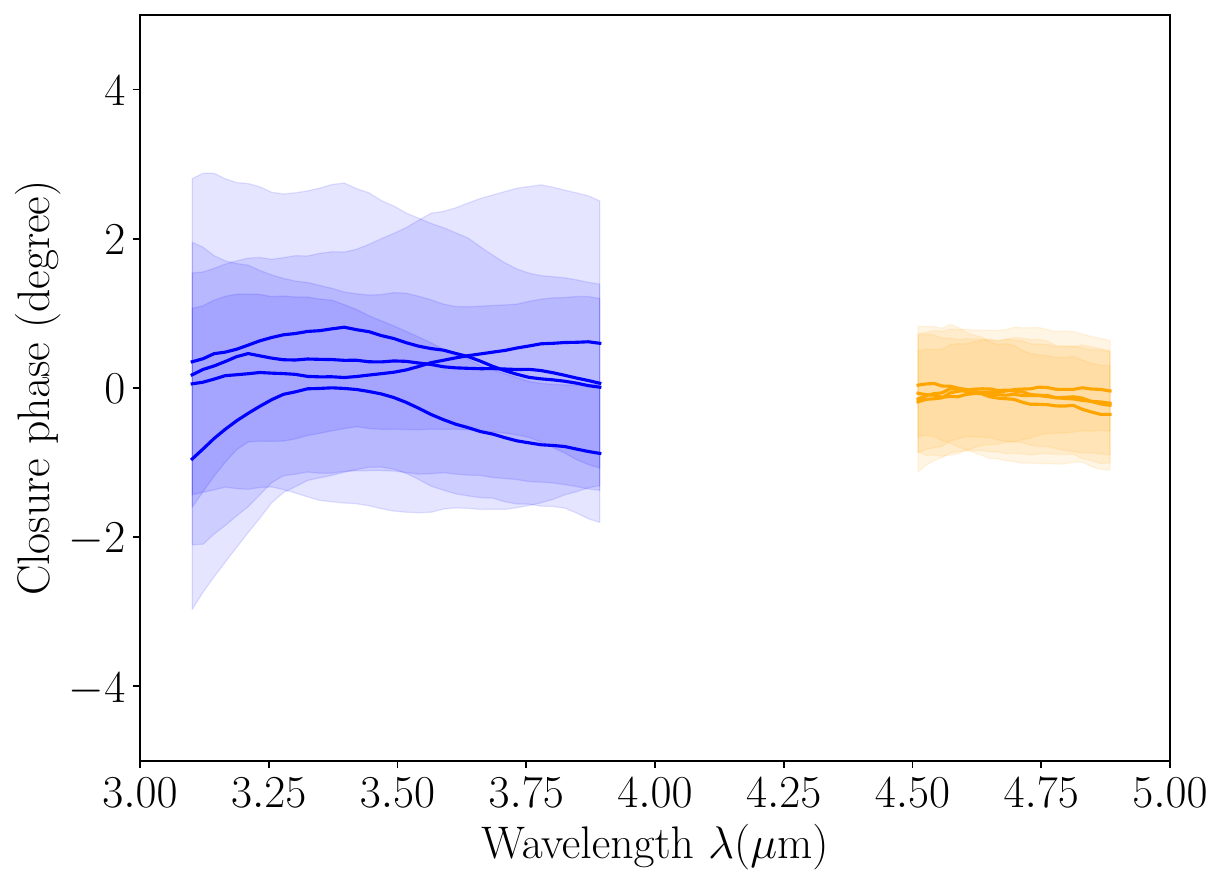}
  \caption{$\eta$ Aql, $LM$}
  \label{fig:1}
\end{subfigure} 
\begin{subfigure}[b]{.24\textwidth}
  \includegraphics[width=\linewidth]{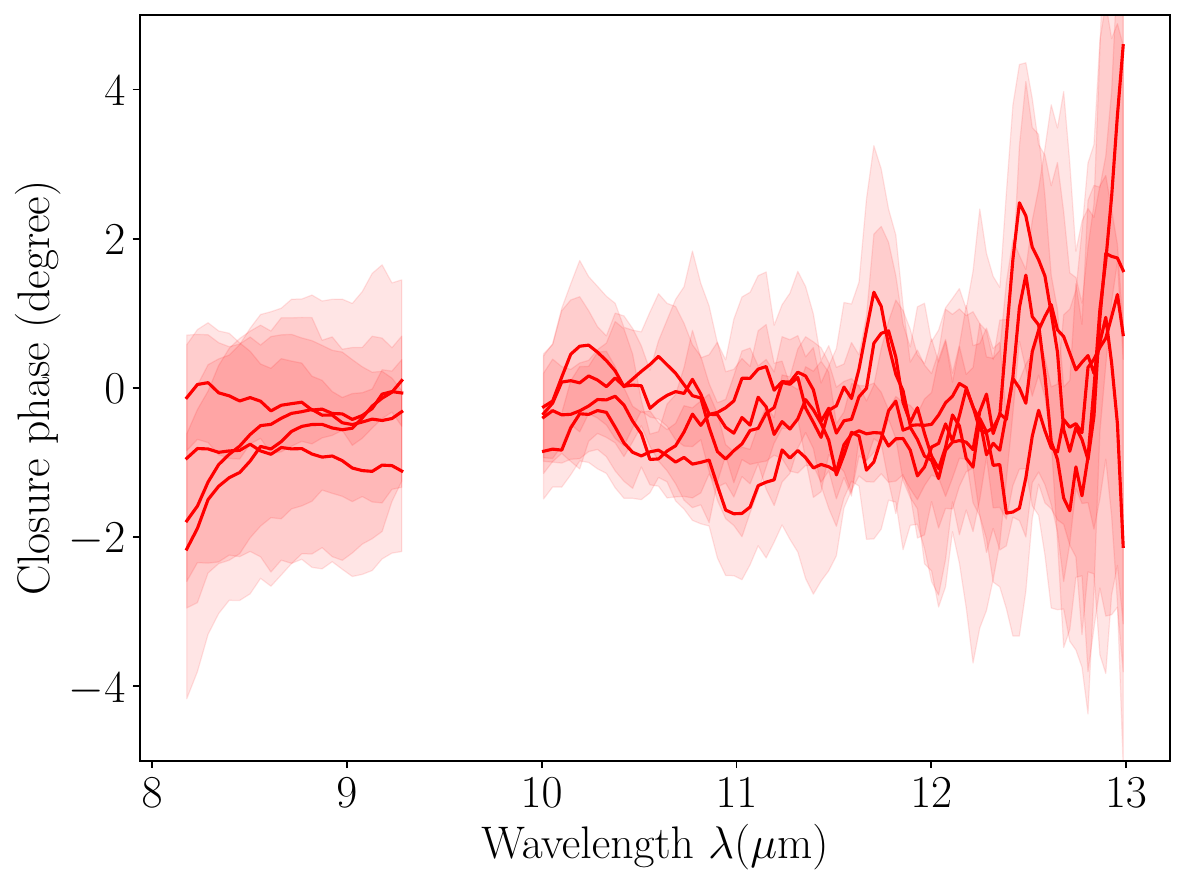}
  \caption{$\eta$ Aql, $N$}
  \label{fig:4}
\end{subfigure}
\begin{subfigure}[b]{.24\textwidth}
  \includegraphics[width=\linewidth]{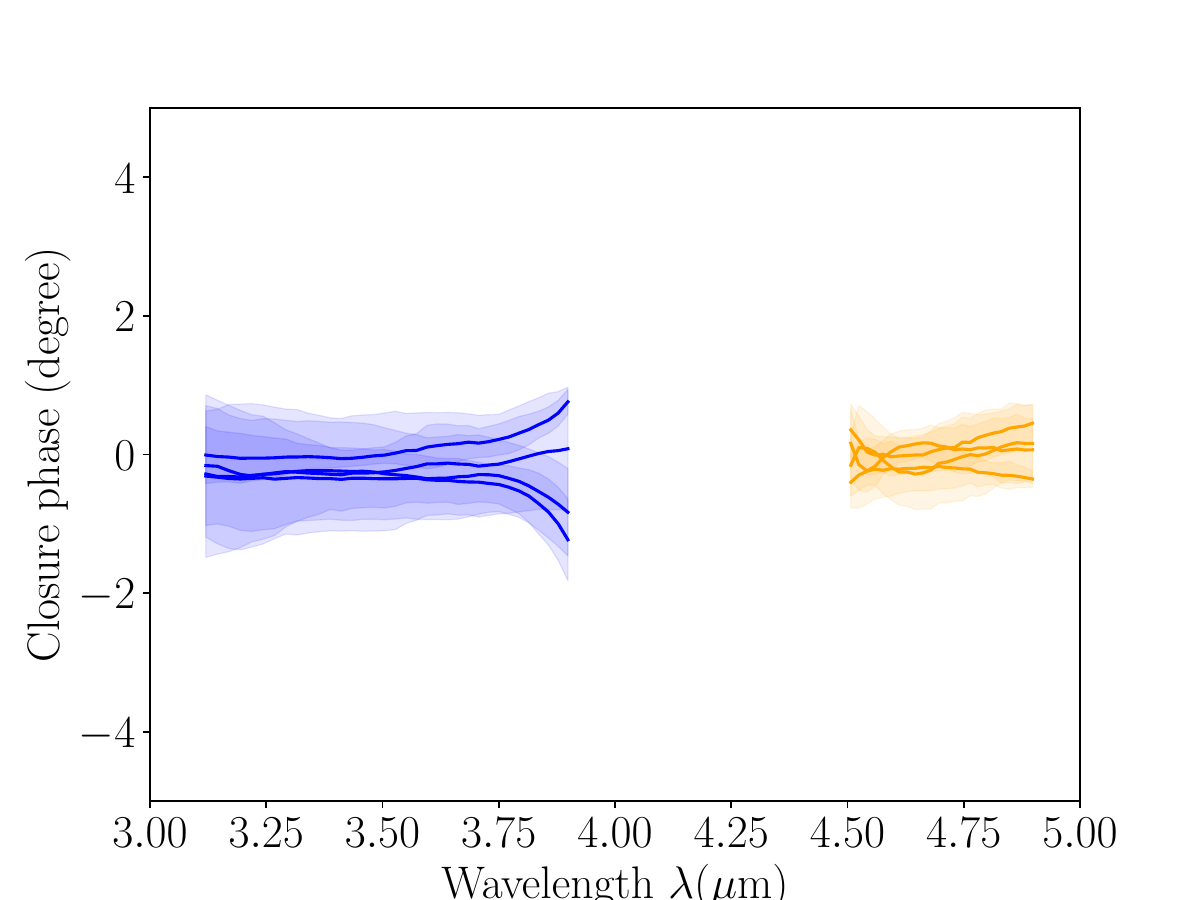}
  \caption{$\beta$ Dor, $LM$}
  \label{fig:4}
\end{subfigure}
\begin{subfigure}[b]{.24\textwidth}
  \includegraphics[width=\linewidth]{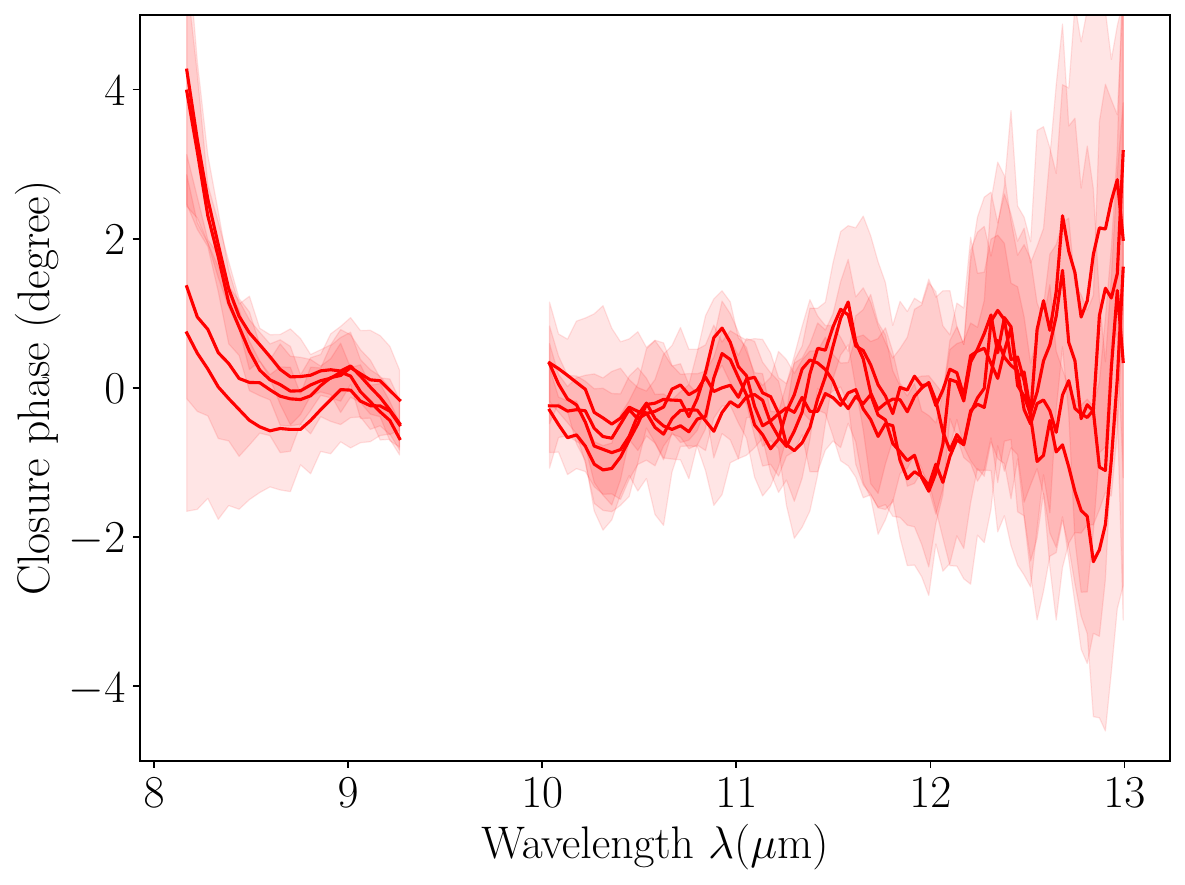}
  \caption{$\beta$ Dor, $N$}
  \label{fig:4}
\end{subfigure}

\begin{subfigure}[b]{0.24\textwidth}
  \includegraphics[width=\linewidth]{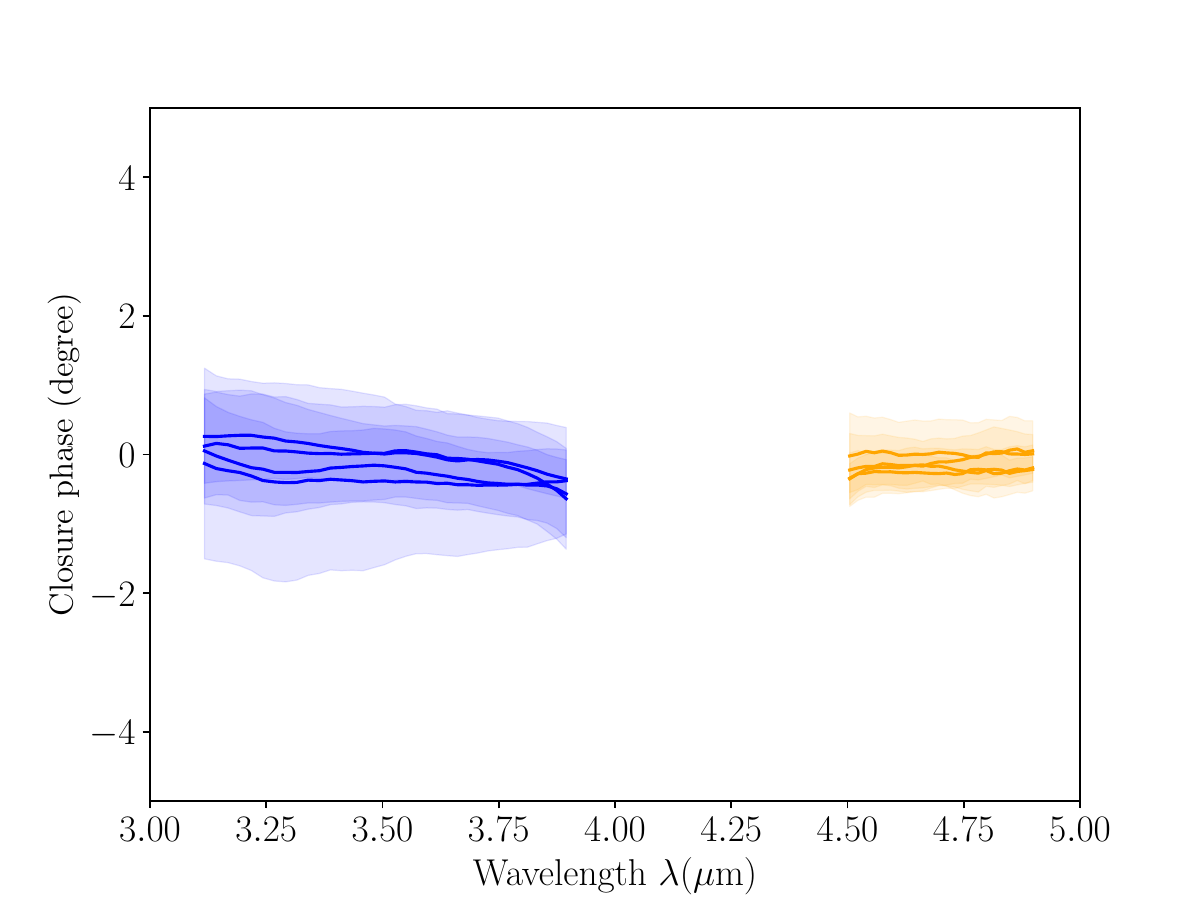}
  \caption{$\zeta$ Gem, $LM$}
  \label{fig:6}
\end{subfigure}
\begin{subfigure}[b]{0.24\textwidth}
  \includegraphics[width=\linewidth]{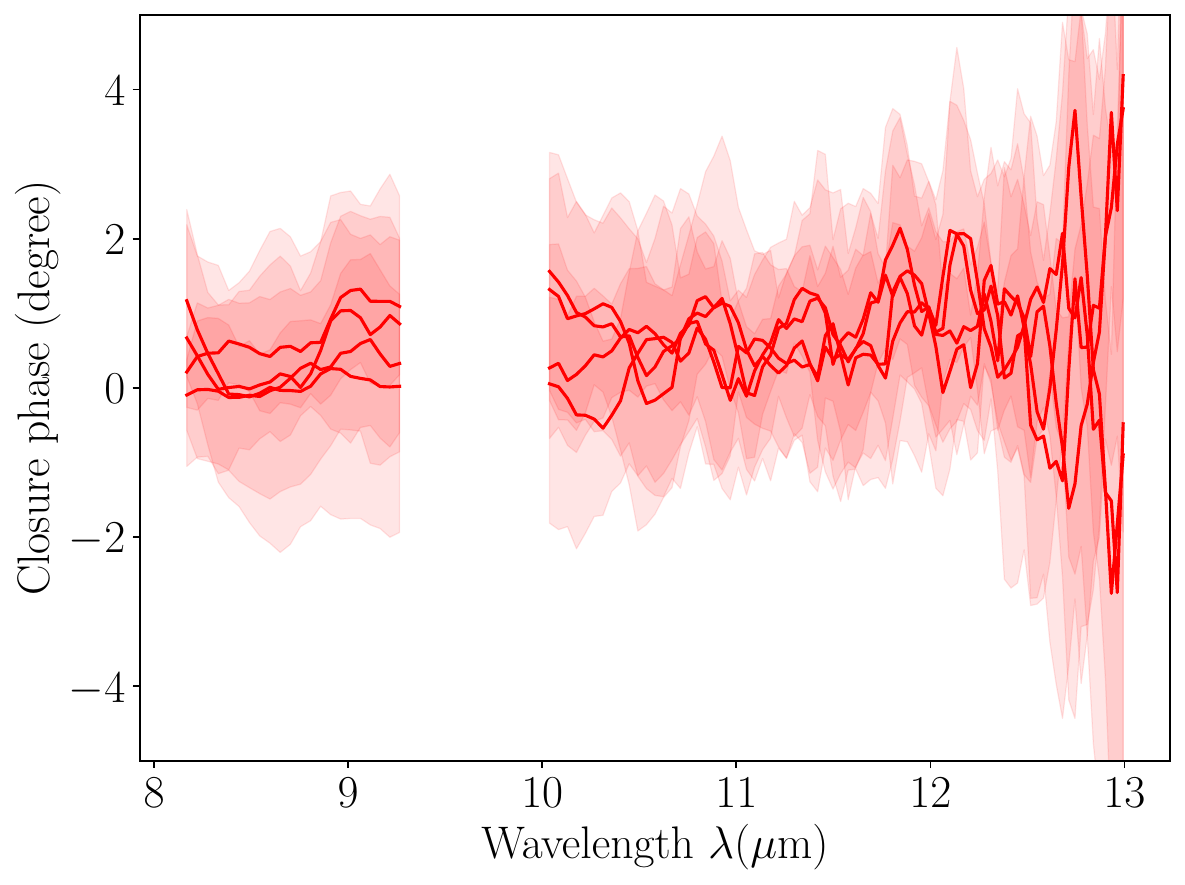}
  \caption{$\zeta$ Gem, $N$}
  \label{fig:6}
\end{subfigure}
\begin{subfigure}[b]{0.24\textwidth}
  \includegraphics[width=\linewidth]{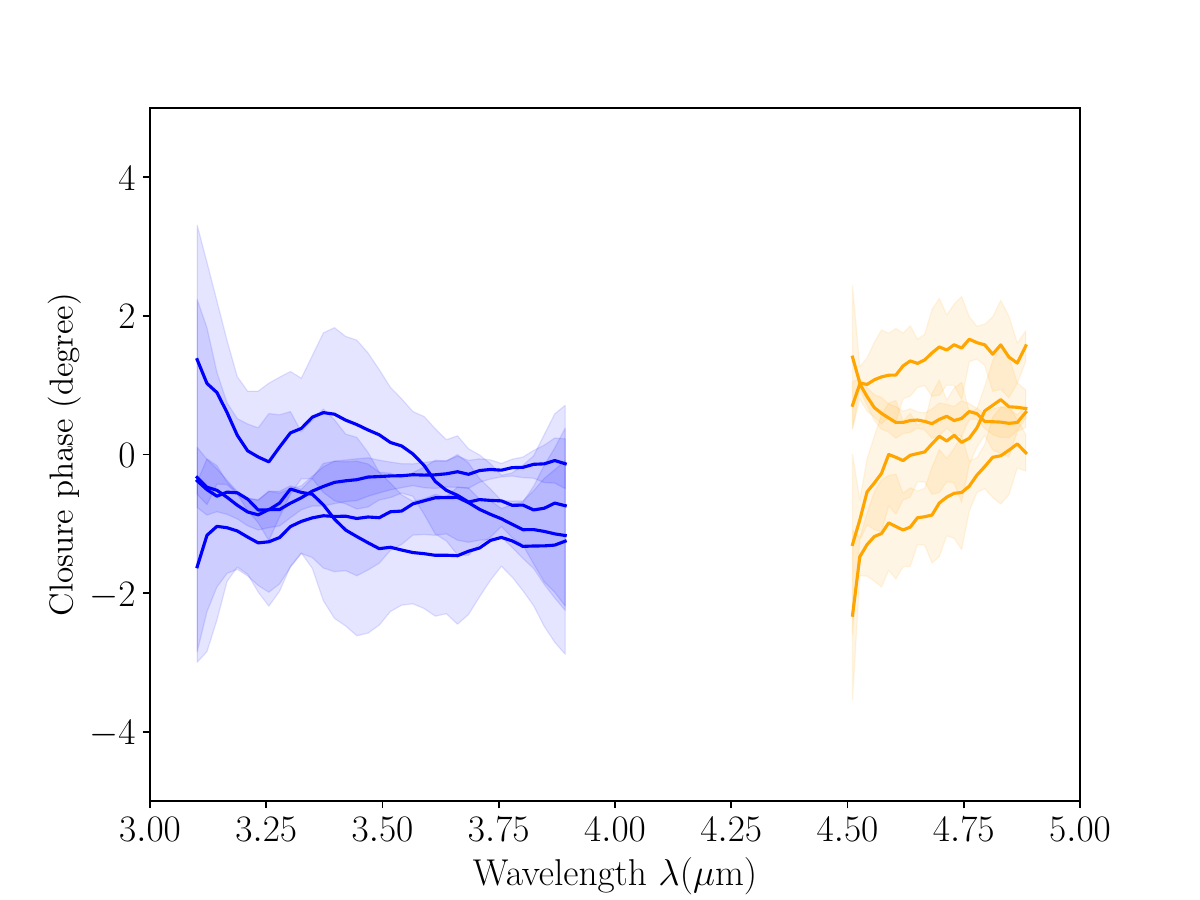}
  \caption{TT Aql, $LM$}
  \label{fig:6}
\end{subfigure}
\begin{subfigure}[b]{0.24\textwidth}
  \includegraphics[width=\linewidth]{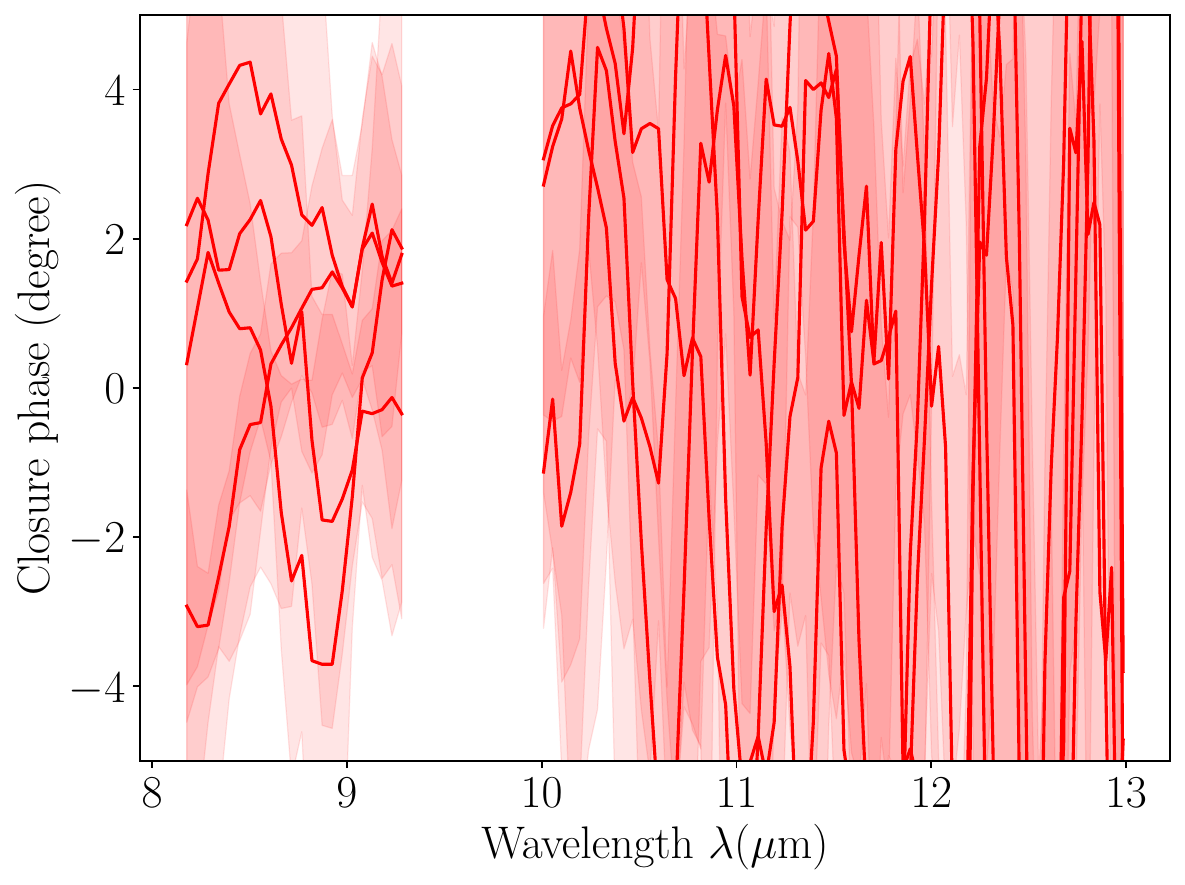}
  \caption{TT Aql, $N$}
  \label{fig:6}
\end{subfigure}

\begin{subfigure}[b]{0.24\textwidth}
  \includegraphics[width=\linewidth]{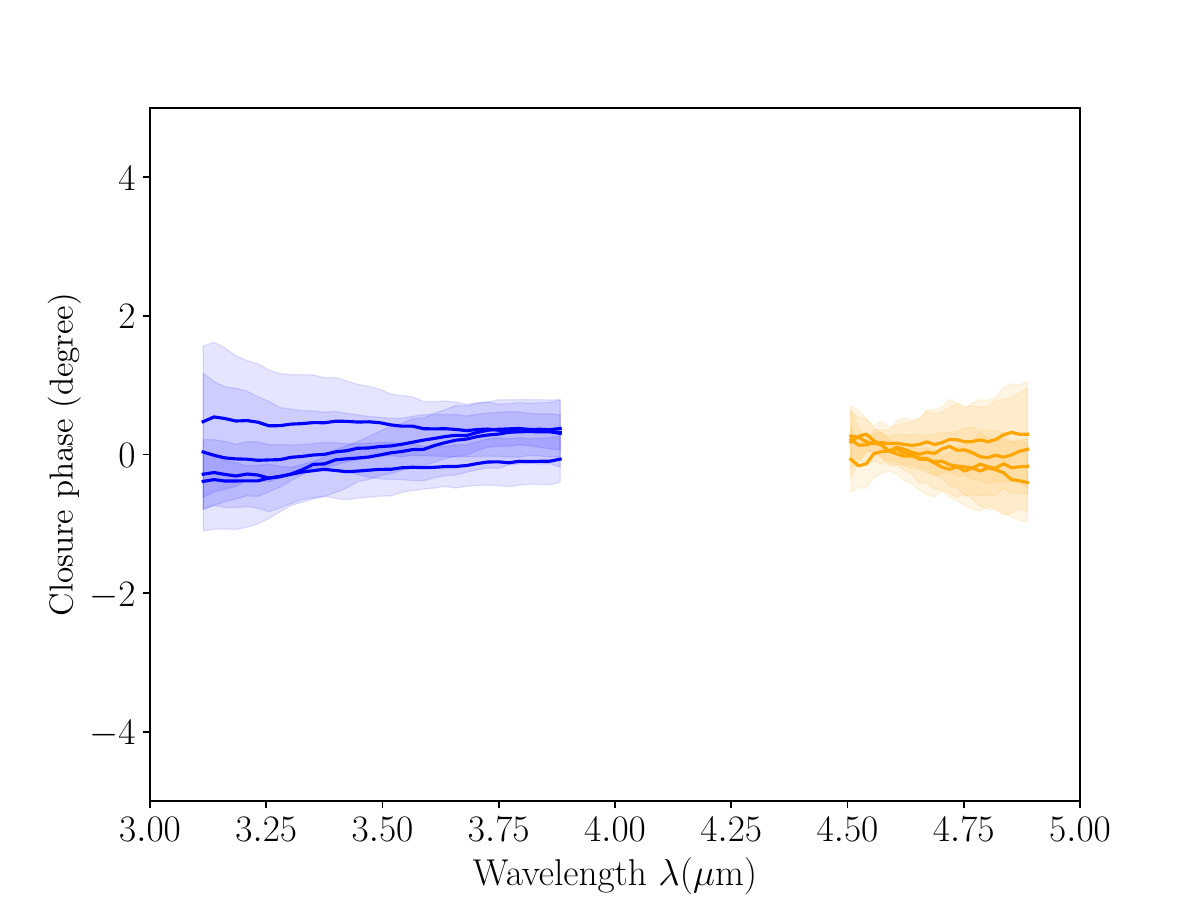}
  \caption{T Mon, $LM$}
  \label{fig:6}
\end{subfigure}
\begin{subfigure}[b]{0.24\textwidth}
  \includegraphics[width=\linewidth]{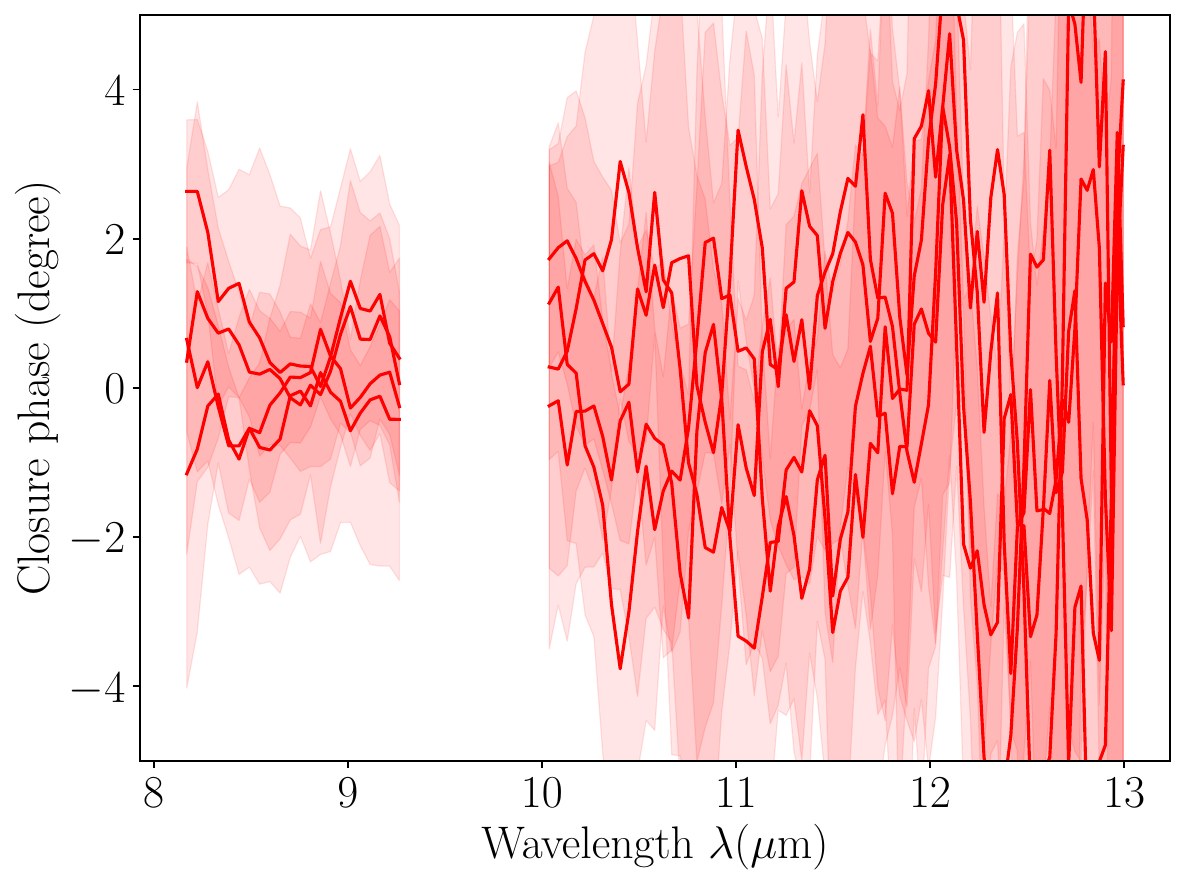}
  \caption{T Mon, $N$}
  \label{fig:6}
\end{subfigure}
\begin{subfigure}[b]{0.24\textwidth}
  \includegraphics[width=\linewidth]{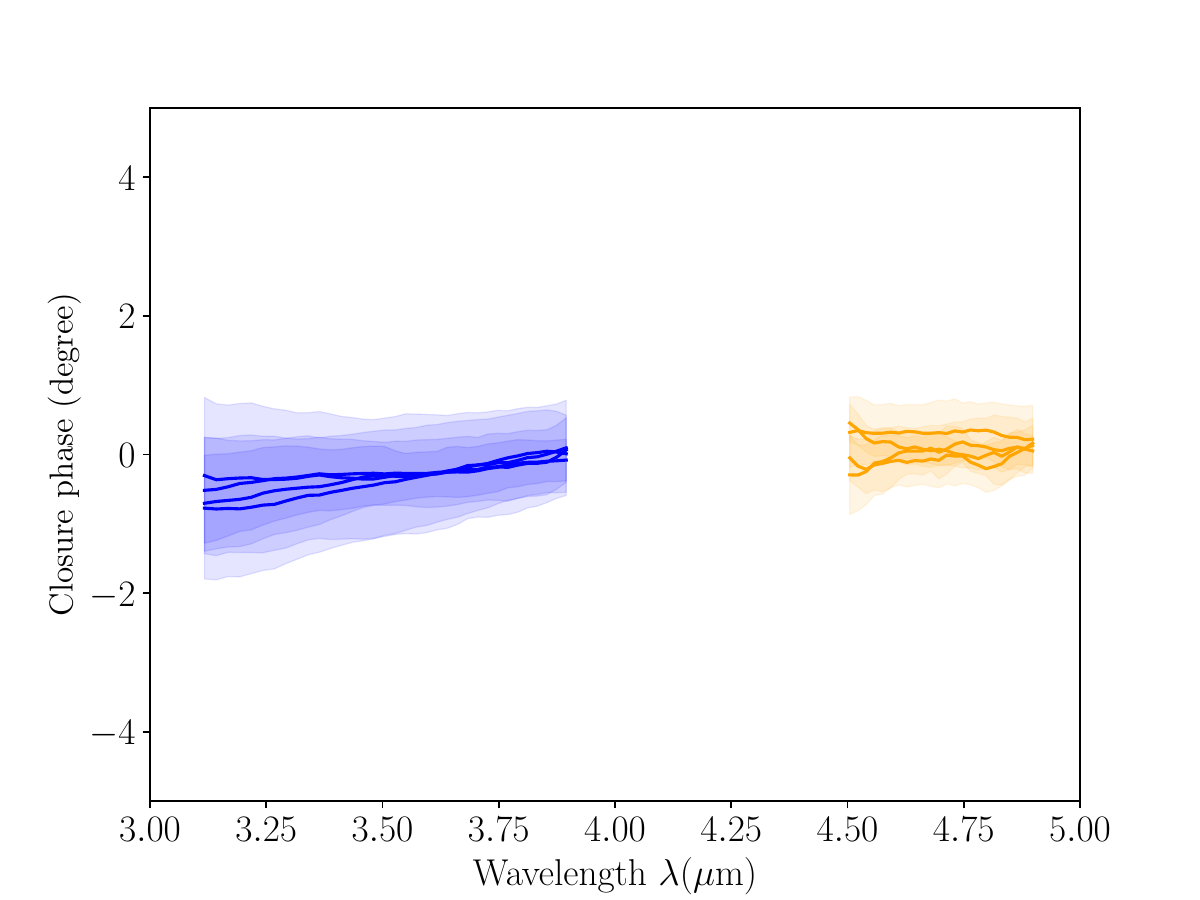}
  \caption{U Car, $LM$}
  \label{fig:6}
\end{subfigure}
\begin{subfigure}[b]{0.24\textwidth}
  \includegraphics[width=\linewidth]{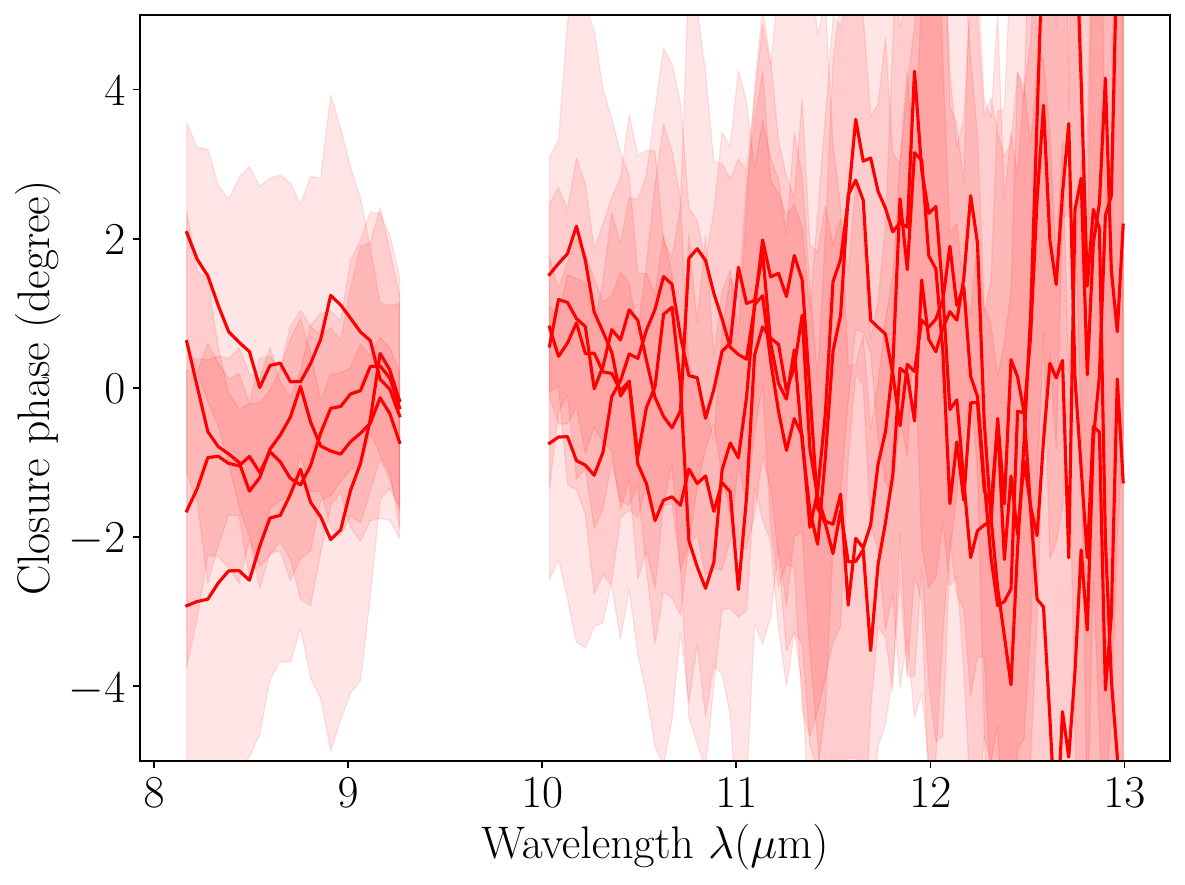}
  \caption{U Car, $N$}
  \label{fig:6}
\end{subfigure}
\caption{Closure phase in $L$,$M$ and $N$ bands for each triplet combination.}
 \label{fig:T3}
\end{figure*}

\end{appendix}

\end{document}